\documentclass[a4paper,fleqn,usenatbib]{mnras}
\pdfoutput=1

\usepackage{newtxtext,newtxmath}
\usepackage[T1]{fontenc}
\usepackage{ae,aecompl}

\usepackage{graphicx}	% Including figure files
\usepackage{amsmath}	% Advanced maths commands
\usepackage{amssymb}	% Extra maths symbols
\usepackage{multirow}
\usepackage{gensymb}
\usepackage{booktabs}
\usepackage{pdflscape}
\usepackage[flushleft]{threeparttable}

\newcommand{\hb}{H$\beta$}
\newcommand{\ha}{H$\alpha$}

\newcommand{\oiii}{[\ion{O}{iii}]}
\newcommand{\oii}{[\ion{O}{ii}]}
\newcommand{\nii}{[\ion{N}{ii}]}
\newcommand{\sii}{[\ion{S}{ii}]}

\newcommand{\esc}{erg cm$^{-2}$ s$^{-1}$}

\newcommand{\myr}{$M_{\odot}$ yr$^{-1}$}
\newcommand{\msun}{$M_{\odot}$}
\newcommand{\mstar}{$M_{\star}$}

\title[Predicting emission line fluxes and number counts of distant galaxies for cosmological surveys
]{Predicting emission line fluxes and number counts of distant galaxies for cosmological surveys}

\author[F. Valentino et al.]{
F. Valentino$^{1,2}$,\thanks{E-mail: francesco.valentino@nbi.ku.dk}
 E. Daddi$^{2}$,
J. D. Silverman$^{3}$,
A. Puglisi$^{4,5}$,
D. Kashino$^{6}$,
\newauthor
 A. Renzini$^{7}$,
A. Cimatti$^{8,9}$,
L. Pozzetti$^{10}$,
G. Rodighiero$^{4}$,
M. Pannella$^{11}$,
R. Gobat$^{12}$,
\newauthor
 G. Zamorani$^{10}$\\
\\ 
$^{1}$Dark Cosmology Centre, Niels Bohr Institute, University of
Copenhagen, Juliane Maries Vej 30, DK-2100 Copenhagen, Denmark\\
$^{2}$Laboratoire AIM-Paris-Saclay, CEA/DSM-CNRS-Universit\'{e} Paris
Diderot, \\
Irfu/Service d'Astrophysique, CEA Saclay, Orme des Merisiers, F-91191
Gif sur Yvette, France\\
$^{3}$Kavli Institute for the Physics and Mathematics of the Universe,
Todai Institutes for Advanced Study, the University of Tokyo, \\
Kashiwa, Japan 277-8583 (Kavli IPMU, WPI)\\
$^{4}$Dipartimento di Fisica e Astronomia, Universit\`{a} di Padova,
Vicolo dell'Osservatorio 2, 35122 Padova, Italy\\
$^{5}$ESO, Karl-Schwarschild-Strasse 2, 85748 Garching bei M\"{u}nchen, Germany\\
$^{6}$Department of Physics, ETH Z\"{u}rich, Wolfgang-Pauli-strasse 27, CH-8093 Z\"{u}rich, Switzerland\\
$^{7}$INAF Osservatorio Astronomico di Padova, Vicolo dell'Osservatorio 5, I-35122 Padova, Italy\\
$^{8}$Department of Physics and Astronomy (DIFA), Universit\`{a} di Bologna, Via Gobetti 93/2- I-40129, Bologna, Italy\\
$^{9}$INAF - Osservatorio Astrofisico di Arcetri, Largo E. Fermi 5, I-50125, Firenze, Italy\\
$^{10}$INAF Osservatorio Astronomico di  Bologna, Via Gobetti 93/3- I-40129, Bologna, Italy\\
$^{11}$Department of Physics, Ludwig-Maximilians-Universit\"{a}t, Scheinerstr. 1, 81679 M\"{u}nchen, Germany\\
$^{12}$School of Physics, Korea Institute for Advanced Study, Hoegiro 85, Dongdaemun-gu, Seoul 02455, Republic of Korea}
 
\date{Accepted XXX. Received YYY; in original form ZZZ}

\pubyear{2017}

\hypersetup{draft}

\begin{document}
\label{firstpage}
\pagerange{\pageref{firstpage}--\pageref{lastpage}}
\maketitle

\begin{abstract}
We estimate the number counts of line emitters at high redshift and
their evolution with cosmic time based on a combination of photometry
and spectroscopy. We predict the \ha, \hb, \oii, and \oiii\ line fluxes
for more than $35,000$ galaxies down to stellar masses of
$\sim10^9$~\msun\ in the COSMOS and GOODS-S fields,
applying standard conversions and exploiting the spectroscopic
coverage of the FMOS-COSMOS survey at $z\sim1.55$ to calibrate the
predictions. We calculate the number counts of \ha, \oii, and \oiii\
emitters down to fluxes of $1\times10^{-17}$~\esc\ in the range $1.4 < z < 1.8$ covered by the
FMOS-COSMOS survey. We model the time evolution of the differential and 
cumulative \ha\ counts, steeply declining at the brightest fluxes. 
We expect $\sim9,300-9,700$ and $\sim2,300-2,900$ galaxies
deg$^{-2}$ for fluxes $\geq1\times10^{-16}$ and
$\geq2\times10^{-16}$~\esc\ over the range
$0.9<z<1.8$. We show that the observed evolution of the Main Sequence 
of galaxies with redshift is enough to reproduce the observed counts 
variation at $0.2<z<2.5$. We characterize the physical properties of
the \ha\ emitters with fluxes $\geq2\times10^{-16}$~\esc\, including
their stellar masses, UV sizes, \nii/\ha\
ratios, and \ha\ equivalent widths. An aperture of $R\sim R_{\rm e}\sim0.5$''
maximizes the signal-to-noise ratio for a detection, while causing a
factor of $\sim2\times$ flux losses, influencing the recoverable number counts,
if neglected. Our approach, based on deep and large photometric
datasets, reduces the uncertainties on the number counts due to the
selection and spectroscopic samplings, while exploring low fluxes. We
publicly release the line flux predictions for the explored
photometric samples.
\end{abstract}


%%%%%%%%%%%%%%%%%%%%%%%%%%%%%%%%%%%%%%%%%%%%%%%%%%

%%%%%%%%%%%%%%%%% BODY OF PAPER %%%%%%%%%%%%%%%%%%

\section{Introduction}
\label{sec:introduction}
As supported by several independent pieces of evidence, mysterious
``dark'' components dominate the mass and energy budget of the
Universe, adding up to $\sim96$\% of the total energy density in the current
$\Lambda$CDM cosmological framework. In particular, a ``dark
energy'' is considered the engine of the accelerated expansion of the
Universe, as suggested by and investigated through the study of supernovae in
galaxies up to $z\sim1$ \citep{riess_1998, schmidt_1998,
  perlmutter_1999}. On the other hand, ``dark matter'' counteracts the
effect of dark energy, braking the expansion via the gravitational
interaction. As a result, the geometry of our Universe is regulated by the
delicate compromise between these two components.\\
The distribution of galaxies on large scales offers crucial insights
on the nature of both these dark components and constitutes a test for
the theory of General Relativity, one of the pillars of modern
physics. In particular, wiggle patterns in galaxy clustering, the so
called Baryonic Acoustic Oscillations (BAOs), provide a standard ruler
to measure the stretch and geometry of the Universe and to put
constraints on dark energy independently of the probe provided by
supernovae. However, the detection of the BAOs is bound to the
precision with which we derive the position of galaxies in the
three-dimensional space and to the collection of vast samples of
objects. The necessity of accurate redshifts to detect BAOs is motivating
the launch of intense spectroscopical campaigns to pinpoint millions of
galaxies in the sky both from the ground (i.e., BOSS, WiggleZ, and the
forthcoming Prime Focus Spectrograph (PFS), Dark Energy Spectroscopic
Instrument (DESI), and Multi-Object Optical Near-infrared Spectrograph (MOONS) surveys, \citealt{dawson_2013,
  blake_2011, takada_2014, levi_2013, cirasuolo_2014}) and in space with dedicated
missions, such as Euclid \citep{laureijs_2009} and WFIRST
\citep{green_2012,spergel_2015}. In particular, taking full advantage of high-precision imaging and
absence of atmospheric absorption, the space missions will probe
critical epochs up to $z\sim2$, when the dark energy starts manifesting its
strongest effects and accurate weak lensing measurements can map the
distribution of dark matter in the Universe. Observationally, these
missions will apply a slitless spectroscopy technique to estimate
redshifts from bright nebular lines and, notably, from \ha\ emission, a primary tracer of hydrogen,
generally ionized by young O- and B-type stars or active galactic nuclei
(AGN). Moreover, even if at low resolution, the spectroscopic characterization
of such a large sample of star-forming and active galaxies will be a gold
mine for the study of galaxy evolution over time.
Therefore, a prediction of the number of potentially observable galaxies is
required to optimize the survey strategies, in order to have the maximal scientific 
return from these missions.\\
As typically done, the
predicted number counts over wide redshift intervals are determined modeling the evolution of the
luminosity function (LF) of \ha\ emitters, reproducing the available
samples of spectroscopic and narrow-band imaging datasets
\citep[][and references therein]{geach_2010, colbert_2013, mehta_2015, sobral_2015, pozzetti_2016}. However, this method
generally relies on empirical extrapolations of the time evolution of
the parameters describing the LF, and it is bound to limited
statistics. Observationally, narrow-band imaging surveys benefit from
the large sky areas they can cover, at the cost of significant
contamination issues and the thin redshift slices probed, making them
prone to the uncertainties due to cosmic variance. On the other hand,
despite the limited covered areas, spectroscopic surveys directly
probe larger redshift intervals, combing large cosmic volumes,
reducing the impact of cosmic variance. Here we propose an alternative method based on photometry of
star forming galaxies (SFGs), covering their whole
Spectral Energy Distribution (SED), in synergy with spectroscopy for a
subsample of them. We
show that spectroscopic observations allow for an accurate calibration
of the \ha\ fluxes expected for
typical Main-Sequence SFGs \citep[MS,][]{noeske_2007,
  daddi_2007}. As a consequence, we can take advantage of much larger photometric samples of galaxies currently available
in cosmological fields to estimate the number counts of line
emitters. We test the validity of this approach exploiting large
photometric samples in the COSMOS and GOODS-S fields, and calibrating the
\ha\ flux predictions against the FMOS-COSMOS survey at $z\sim 1.55$
\citep{silverman_2015}. Flux predictions for the \ha\ and other
relevant emission lines (\oii$\lambda3727$~\AA, \hb$\,\lambda4861$~\AA,
and \oiii$\lambda5007$~\AA) and the photometric properties of this
sample are released in a catalog.
We, then, compute the number counts
of \ha, \oii, and \oiii\ emitters in the redshift range $1.4<z<1.8$
covered by the FMOS-COSMOS survey and we predict the evolution of the \ha\ counts with redshift,
modeling the evolution of the
normalization of the MS and including the effect of the luminosity
distance. We argue that this is enough to reproduce the observed
trends over the redshift range $0.2<z<2.5$.
Admittedly, this process relies on a few
assumptions and is affected by uncertainties and
limitations we discuss in the article, but it is physically motivated
and it has the general advantage of sensibly decreasing the errors due to low
number statistics, overcoming some of the observational limitations
of current spectroscopic surveys from the ground. It also benefits
from a better control of selection effects than studies based on
the detection of emission lines only. Coupled with the canonical approach based on the evolution of
the \ha\ LF, our method
strives to obtain a more solid estimate of the integrated \ha\
counts. 
Finally, we present a detailed physical characterization of the
brightest \ha\ emitters in terms of stellar mass, redshift distribution, dust extinction, nebular
line ratios, and \ha\ equivalent widths, key elements to prepare
realistic simulations of the primary population of galaxies
observable by forthcoming wide spectroscopic surveys. 

This paper is organized as follows: in Section \ref{sec:sample} we
present the photometric and the FMOS-COSMOS spectroscopic samples used to
estimate the number counts of emitters and calibrate the
prediction of line fluxes, respectively. In
Section \ref{sec:linepredictions} we introduce the procedure 
to calculate \ha, \hb, \oii, and \oiii\ fluxes. We characterize the photometric and spectroscopic
properties of a sample of bright \ha\ emitters visible in future surveys in
Section \ref{sec:properties}. In Section \ref{sec:hacounts} we compute
the number counts of \ha, \oii, and \oiii\ emitters for the redshift range
covered by FMOS-COSMOS. In the same Section we extend the predictions
on the \ha\ number counts to broader redshift intervals probed by
the forthcoming cosmological missions. Finally, we discuss our
results, caveats, and possible developments in Section \ref{sec:discussion},
presenting the concluding remarks in Section \ref{sec:conclusions}. Unless stated otherwise, we assume a $\Lambda$CDM cosmology with
$\Omega_{\rm m} = 0.3$, $\Omega_{\rm \Lambda} = 0.7$, and $H_0 = 70$
km s$^{-1}$ Mpc$^{-1}$ and a Salpeter initial mass function
\citep[IMF,][]{salpeter_1955}. All magnitudes are expressed in the AB system.  
\begin{figure*}
  \centering
  \includegraphics[width=\textwidth]{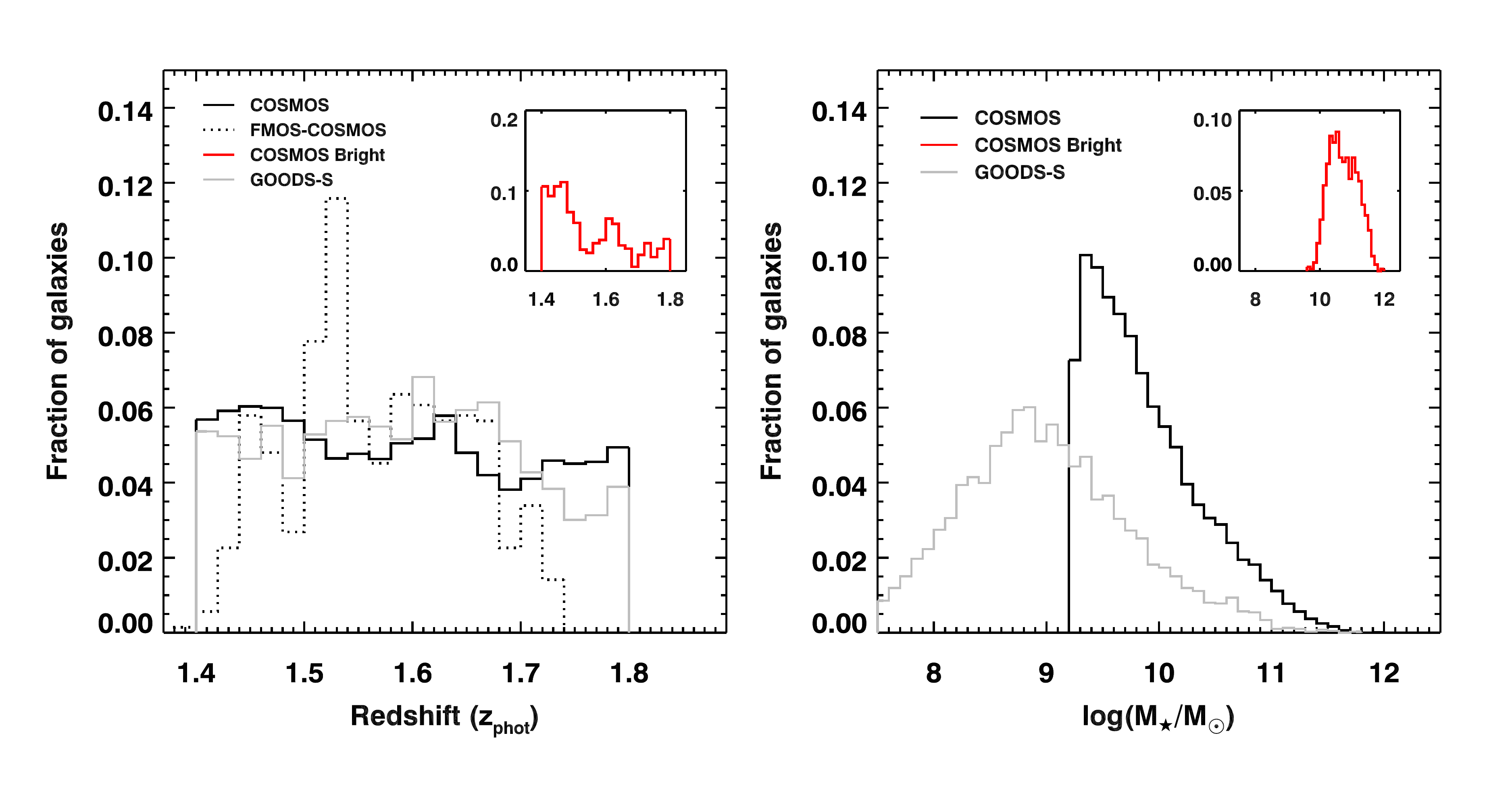}
  \caption{\textbf{Properties of the photometric samples.} \textit{Left:}
    The solid black and grey histograms show the photometric redshift distributions of
    the SFGs we selected in COSMOS and GOODS-S,
    respectively. The black dotted histogram shows the
      FMOS-COSMOS spectroscopic redshift distribution of \ha\
      emitters. The
    histograms are normalized to the total number of objects in each
    sample . The red histogram in the inset shows the normalized
    distribution for a subsample of $750$ galaxies
    with predicted \ha\ fluxes $\geq 2\times10^{-16}$~\esc\ in COSMOS (Section
    \ref{sec:properties}). \textit{Right:} Stellar mass distributions
    for the same COSMOS (black) and GOODS-S (grey) samples, and for the subsample of
    bright \ha\ emitters in COSMOS (red histogram).}
  \label{fig:cosmos_sample}
\end{figure*}

\section{Data and sample selection}
\label{sec:sample}
In this section, we introduce the photometric samples of SFGs
  drawn from the COSMOS and GOODS-S fields. We further present the
  FMOS-COSMOS spectroscopic survey dataset used to calibrate the predictions
  of \ha\ fluxes, the latter being based on the star formation rates
  (SFRs) from SED fitting. Unless specified otherwise, the ``COSMOS''
  and ``GOODS-S photometric'' samples will be treated separately and
  compared when possible. We will refer to the calibration dataset as
  the ``FMOS-COSMOS'' or the ``spectroscopic'' sample.

\subsection{The COSMOS photometric sample}
\label{sec:photometric_sample}
We selected the target sample from the latest COSMOS photometric
catalog by \cite{laigle_2016}, including the UltraVISTA-DR2
photometry. We identified star-forming galaxies
according to the \textit{NUV}-\textit{r}, \textit{r}-\textit{J}
criterion \citep{williams_2009, ilbert_2013}, and
retained only the objects falling in the photometric redshift range
$1.4< z < 1.8$, resulting in a sample of $31,193$ galaxies with
stellar masses of $M_\star\geq10^{9.2}$~\msun. X-ray detected AGN from
  \cite{civano_2016} were flagged (PHOTOZ=9.99 in
  \citealt{laigle_2016}) and removed from our sample, since
  we could not reliably predict their line fluxes. We show the photometric redshift and
the stellar mass distributions in Figure \ref{fig:cosmos_sample}. The
distribution of $z_{\rm phot}$ is flat in the redshift range we
considered. On the other hand, the $M_\star$ distribution shows a
substantial drop at $M_\star\sim10^{9.2}$~\msun. The COSMOS sample is
formally $\sim90$\% complete down to $M_\star\geq10^{9.8}$~\msun\ in this
redshift range, corresponding to a cut at $K_{\mathrm{s}} = 24$~mag in the shallowest
regions covered by UltraVISTA \citep{laigle_2016}. However, Figure
\ref{fig:cosmos_sample} shows that the completeness limit can be
pushed to a lower value for the sample of SFGs we selected, simply
because low mass galaxies are generally blue. In this case, this extended photometric
sample allows for putting 
constraints on the number counts at low fluxes (Section
\ref{sec:hacounts}), a regime usually inaccessible for purely
spectroscopic analyses. Notice that we limit 
the number counts to a flux of $5\times10^{-17}$~\esc, above which the sample of \ha\ emitters
is virtually flux complete. Seventy-eight percent of the whole sample above this flux threshold have
a stellar mass above the mass completeness limit, and this fraction
rises to 95\% for \ha\ fluxes above $1\times10^{-16}$~\esc\ used as a reference for
the differential number counts in Section
\ref{sec:redshiftevolution}. Therefore, the results on the brightest tail
of emitters are not affected by the drop of the stellar mass distribution in the sample.

We selected the $1.4<z<1.8$ redshift interval to match the one of the
FMOS-COSMOS survey (Section \ref{sec:fmos_survey}). 
We adopted the stellar masses from the catalog by \cite{laigle_2016}, computed with
\textsc{LePhare} \citep{ilbert_2006} and assuming \cite{bruzual_2003} stellar
population synthesis models, a composite star formation history
($\mathrm{SFR}\propto\tau^{-2}te^{-t/\tau}$), solar and half-solar
metallicities, and \cite{calzetti_2000} or \cite{arnouts_2013}
extinction curves. We homogenized the IMFs applying a 0.23~dex
correction to the stellar masses in the catalog, computed with the prescription by \cite{chabrier_2003}.
We then re-modeled the SED from the rest-frame UV to the
  \textit{Spitzer}/IRAC 3.6~$\mu$m band with the code \textsc{Hyperz}
\citep{bolzonella_2000}, using the same set of stellar population
models and a \cite{calzetti_2000} reddening law, but assuming constant
SFRs. We chose the latter since they proved to
reconcile the SFR estimates derived independently from different
indicators and to consistently represent the main sequence of SFGs
\citep{rodighiero_2014}. We checked the resulting SFRs and dust
attenuation $A_{\rm V}$ from SED modeling against estimates from the luminosity at 1600~\AA\ only
\citep{kennicutt_1998} and UV $\beta$-slope \citep{meurer_1999}. In
both cases, we obtain consistent results within the scatter and the
systematic uncertainties likely dominating these estimates. A tail of
$\sim8$\% of the total COSMOS sample shows SFRs(UV)$\sim0.15$~dex
lower than SFR(SED), but at the same time they exhibit $A_{\rm
  V}$(UV)$\sim0.1$ mag lower than $A_{\rm V}$(SED). However, these
objects do not deviate anyhow appreciably from the distribution of
predicted \ha\ fluxes computed in Section \ref{sec:linepredictions},
nor in stellar masses or photometric redshifts, as confirmed by a
Kolmogorov-Smirnov test. We, thus, retain these galaxies in the
analysis. SFRs derived from
the rest-frame UV range only and dust
extinctions from the modeling of the full SED extended to the \textit{Spitzer}/IRAC
$3.6$~$\mu$m band proved to robustly
predict \ha\ fluxes, not requiring any secondary corrections. We adopt these estimates in the rest of this work.
\begin{figure}
  \includegraphics[width=0.5\textwidth]{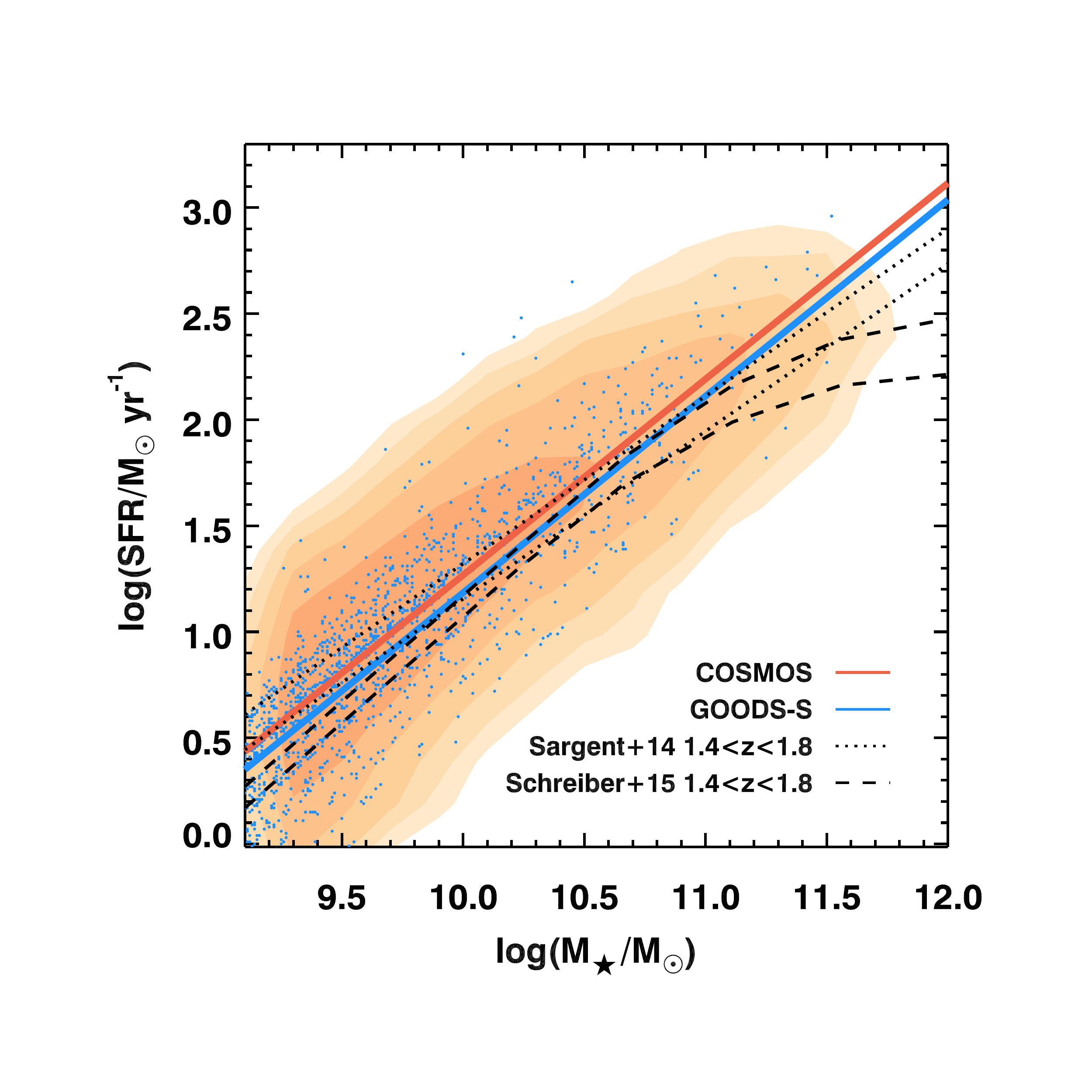}
  \caption{\textbf{Main sequence of star forming galaxies at
      $z\sim1.5$.} Orange contours mark density contours of our sample
    of NUVrJ-selected SFGs at
    $1.4<z<1.8$ and with $M_\star \geq 10^{9.2}$~\msun\ from the COSMOS
    field. Objects similarly selected and modeled in GOODS-S are
    indicated with blue points (Pannella et al., private
    communication). Best fit to the COSMOS and GOODS-S data are
    shown with orange and blue solid lines, respectively.
    Analytical parametrizations of the MS by \citet{sargent_2014} and
    \citet{schreiber_2015} for $z=1.4$ and $z=1.8$ are marked by
    dotted and dashed dark lines, respectively.}
  \label{fig:ms_comparison}
\end{figure}

\subsubsection{A control sample in GOODS-South}
\label{sec:photometric_sample_goods}
We further check the consistency of our compilation of stellar masses
and SFRs in COSMOS comparing it with a sample of SFGs in GOODS-S.
This field benefits from a deeper coverage of the rest-frame UV
range, allowing for a better constraint of the SFRs down to lower
levels, and to put constraints on the tail of \ha\ emitters at low
fluxes and masses, not recoverable in COSMOS.
We, thus, selected a sample of $3,858$ galaxies with $M_\star
\geq10^{7.5}$~\msun\ at $1.4<z<1.8$ applying the same criteria listed
above. The $90$\% mass completeness limit is $M_\star
=10^{9}$~\msun\ and $1,813$ galaxies fall above this threshold.
We show the normalized redshift and stellar mass distribution of the
GOODS-S in Figure \ref{fig:cosmos_sample}. A two-tail
Kolmogorov-Smirnov test shows that the redshift distributions are
compatible. The different mass completeness limits between COSMOS and
GOODS-S are evident from the right panel, with a tail of GOODS-S
objects extending below  $M_\star=10^{9}$~\msun. A Kolmogorov-Smirnov
test on the raw data shows that the distributions are consistent with
the hypotesis of being drawn from the same parent sample, especially
when limiting the analysis to the COSMOS mass completeness threshold. 
We then modeled the SEDs of objects in GOODS-S applying the same
recipes we adopted for the COSMOS sample (Pannella et al., private
communication). As shown in Figure \ref{fig:ms_comparison}, we
consistently recover the MS of galaxies in COSMOS and GOODS-S. We also
find a good agreement with the analytical parametrizations of the MS
by \cite{sargent_2014} and \cite{schreiber_2015}.

\subsection{The FMOS-COSMOS survey}
\label{sec:fmos_survey}
The FMOS-COSMOS survey is a near infrared spectroscopic survey
designed to detect \ha\ and \nii$\lambda\lambda6549,6584$~\AA\ in galaxies at $1.43<z<1.74$ in the \textit{H} band with
the Fiber Multi-Object Spectrograph \citep[FMOS,][]{kimura_2010} on the Subaru
Telescope. An integration of five hours allows
for the identification of emission lines of total flux down to $4\times10^{-17}$
\esc\ at $5\sigma$ with the \textit{H}-long grism ($R\sim 2600$). Galaxies with
positive \ha\ detections have been re-imaged with the \textit{J}-long
grism ($R\sim 2200$) to detect \oiii$\lambda\lambda4959,5007$~\AA\ and
\hb\ emission lines to characterize the properties of the ionized
interstellar medium \citep[ISM,][]{zahid_2014_fmos, kashino_2017}. 
For a detailed description of the
target selection, observations, data reduction, and the creation of
the spectroscopic catalog, we refer the reader to \cite{silverman_2015}. For the scope of this work, i.e., the calibration of
the \ha\ fluxes predictions from the photometry, we selected only the
objects with a signal-to-noise ratio $\geq5$ on the observed \ha\
flux. Their spectroscopic redshifts distribution is consistent with the
  one of photometric redshifts of the COSMOS sample discussed in
  Section \ref{sec:photometric_sample} (Figure \ref{fig:cosmos_sample}).
We mention here that the primary selection relies on \ha\ flux
  predictions based on continuum emission similar to the ones reported
  in the next section. This strategy might result in a bias against
  starbursting sources with anomalously large line EWs, strongly
  deviating from the average stellar mass, SFR, and extinction
  trends. While this is unlikely to affect the most massive galaxies,
  given their large dust content, we could miss starbursting galaxies
  at the low mass end ($M_\star \lesssim10^{9.5}$~\msun), where the
  survey is not complete (Section \ref{sec:discussion:sb}). Moreover, since we preferentially targeted
  massive galaxies and J-band observations aimed at identifying the
  \oiii\ emission followed a positive \ha\ detection, we lack
  direct observational probe of sources with large \oiii/\ha\ ratios
  at low masses and \ha\ fluxes. However, as we further discuss in
  Section \ref{sec:oiii}, this potential bias is likely mitigated by the
  extrapolation of the analytical form we adopt to model the line
  ratios and predict \oiii\ fluxes. 

Note that $\sim44$\% of the initial FMOS-COSMOS targets were eventually
assigned a spectroscopic redshift \citep{silverman_2015}. The success
rate when predicting line fluxes and redshifts is likely higher
considering that $\sim25$\% of the wavelength range is removed by the
FMOS OH-blocking filter. The remaining failures can be ascribed to bad
weather observing conditions; telescope tracking issues and fiber flux losses; high instrumental
noise in the outer-part of the spectral range; errors on photometric
redshifts (11\% of objects are missed due to stochastic errors); the
uncertainties on the dust content of galaxies; significant
intra-population surface brightness variations. We also note that the
misidentification of fake signal and/or non-\ha\ line may occur in
$\sim10$\% of the all line detections \citep{kashino_2017b}. The
latter is a rough estimate based
on $4$ discordant spectroscopic redshift between the FMOS-COSMOS and
the zCOSMOS(-deep) surveys \citep{lilly_2007} out of $28$ galaxies in
common, assuming that the zCOSMOS determinations are correct. This
line misidentification fraction may be overestimated, given the small
sampling rate of zCOSMOS-deep at the range of the FMOS-COSMOS
survey. Since we use the spectroscopic observations mainly to
calibrate the flux predictions from photometry (Section
\ref{sec:linepredictions}), line misidentification does not strongly affect our
results. In fact, either they cause flux predictions to be widely
different from observations and, thus, they are excluded from the
calibration sample (Figure \ref{fig:hapredictions}); or, if by a lucky
coincidence, the predicted \ha\ fluxes fall close to the observed
values of a different line, they spread the distribution of the
observed-to-predicted flux ratios (Figure \ref{fig:hapredictions}),
naturally contributing to the final error budget we discuss later
on. Notice also that the success rate increases up to $\sim60$\% for
predicted \ha\ fluxes $\geq2\times10^{-16}$~\esc, a relevant flux
regime further discussed in detail in the rest of the article.

\section{Prediction of line fluxes from photometry}
\label{sec:linepredictions}
In this section we introduce the method we applied to predict the
nebular line emission from the photometry of the samples presented above.
The expected line fluxes are released in a publicly available catalog.

\subsection{\ha\ fluxes}
\label{sec:hapredictions}
\begin{figure*}
  \includegraphics[width=0.49\textwidth]{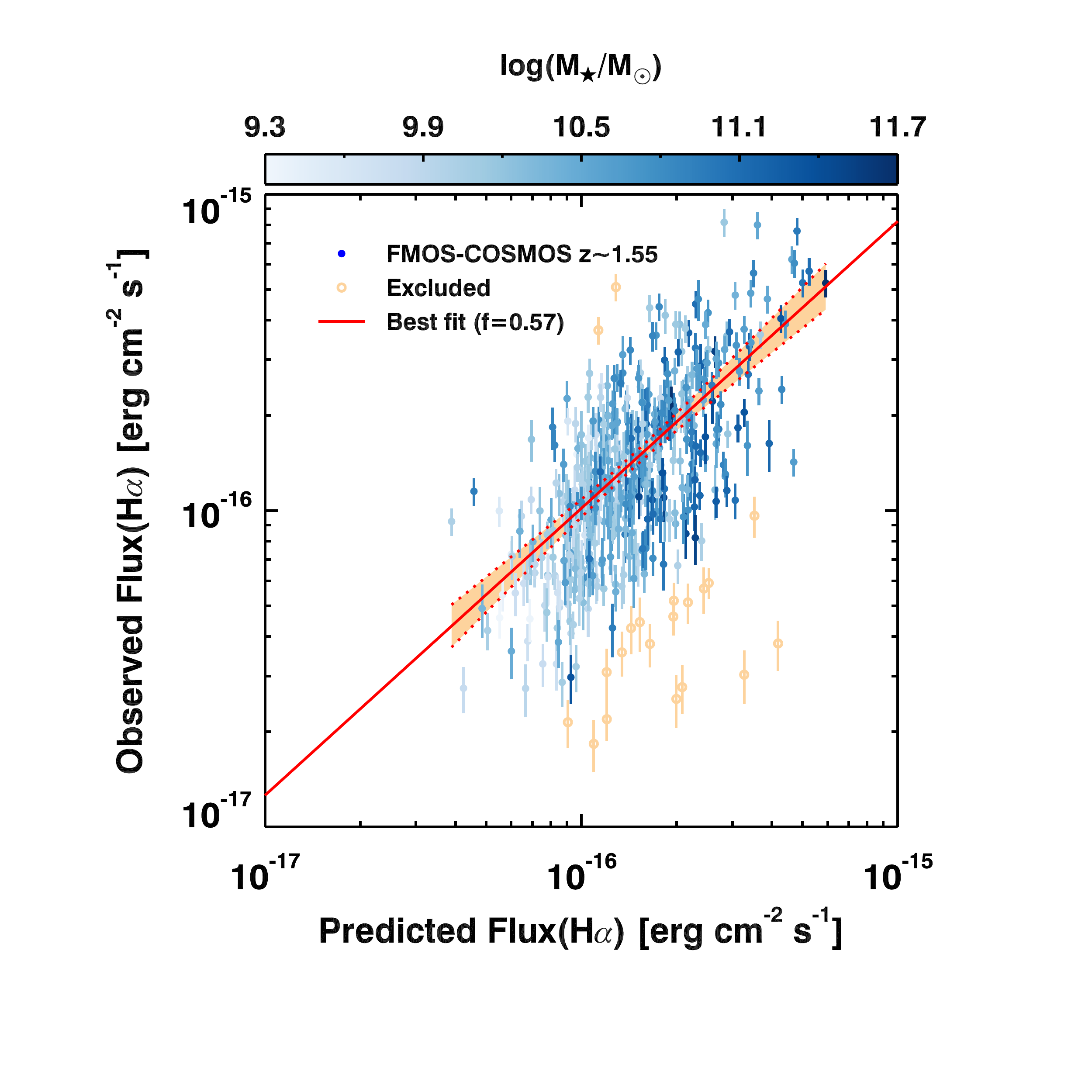}
  \includegraphics[width=0.49\textwidth]{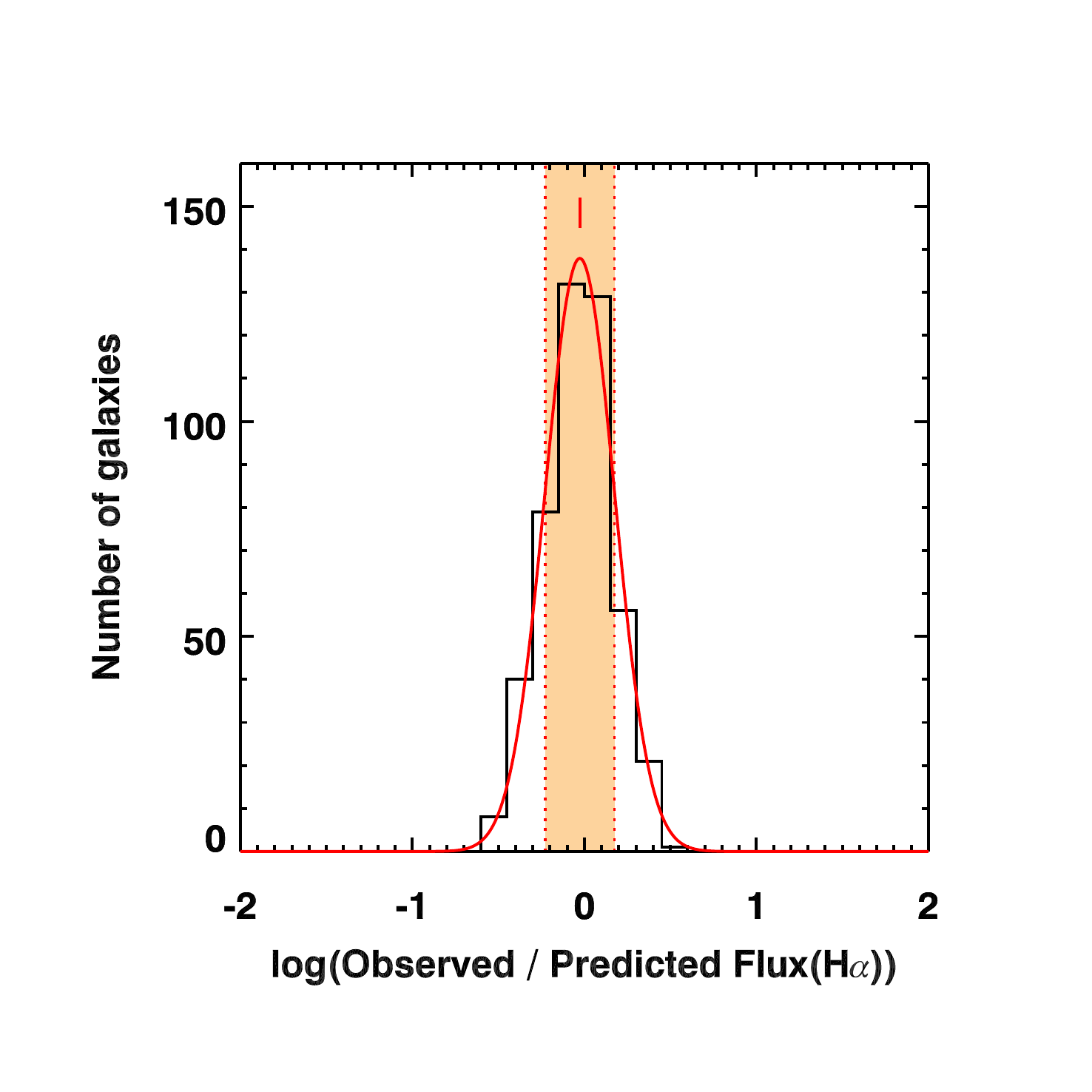}
  \caption{\textbf{Prediction of total \ha\ fluxes for star-forming
      galaxies in COSMOS.} \textit{Left:} Predicted and observed
    total, aperture corrected, \ha\ fluxes for $486$ sources
    detected at $\geq 5\sigma$ in the
  spectroscopic follow-up with FMOS. Blue circles mark the $440$ sources
  used to optimize the $f$ value. Blue colors scale
  as the stellar mass. Orange empty circles represent sources with widely
  different predictions and observations, excluded from the
  calculation of the $f$ factor. The red solid and dotted
  lines represent the best fit to the logarithmic data and the associated
  $95$\% confidence interval. \textit{Right:} The black line shows the distribution
  of the observed-to-predicted \ha\ flux ratios in
  logarithmic scale. The best gaussian fitting is overplot in
  red. The red tick and the orange shaded area mark the mean and the
  $1\sigma$ standard deviation of the best gaussian fit.}
\label{fig:hapredictions}
\end{figure*}
For each source in the photometric sample we computed the expected
\textit{total} observed \ha\ flux based on 
SFRs and dust attenuation estimated in Section \ref{sec:photometric_sample}.
We converted the SFR into \ha\ flux following \cite{kennicutt_1998},
and we applied a reddening correction
converting the $E_{\mathrm{star}}(B-V)$ for the stellar component into
$E_{\mathrm{neb}}(B-V)$ for the nebular emission by dividing by
$f=E_{\mathrm{star}}(B-V) / E_{\mathrm{neb}}(B-V)$. 
We computed $f$ minimizing \textit{a
  posteriori} the difference between the observed and expected
total \ha\ fluxes from the FMOS-COSMOS survey presented in
\cite{kashino_2017}. Therefore, here $f$ assumes the role of a fudge factor
to empirically predict \ha\ fluxes as close as possible to observations. Assigning
a physical meaning to $f$ is prone to several uncertainties
\citep{puglisi_2016}, and it is beyond the scope of this work. The minimization is based on $486$ galaxies in
the spectroscopic sample with an observed \ha\ flux
$\gtrsim2\times10^{-17}$~\esc\ detected at $\geq5\sigma$ (Figure
\ref{fig:hapredictions}). We verified that the value of $f$ is
not biased by low SN detections or by a small subset of very bright
sources, excluding objects in the 10$^{\mathrm{th}}$ and
90$^{\mathrm{th}}$ percentiles of the distribution of predicted \ha\
fluxes. Moreover, the results do not change imposing $FLAG\geq2$ and a
lower signal-to-noise cut of $3$ on the observed \ha\ fluxes from FMOS spectroscopy.
Sources with divergent predictions and observations were excluded
by applying a $2.5\sigma$ clipping on the 
ratios between observed and predicted \ha\ fluxes, leaving $440$
galaxies available for the minimization procedure. These ratios are
log-normally distributed, with a standard deviation of
$0.19$~dex (Figure \ref{fig:hapredictions}). The dispersion is widely dominated
by the $\sim50$\% fiber losses and the ensuing uncertainties on
the aperture corrections for the FMOS observations
\citep{silverman_2015}. A $0.17$~dex dispersion is ascribable to this
effect, while the remaining $0.1$~dex is partly
intrinsic, due to the different star formation timescales traced by UV
and \ha\ light, and partly owing to the systematic
uncertainties of the SED modeling.\\

Applying this technique, we obtain
$f = 0.57\pm 0.01$, with a scatter of 0.23.
A consistent result is retrieved comparing the observed SFR(UV)
and SFR(\ha) \citep{kashino_2013}.
The value of $f$ is higher than the one normally applied for local galaxies
\citep[$f=0.44\pm0.03$,][]{calzetti_2000}, consistently with recent results
for high-redshift galaxies \citep{kashino_2013,
  pannella_2015, puglisi_2016}. Note that we estimated $E_{\mathrm{star}}(B-V)$
using the \cite{calzetti_2000} reddening law, while we adopted the
\cite{cardelli_1989} prescription with $R_{\rm V} = 3.1$ to compute
$E_{\mathrm{neb}}(B-V)$, analogously to what reported in the original
work by \citet{calzetti_2000}, where they used the similar
law by \cite{fitzpatrick_1999}. Using the \cite{calzetti_2000}
reddening curve to compute both the stellar and nebular extinction
would result in higher values of $f$ for local ($f=0.58$) and
$z\sim1.55$ galaxies ($f=0.76\pm0.01$).\\ 

\begin{figure*}
  \includegraphics[width=\textwidth]{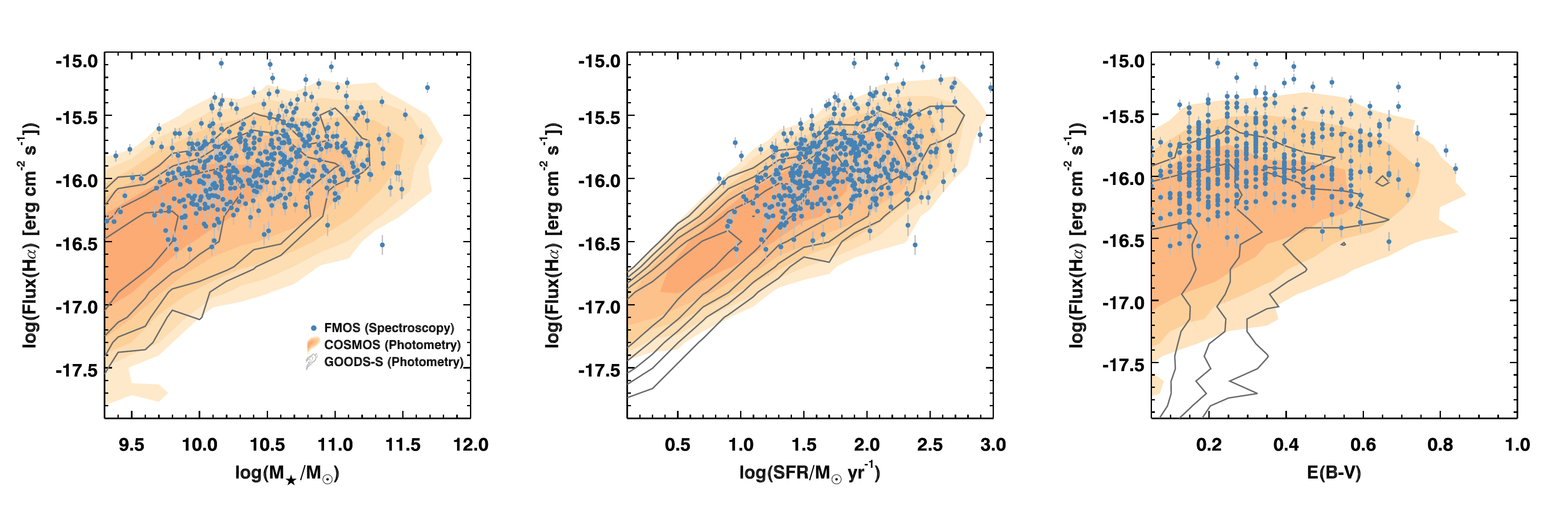}
  \caption{\textbf{Predicted \ha\ fluxes as a function of SED derived
      quantities.} The orange and grey contours show the density contours of
    the COSMOS and GOODS-S photometric samples, respectively. The blue
    points mark the position of spectroscopically confirmed objects in
    the FMOS-COSMOS survey. The \ha\ fluxes are integrated and not
    corrected for reddening. \textit{Left:} Stellar mass vs \ha\
    fluxes. \textit{Centre:} SFR vs \ha\ fluxes. \textit{Right:} $E(B-V)$ vs \ha\
    fluxes.}  
\label{fig:physical}
\end{figure*}

Adopting $f=0.57$, the best fit to the logarithmic
data is $\rm{log(H\alpha_{obs})} = (0.91\pm0.01)\,
\rm{log(H\alpha_{pred})} + (-1.48\pm0.19)$ with a correlation
coefficient $\rho=0.9998$. The uncertainties represent the statistical error in
the fitting procedure, while the scatter of the relation is
$\sigma=0.19$~dex (Figure \ref{fig:hapredictions}). Assuming a
fixed slope of 1, the best fit is $\rm{log(H\alpha_{obs})} = 
\rm{log(H\alpha_{pred})} + (-0.009\pm0.002)$.
Secondary corrections as a function of
$M_\star$ or $E(B-V)$ are not necessary, since the
$\rm{log(H\alpha_{obs}/H\alpha_{pred})}$ ratio is constant and
consistent with 0 over the ranges probed by the FMOS-COSMOS detections
($10^{9.3} \leq M_\star \leq 10^{11.7}$~\msun, $E(B-V)\leq0.84$~mag).
Eventually, we adopted $f=0.57$ to predict the \ha\ and other line fluxes (see
below) both in COSMOS and GOODS-S, assuming its validity over the
entire stellar mass and reddening ranges covered by these samples. We
also assume that the uncertainties on the predicted \ha\ fluxes derived for the
FMOS-COSMOS sample are applicable for galaxies in GOODS-S.
In Figure \ref{fig:physical} we show the correlations among the
predicted \ha\ fluxes and the SED-derived stellar masses, SFRs, 
and reddening $E(B-V)$ for the COSMOS and GOODS-S photometric
samples. We also plot the spectroscopically confirmed objects from the
FMOS-COSMOS survey. The large $E(B-V)$ at high stellar masses compensates the
increase of the SFR on the Main Sequence, so that the $M_\star$ -
observed \ha\ flux relation is flat above $M_\star\sim10^{10}$~\msun,
ensuring high stellar mass completeness above this threshold when
observing down to \ha\ fluxes of $1\times10^{-17}$~\esc. 
Notice that the FMOS-COSMOS observations are biased towards the lower
$E(B-V)$, as expected from the initial selection (Section
\ref{sec:fmos_survey}) and the fact that less dusty objects are
naturally easier to detect. Finally, the uncertainties on $E(B-V)$
are included in the correlation of SFR into observed \ha\ fluxes shown
in the central panel. 

\subsection{H$\beta$ fluxes}
\label{sec:hb}
 We computed \hb\ fluxes rescaling the \ha\ values for the
different extinction coefficients $k_\lambda$ and assuming the intrinsic ratio
$\rm{H}\beta=\rm{H}\alpha/2.86$ \citep{osterbrock_2006}. Note that the stellar
Balmer absorption might impact the final observed \hb\
flux. We, thus, compute a stellar mass dependent correction following \cite{kashino_2017}:
\begin{equation}
f_{\mathrm{corr}} = \mathrm{max}[1,1.02+0.30\,\mathrm{log}(M_\star/10^{10}\,M_\odot)]
\end{equation}
where $f_{\mathrm{corr}}$ corresponds to a correction up to
$50$\%. We report this term in the released catalog for
  completeness so to compute the observed, Balmer-absorbed fluxes, if
  needed. However, the correction is not applied to the total \hb\
  fluxes shown in the rest of this work.
\begin{figure}
  \includegraphics[width=0.5\textwidth]{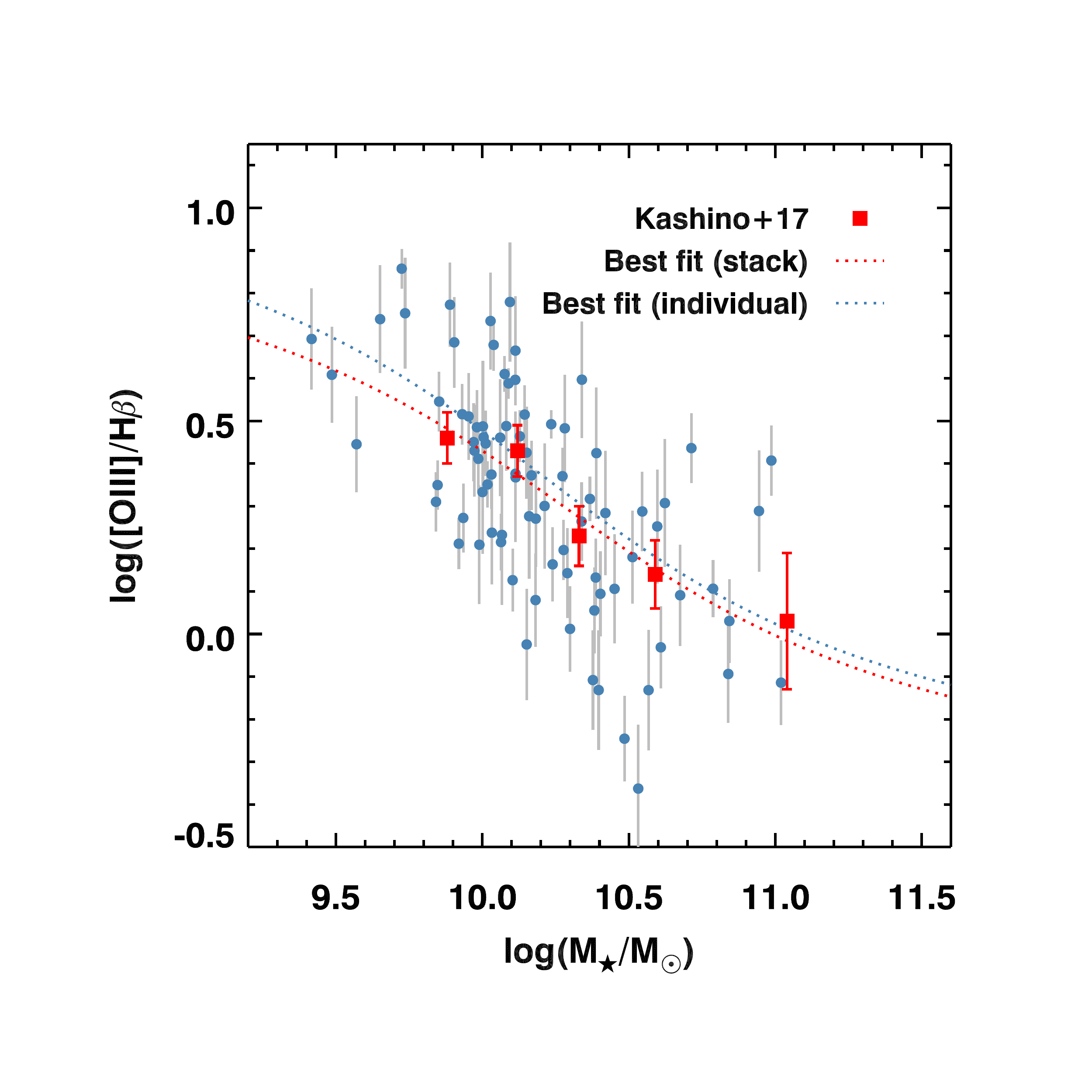}
  \caption{\textbf{\textsc{[OIII]}/\hb\ ratio as a function stellar mass for the
      FMOS-COSMOS survey at $z\sim1.5$.} Blue points mark the observed
    galaxies in the FMOS-COSMOS survey with $3\sigma$ detections of both
    the \oiii\ and \hb\ lines. Grey error bars represent the $1\sigma$
    uncertainties on the ratio estimates. Red squares mark the
    average values for SFGs in the FMOS-COSMOS survey as derived
    in \citealt{kashino_2017}. The red and blue lines indicate
    the best fit to the stacked values and individual sources, respectively.}  
\label{fig:oiiihb}
\end{figure}

\subsection{\textsc{[OIII]} fluxes}
\label{sec:oiii}
We predict \oiii\ fluxes adopting a purely empirical approach
calibrated against the average spectra of the FMOS-COSMOS
SFGs described in \cite{kashino_2017}. The observed $\rm{log}$(\oiii$/$\hb) ratio
anticorrelates with $\rm{log}(M_\star)$, as shown in Figure
\ref{fig:oiiihb} \citep[Mass-Excitation diagram,
][]{juneau_2011}. Being close in wavelength, this line ratio is not
deeply affected by reddening corrections. Here we predict \oiii\
fluxes from \hb\ forcing the line ratio to
follow a simple arctangent model fitting the stacked values. The best fit
model is: $\rm{log}$(\oiii$/$\hb$) =
(0.30\pm0.37)+(0.48\pm0.12)\,\mathrm{arctan}\{-[\mathrm{log}(M_\star/M_\odot)-(10.28\pm0.84)]\}$. 
Fitting the individual sources does not impact the main conclusions of
this work. Note that these predictions are valid only for the
redshift window $1.4<z<1.8$, where a significant evolution of the
\oiii/\hb\ ratio is not expected \citep{cullen_2016}. Notice also that
the number of secure individual $3\sigma$ detections of both \oiii\
and \hb\ is restrained ($84$ galaxies) and that the line ratio suffers
from a significant scatter.\\

The comparison between predicted and
observed \oiii\ fluxes is shown in Figure \ref{fig:oxygen}. The best
fit to the logarithmic data is $\rm{log(}$\oiii$_{\mathrm{obs}}) = (1.00\pm0.03)\,$
$\rm{log(}$\oiii$_{\mathrm{pred}}) + (0.08\pm0.45)$ with a correlation
coefficient $\rho=0.99995$. The best model is derived from
  $181$ galaxies with a $\geq3\sigma$ detection of \oiii\ from our
  FMOS-COSMOS sample, after applying a $2\sigma$ clipping to remove $22$
  strong outliers. Note that the flux range covered by FMOS
  \oiii\ observations is more limited than for \ha. The
  distribution of observed-to-predicted
\oiii\ fluxes has a width of $\sigma=0.25$~dex, 
dominated by the uncertainties on FMOS aperture corrections, as for
the \ha\ line. Figure \ref{fig:oiii_systematics} shows that we underpredict the \oiii\ flux by up
to $\sim0.1$~dex for galaxies with low SFR ($\lesssim 30$~\myr) and
low $A_{\rm V}$ ($\lesssim 0.8$~mag) from the SED fitting, but we do
not find any evident dependence on stellar mass, even if FMOS-COSMOS
\oiii\ observations probe only the $M_\star \gtrsim 10^{9.5}$~\msun\
regime. Since we allowed for a lower signal-to-noise
  ratio to detect \oiii\ emission than \ha\ fluxes in order
  to increase the sample statistics, here we adopted a stricter
  clipping threshold to eliminate outliers. In particular, AGN
  contamination likely boosts \oiii\ fluxes in the latter, massive
  objects (median $M_\star = 10^{10.8}$~\msun), causing systematically
  larger observed fluxes than predicted for inactive SFGs. We applied
the same calibration to the galaxies in GOODS-S,
and assumed that the uncertainties derived from the spectroscopic
sample in COSMOS applies to GOODS-S, too.
Note that the \oiii\ flux and the \oiii/\hb\ ratio are
sensitive to the presence of AGN. Moreover, the number of
bright \oiii\ emitters with low masses is significantly larger than
for the \ha\ line, since the \oiii/\hb\ increases for decreasing
masses. This is particularly relevant for the GOODS-S sample.
As mentioned in Section \ref{sec:fmos_survey}, the FMOS-COSMOS
  survey does not probe the low-mass, high \oiii/\hb\ regime,
  where line ratios up to $0.8-1$ are typically observed
  \citep{henry_2013}. However, extrapolating the best fit models shown
  in Figure \ref{fig:oiiihb} down to $M_\star\sim10^{8}$~\msun, we
  cover the range of observed ratios, likely mitigating a potential
  bias against large \oiii\ fluxes. 

\subsection{\textsc{[OII]} fluxes}
\label{sec:oii}
\begin{figure*}
  \includegraphics[width=\textwidth]{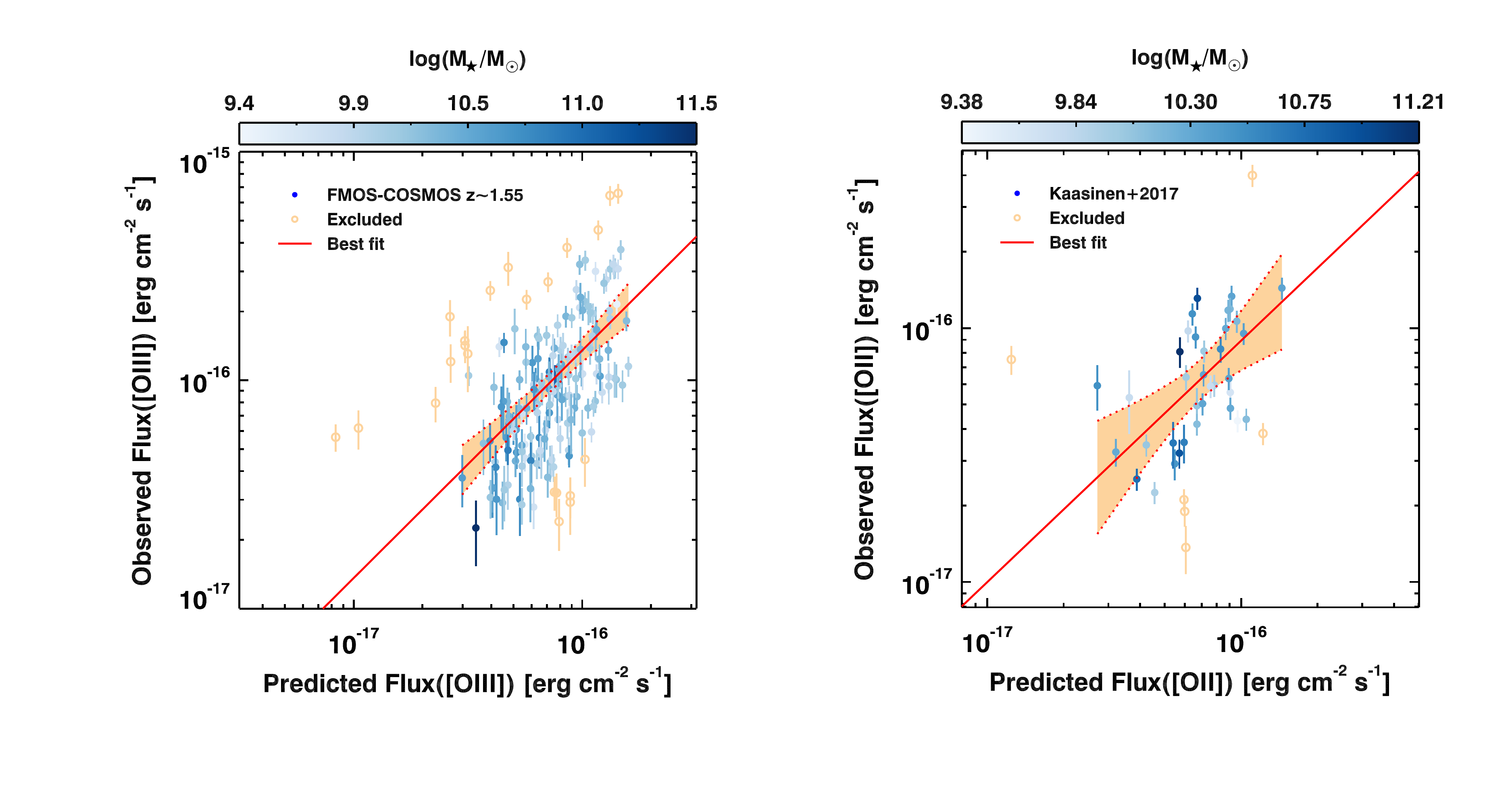}
  \caption{\textbf{Oxygen lines flux predictions.} \textit{Left}: Blue
    circles mark predicted and observed \oiii\
    fluxes from a sample of $159$ galaxies with $3\sigma$ detections
    in the FMOS-COSMOS survey. Symbols are color coded according to
     stellar masses. The red solid line and orange shaded area
    indicate the best fit to the data and its $95$\% confidence
    interval. Orange empty circles have been excluded from the
    fit. \textit{Right}: Blue circles mark predicted and observed \oii\
    fluxes from a sample of $37$ galaxies with $3\sigma$ detections
    from \citealt{kaasinen_2017}. Symbols are color coded according to
     stellar masses. The red solid line and orange shaded area
    indicate the best fit to the data and its $95$\% confidence
    interval. Orange empty circles have been excluded from the fit.}  
\label{fig:oxygen}
\end{figure*}

\begin{figure*}
  \centering
  \includegraphics[width=0.32\textwidth]{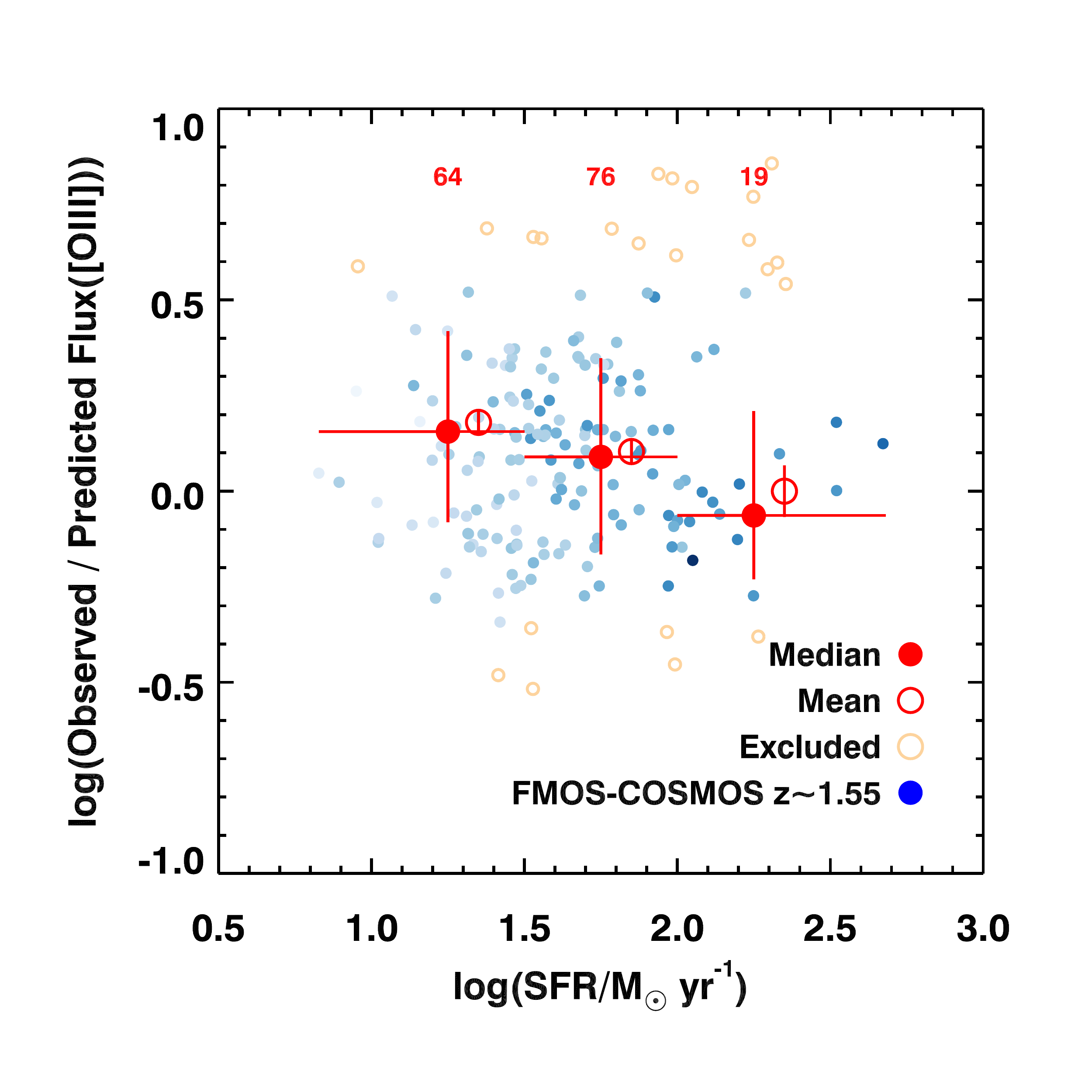}
  \includegraphics[width=0.32\textwidth]{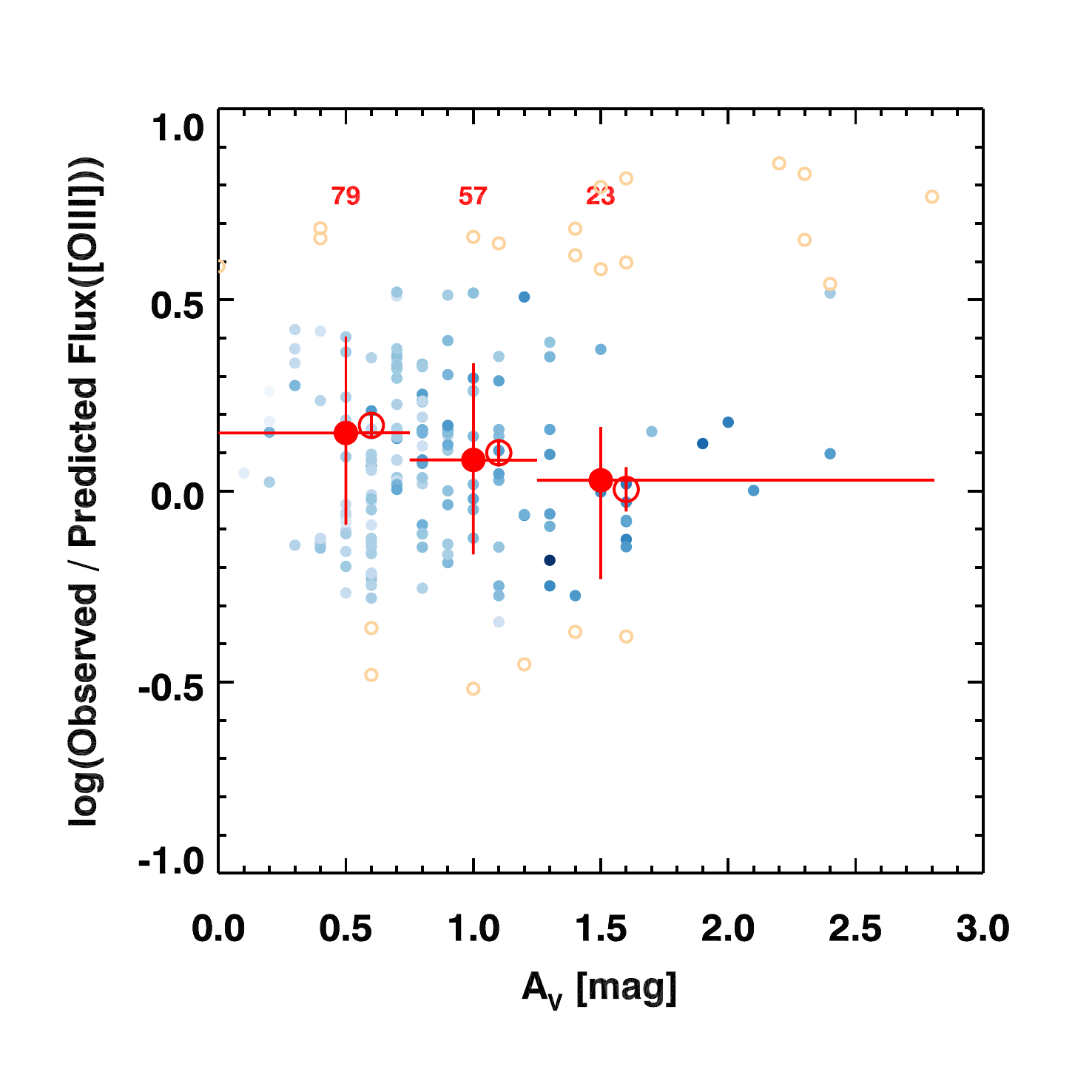}
  \includegraphics[width=0.32\textwidth]{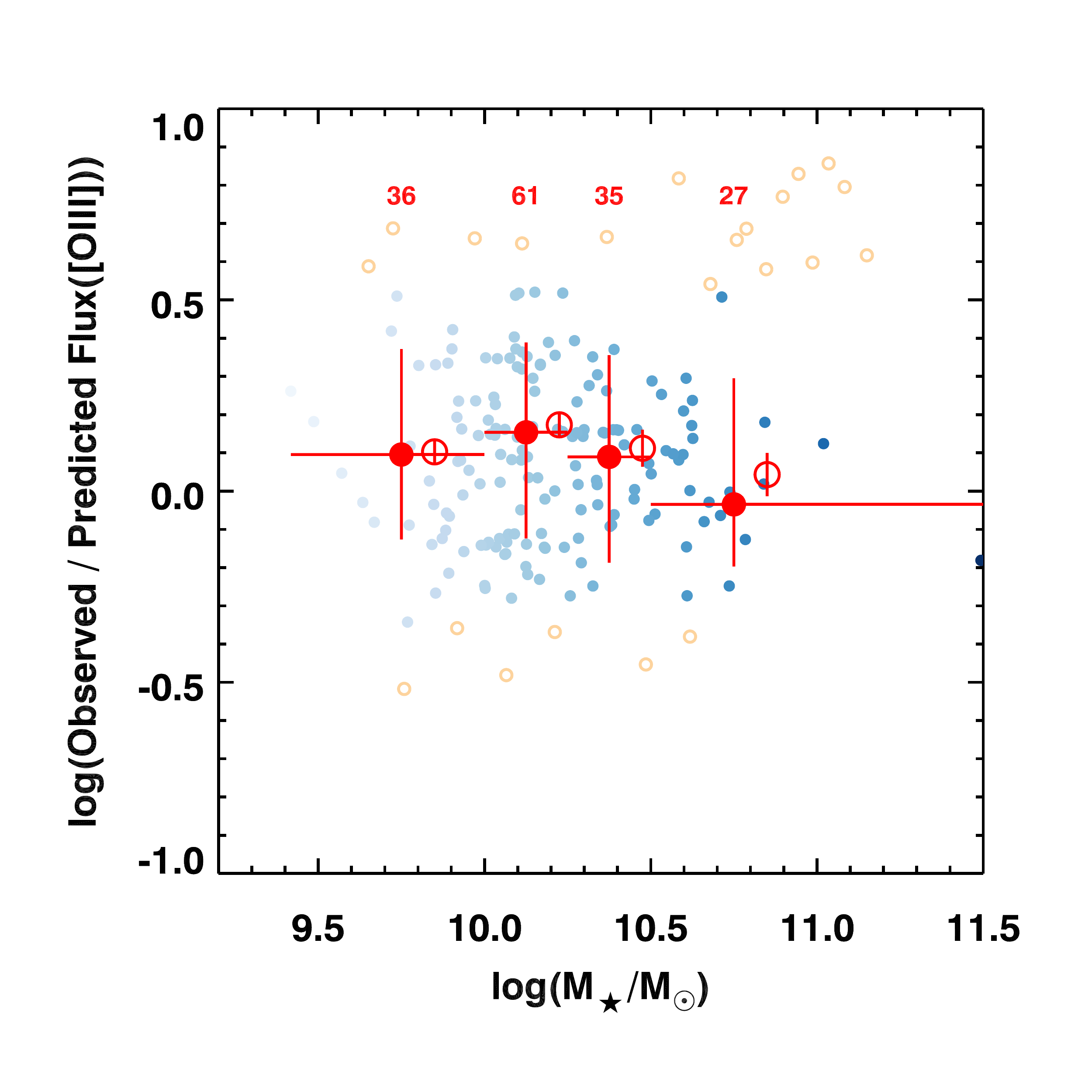}
  \caption{\textbf{Observed-to-predicted} \oiii\ \textbf{flux ratios as a function of
      SED-derived quantities.} In each panel, blue filled circles
    show the \oiii$_{\mathrm{obs}}$/\oiii$_{\mathrm{pred}}$ ratios
    against SED-derived SFRs (\textit{left}), $A_{\rm
      V}$ (\textit{center}), and stellar masses \mstar\
    (\textit{right}) for the sample of
    galaxies with an \oiii\ flux measurement from the FMOS-COSMOS
    survey. Symbols are color coded according to
     stellar masses as in Figure \ref{fig:oxygen}. Orange empty
     circles have been excluded applying the
    $2\sigma$-clipping described in Section \ref{sec:oiii}. The red filled circles and vertical bars represent the
    median of \oiii$_{\mathrm{obs}}$/\oiii$_{\mathrm{pred}}$ ratios in
    subsequent bins and the $\pm1\sigma$ percentiles ($15.84$,
    $84.16$\%). The horizontal bars show the width of each bin,
    selected based on the enclosed number of objects (reported in red in
    the three panels) and the typical systematics affecting SED
    modeling. Red
    open circles and bars represent the mean of line ratios in each
    bin and its standard error ($=\sigma/\sqrt{N}$), where $N$ is the
    number of objects per bin.}
  \label{fig:oiii_systematics}
\end{figure*}

\oii\ might be used as a SFR tracer \citep{kennicutt_1998,
  kewley_2004, talia_2015}, even if its calibration depends on secondary
parameters such as the metal abundance. Here we simply assume $L($\oii$) =
L($\ha$)$ \citep{kewley_2004} and the extinction coefficient
$k($\oii$)=4.771$ from the \cite{cardelli_1989} reddening curve ($R_{\rm
  V}=3.1$). In Figure \ref{fig:oxygen} we show the predicted
  \oii\ fluxes against a sample of $43$ spectroscopic measurements in
  COSMOS from \cite{kaasinen_2017} in common with our catalog. After applying a $2\sigma$
  clipping to the $\rm{[OII]_{obs}/[OII]_{pred}}$ flux ratios, the best
  fit to the relation between these two quantities is $\rm{log([OII]_{obs})} = (0.95\pm0.06)\,
\rm{log([OII]_{pred})} + (-0.83\pm0.92)$, with a correlation
coefficient $\rho=0.99996$. The width of the distribution of the
ratios $\rm{[OII]_{obs}/[OII]_{pred}}$ is $\sigma\sim0.22$~dex. We
applied the same method to the sample in GOODS-S. Also in this
  case, the stricter clipping threshold than for \ha\ fluxes (Section
  \ref{sec:hapredictions}) compensates for the lower signal-to-noise
  limit allowed for \oii\ detections, so to increase the size
  of the available sample. Applying a $5\sigma$ detection threshold
  and a $2.5\sigma$ clipping to \oii\ observed fluxes results
  in a similar final object selection to the one presented above.\\

\noindent
We note that a similar approach was applied by \cite{jouvel_2009} to simulate emission
lines for a mock sample of objects based on the observed SEDs of galaxies
in COSMOS. In their work, \cite{jouvel_2009} based the flux predictions assuming
\oii\ as a primary tracer of SFR and on a set of fixed line ratios. However, \oii\ 
shows secondary dependencies on other parameters such as metallicity, even if in 
first approximation it traces the current SFR. Moreover, the line ratios significantly 
change with redshift. Furthermore, a proper treatment of the dust extinction
is fundamental to derive reliable nebular line fluxes, introducing a 
conversion between the absorption of the stellar continuum and of the emission
lines. Here we exploited the updated photometry in the same field and GOODS-S, 
and we tied our predictions to direct spectroscopic observations of a large sample of  
multiple lines in high-redshift galaxies, the target of future surveys. We primarily estimated the \ha\ fluxes,
a line directly tracing hydrogen ionized by young stars and brighter than \oii, 
thus accessible for larger samples of galaxies spanning a broader range of SFRs
and masses. Predictions for oxygen lines emission were directly compared to 
observations as well.

\section{A sample of bright \ha\ emitters at \textit{\MakeLowercase{z}}$\,\sim1.5$}
\label{sec:properties}
\begin{figure}
  \includegraphics[width=0.5\textwidth]{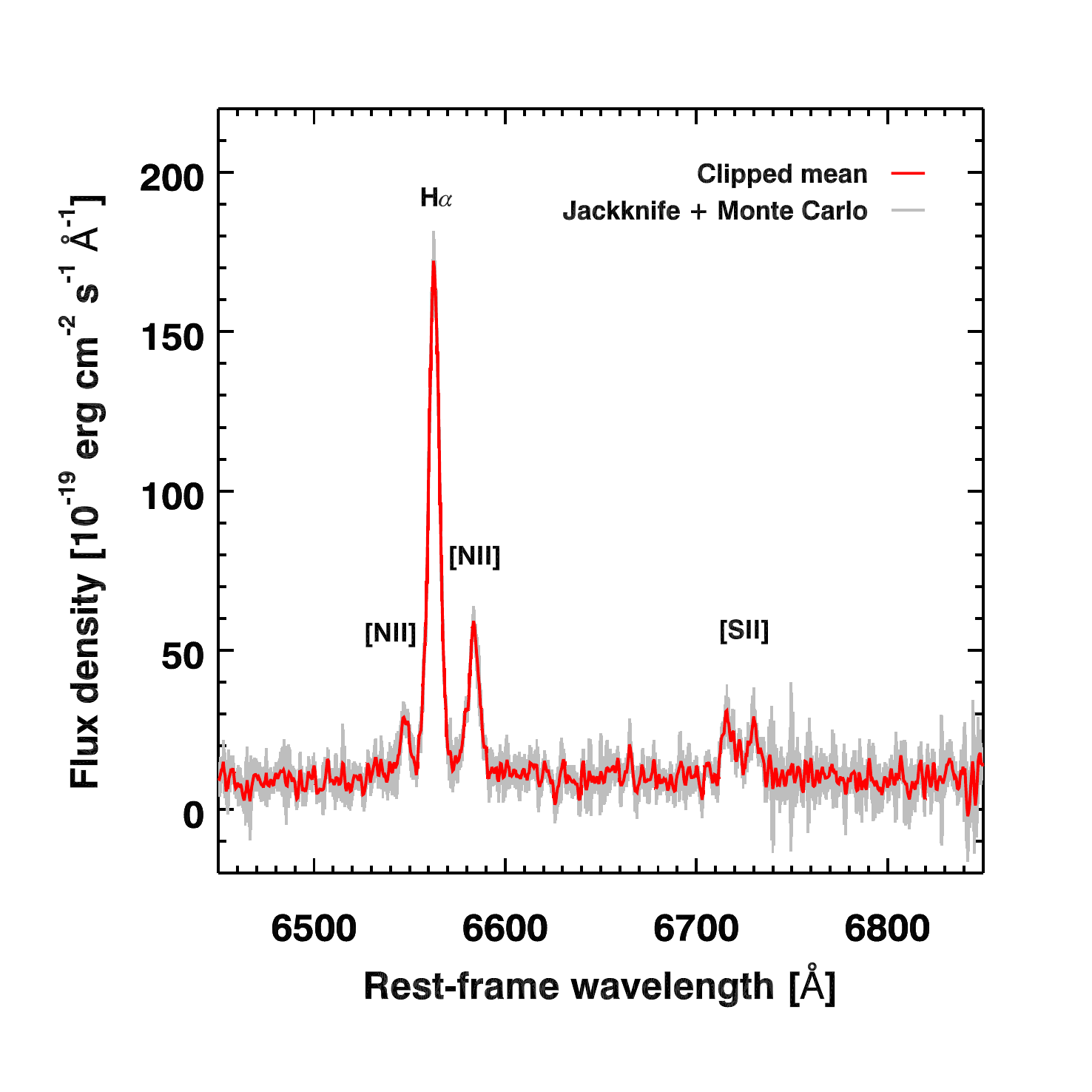}
  \caption{\textbf{Average spectrum of bright \ha\ emitters
      from the FMOS-COSMOS survey.} The red line marks the
    clipped average spectrum of $135$
    individual line emitters with aperture corrected, observed \ha\
    fluxes $\geq 2\times10^{-16}$~\esc from the catalog by
    \citet{kashino_2017}. The grey line shows the associated
    uncertainty estimated with Monte Carlo and Jackknife techniques.}
\label{fig:stacked_spectrum}
\end{figure}
 The sensitivity to emission lines achieved by the FMOS-COSMOS and
 similar spectroscopic surveys
is an order-of-magnitude deeper than what expected for forthcoming
large surveys (i.e., Euclid wide survey: $\geq2\times10^{-16}$~\esc,
$3.5\sigma$; WFIRST: $\geq0.5-1\times10^{-16}$~\esc\ for extended sources,
$3\sigma$, Figure 2-15 of \citealt{spergel_2015}). Therefore, the
physical characterization of the population of bright \ha\
emitters is a key feature in the current phase of preparation for these
missions. Here we have the opportunity to achieve this goal for a
fairly large sample of galaxies, exploiting both photometric and
spectroscopic data.
\begin{figure*}
  \includegraphics[width=\textwidth]{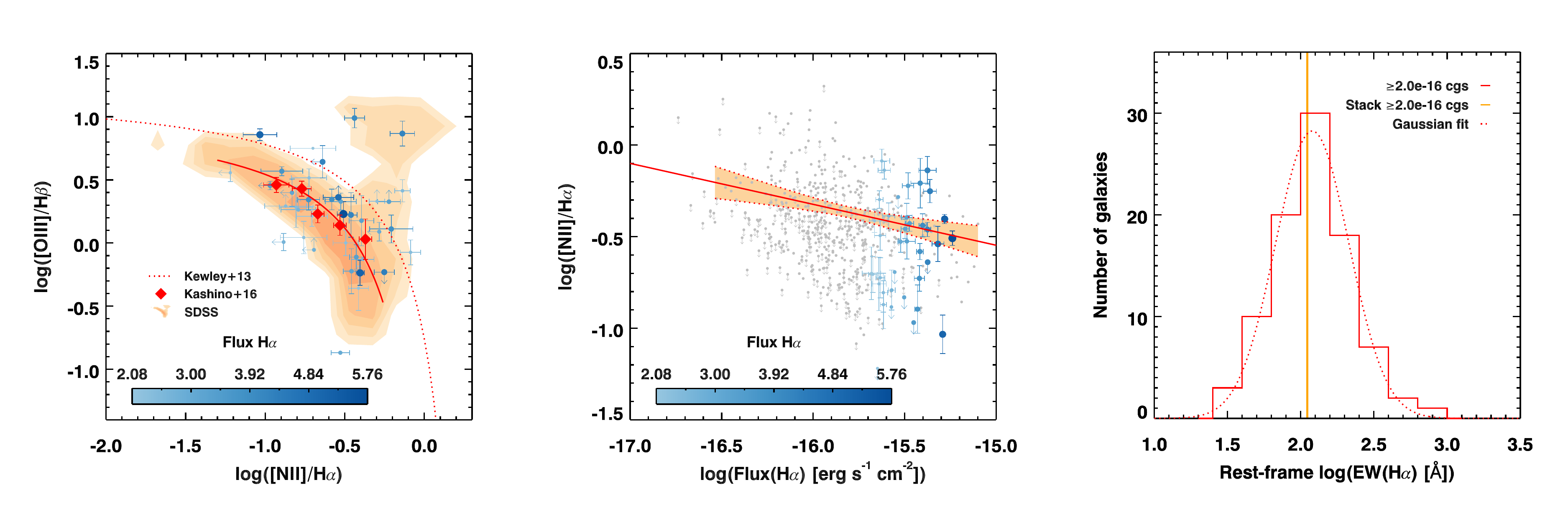}
  \caption{\textbf{Spectroscopic properties of bright \ha\ emitters at
      $z\sim1.5$.} \textit{Left:} BPT diagram for spectroscopically
    confirmed emitters with \ha\ flux $\geq2\times10^{-16}$~\esc\ from
    the FMOS-COSMOS survey. The blue solid circles mark bright
    emitters. The color intensity scales as the \ha\ flux in units of $10^{-16}$~\esc\ reported in
    the color bar. The red
    diamonds and solid line mark the average location of the 
    FMOS-COSMOS sample of SFGs with total \ha\ flux $\geq4\times10^{-17}$~\esc\ and the best fit by
    \citet{kashino_2017}. The red dotted line indicates the
     limiting curve dividing SFGs and AGN at $z=1.55$ as parametrized in \citet{kewley_2013p}.
    The orange shaded area marks the location of
    SDSS galaxies with an intrinsic \ha\ luminosity corresponding to
    a total \ha\ flux of $\geq4\times10^{-17}$~\esc\ at
    $z=1.55$. \textit{Centre:} \nii/\ha\ ratios as a function
      of the total observed \ha\ flux for the spectroscopic
      FMOS-COSMOS sample. The blue solid circles mark bright \ha\
      emitters in the BPT in the left panel. The color intensity scales as the \ha\ flux in units of $10^{-16}$~\esc\ reported in
    the color bar. Grey dots and arrows mark the position of the rest
    of the FMOS-COSMOS spectroscopic sample described in
    \citet{kashino_2017}. The red solid line indicates the best 
    fit to the data. The orange area and the red dotted lines mark the
    $95$\% confidence limits of the fit. \textit{Right:} The red histogram shows the distribution
    of rest-frame $\rm{log}[$EW$($H$\alpha)]$ for spectroscopically
    confirmed emitters with \ha\ flux
    $\geq2\times10^{-16}$~\esc. The red dotted line marks the best
    Gaussian fit to the distribution. The orange band indicates the
    $1\sigma$ confidence limit around the value estimated from the
    stacked spectrum in Figure \ref{fig:stacked_spectrum}.}
\label{fig:specprop}
\end{figure*}

\subsection{Spectroscopy: line ratios and equivalent widths}
\label{sec:properties:spectroscopy}
The general spectroscopic properties of the FMOS-COSMOS sample are
 detailed in \cite{kashino_2017}. Here we focus on a subset of $135$
 bright sources with total, observed (i.e., corrected for aperture effects, but not for
extinction) \ha\ fluxes $\geq
2\times10^{-16}$~\esc\ from their catalog. 
First, we visually inspected and manually re-fitted the FMOS spectra of these sources.
We, then, stacked the individual spectra, applying a $5\sigma$ clipping at each
wavelength. The clipping does not introduce 
evident biases: the resulting spectrum is fully consistent
both with an optimally weighted average and a median spectrum. The
average spectrum and the associated uncertainty,
estimated through Jackknife and Monte Carlo techniques, are shown in
Figure \ref{fig:stacked_spectrum}. From this spectrum we derived \ha,
\nii, \sii$\lambda\lambda6717,6731$~\AA, and continuum emission
fluxes for the population of bright emitters. Note that \sii\ lines
are not in the observed wavelength range for galaxies at
$1.67<z<1.74$.\\ 

The left panel of Figure \ref{fig:specprop} shows the
 BPT diagram for a subsample of $39$ bright emitters in the
   FMOS-COSMOS sample with coverage of
 \hb\ and \oiii. The bright emitters at lower \nii/H$\alpha$ ratios
 are mainly distributed around the
 average locus of the FMOS-COSMOS sample down to the detection limit
 of $\geq4\times10^{-17}$~\esc\ \citep{kashino_2017}. At ratios above
 log(\nii/\ha)$\gtrsim-0.5$, bright \ha\ emitters show higher \oiii/\hb\
 ratios, possibly due to contamination by AGN, which dominate the
 line emission in some extreme cases. However, there are not evident trends between the
 position in the BPT and the \ha\ flux of these bright emitters, as
 shown by the color bar. The sample is also offset with respect to the
 average locus of a sample of $6,638$ low-redshift galaxies
 ($0.04<z<0.2)$ selected from the Sloan Digital Sky Survey DR7
 \citep{abazajian_2009} with well-constrained \oiii/\hb\ and \nii/\ha\
 ratios \citep{juneau_2014} and with an intrinsic \ha\ luminosity corresponding to fluxes
 $\geq4\times10^{-17}$~\esc\ at $z=1.55$. This shows that the offset
 in the BPT diagram is not merely due to selection effects
 \citep{juneau_2014, kashino_2017}. Nine out of $39$ emitters ($\sim23$\%) are classified as AGN
 according to the criterion by \cite{kewley_2013p} at $z\sim1.55$, and this partly results
 from the selection of \textit{Chandra} detected sources to complement
 the main color selection for the FMOS-COSMOS survey
 \citep{silverman_2015}. In Figure \ref{fig:specprop} we show how 
log(\nii/\ha) apparently anticorrelates with observed \ha\ fluxes. The best fit is 
log(\nii/\ha)$ = (-0.22\pm0.02)\rm{log(H\alpha)} - (3.90\pm0.26)$
(correlation coefficient $\rho=0.99983$). However, this correlation is
naturally affected by observational biases and disappears when stacking \nii\ non-detections
\citep{kashino_2017}. The mean ratio log(\nii/\ha)
of the subsample of $91$ sources with \nii\ $3\sigma$ detections is log(\nii/\ha)$=-0.47\pm0.02$,
compatible with the value obtained from the stacked spectrum of the
whole sample of $135$ bright spectroscopic emitters (log(\nii/\ha)$=-0.52\pm0.01$). 
Finally, we computed the distribution of rest-frame equivalent widths
of \ha\ (EW(\ha)) and its mean (Figure
 \ref{fig:specprop}), obtaining log[EW(\ha)/\AA]$=2.08\pm0.03$,
 similar to the result from stacking
 (log[EW(\ha)/\AA]$=2.05\pm0.01$). Adopting the median, a gaussian
 model of the distribution, or a $3\sigma$-clipped average does not
 impact the results. These values are consistent with
 recent compilations of high-redshift galaxies at similar masses
 \citep[i.e.,][]{fumagalli_2012, marmol-queralto_2016}. 
\begin{figure*}
  \includegraphics[width=\textwidth]{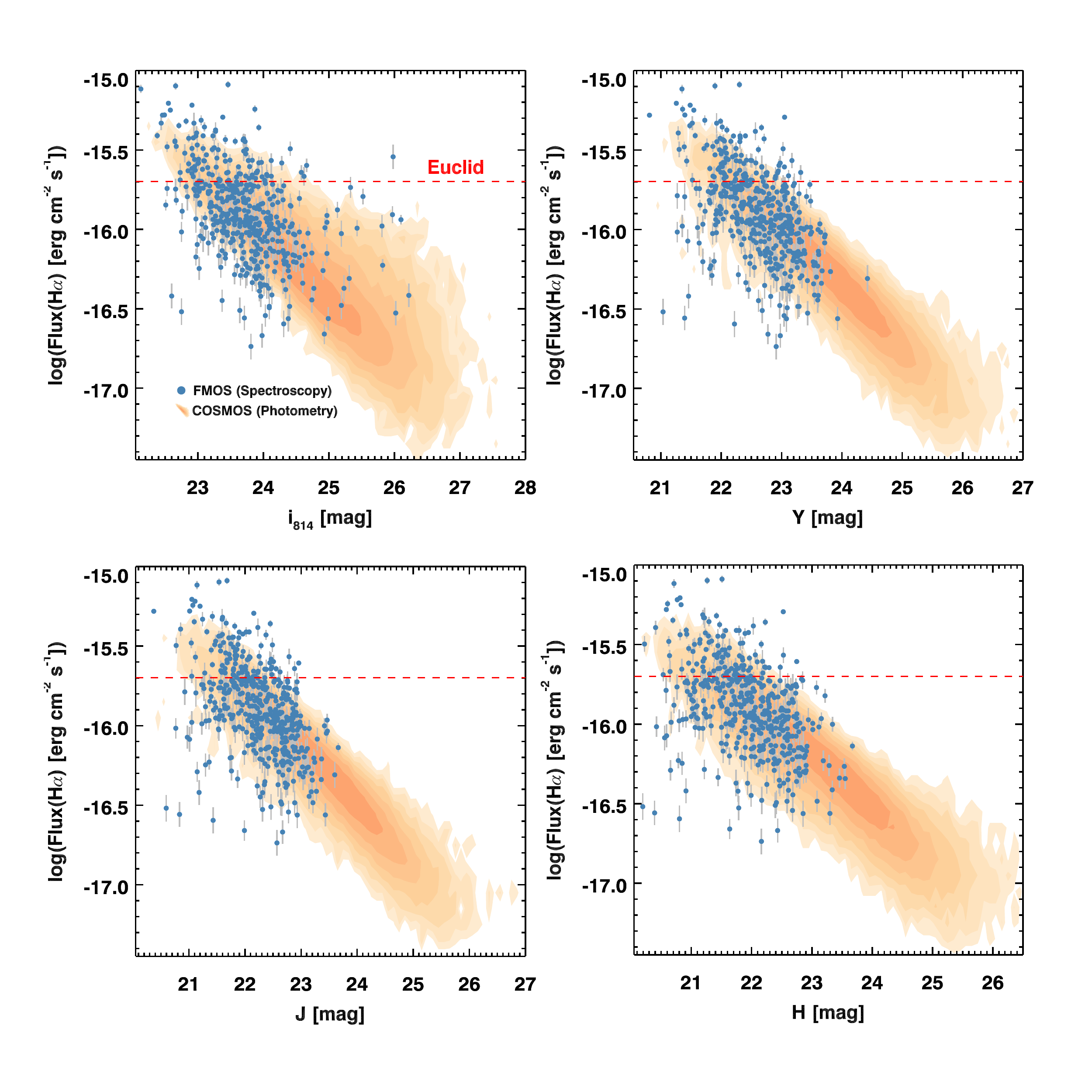}
  \caption{\textbf{Photometric properties of the COSMOS sample of
      star-forming galaxies at $z\sim1.5$.} The panels show the
      relation between the \ha\ fluxes and the \textit{HST}/ACS \textit{i} band (top left),
      \textit{Y} band (top right), \textit{J} band (bottom
      left), and \textit{H} band (bottom right) magnitudes from UltraVISTA-DR2. Orange contours
      represent the whole photometric COSMOS sample and the predicted
      \ha\ fluxes. Blue points indicate the subset of objects 
      confirmed by FMOS and their spectroscopic \ha\ fluxes. Grey
      bars mark the $1\sigma$ uncertainties on the observed \ha\
      fluxes. The red dashed line marks the limit of
        $2\times10^{-16}$~\esc\ expected for
      the Euclid wide survey.}
\label{fig:distr}
\end{figure*}

\subsection{Optical and near-IR photometry}
\label{sec:properties:photometry}
The tail of bright \ha\ emitters from the FMOS-COSMOS sample is fairly
bright in the observed optical and near-IR bands. In Figure \ref{fig:distr} we
show the relation between the \ha\ fluxes and \textit{HST}/ACS
\textit{i}$_{814}$, and the UltraVISTA-DR2 \textit{Y, J, H}
 band \texttt{MAG\_AUTO} magnitudes for the COSMOS photometric sample
\citep{laigle_2016} and the subset of objects spectroscopically confirmed with FMOS. For
 reference, the emitters with expected \ha\ fluxes
 $\geq2\times10^{-16}$~\esc\ in the COSMOS field have $H<22.5$~mag.
The contours representing the whole photometric sample of SFGs in COSMOS show that our flux predictions
 capture the scatter of the spectroscopic observations, while
 correctly reproducing the slope of the relations in each band. 
Note that, by construction, the FMOS-COSMOS selection prioritizes
bright galaxies to ensure a high detection rate of emission lines.

\subsection{Rest-frame UV sizes}
\label{sec:properties:sizes}
We further attempted to estimate the typical sizes of bright \ha\
emitters. In order to increase the statistics of bright emitters and
not to limit the analysis to spectroscopically confirmed objects, we selected a
subsample of $750$ SFGs in COSMOS with \textit{predicted} \ha\ fluxes 
 $\geq2\times10^{-16}$~\esc\ ($2$\% of the total photometric sample).
 The insets in Figure \ref{fig:cosmos_sample} show the normalized distributions of photometric
 redshifts and stellar masses for this subsample. Bright emitters follow the same redshift distribution
 of the whole population, while being fairly massive ($\langle
 \rm{log}(M_\star/M_\odot) \rangle = 10.7\pm 0.4$). Note
 that all bright emitters in COSMOS lie well above the stellar mass completeness
 threshold. This is consistent with the fact that we do not find any SFG
 on the main sequence in GOODS-S with a predicted \ha\ flux
 $\geq2\times10^{-16}$~\esc\ at any mass below our COSMOS completeness
 limit of $M_\star = 10^{9.8}$~\msun.\\

\begin{figure}
  \includegraphics[width=0.5\textwidth]{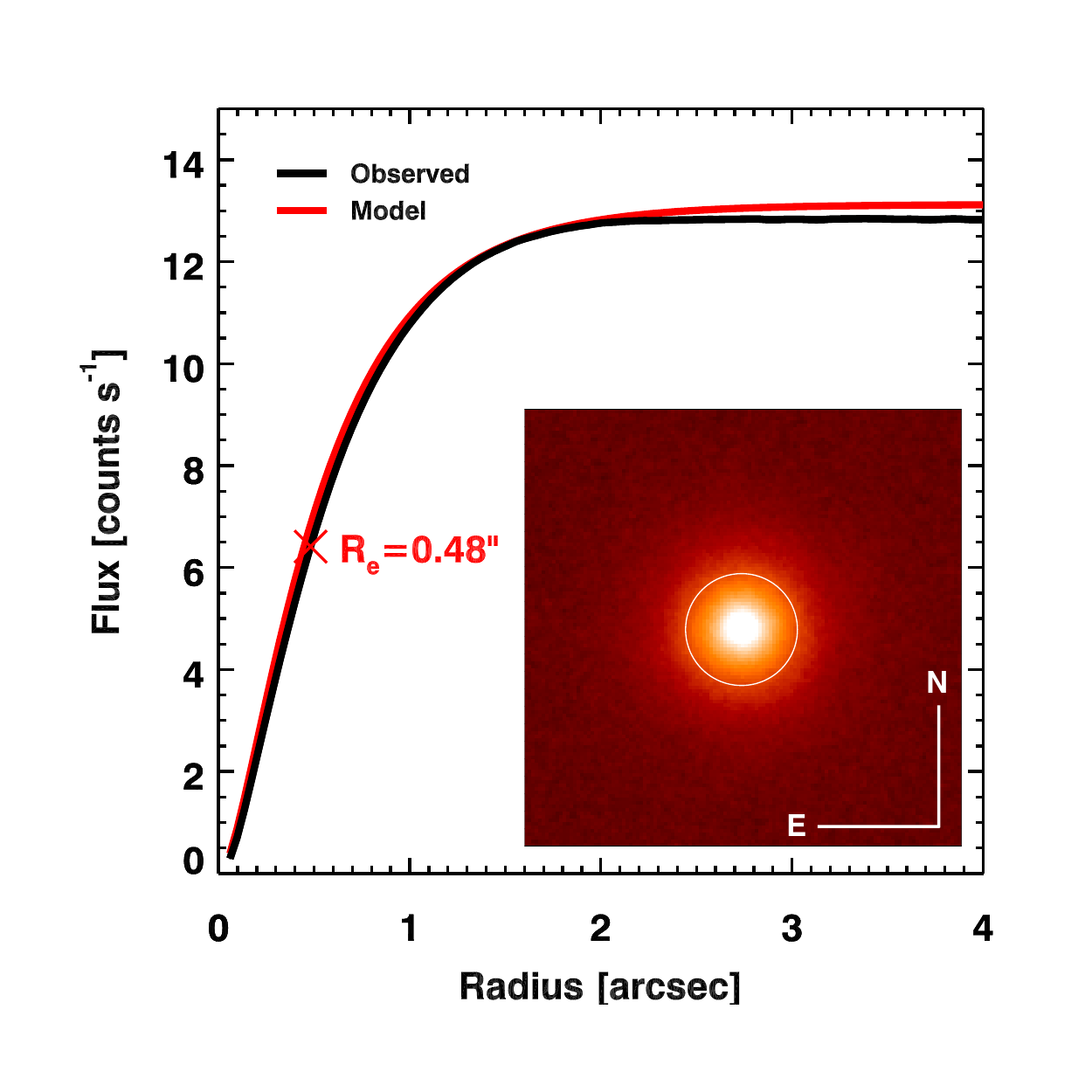}
  \caption{\textbf{Median \textit{HST}/ACS $i_{814}$ image of bright
      \ha\ emitters and its curve of growth.} The black line represents the
    curve-of-growth of the median image of $649$ galaxies from the
    FMOS-COSMOS survey with predicted \ha\ fluxes
    $\geq2\times10^{-16}$~\esc. The red line shows the result
      of the 2D light decomposition with \textsc{GALFIT}, including
      the deconvolution of the PSF. The red cross marks the effective
    radius. The inset shows the
    $3.75"\times3.75"$ median stacked image. 
    The white circle
    indicates the effective radius $R_{\rm e} = 0.48"$. It also
    roughly corresponds to the ``optimal'' aperture size maximizing
    the signal-to-noise ratio for the detection ($R=0.43"$, Section \ref{sec:optimization}).} 
\label{fig:stack}
\end{figure}

Since we do not have direct access to the spatial distribution
of the \ha\ flux, we measured the sizes in the
\textit{HST}/ACS \textit{i}$_{814}$ band, corresponding to
rest-frame $\sim3100$~\AA\ at $z=1.55$. Note that given the
  result on $f$, the attenuation of \ha\ and in the \textit{i}$_{814}$
  band are expected to be nearly identical. We present the analysis for
the $750$ emitters with \textit{predicted} \ha\ flux
$\geq2\times10^{-16}$~\esc, but the results do not change if we
consider only the spectroscopic subsample from the FMOS-COSMOS
survey. First, we extracted $15"\times15"$ cutouts from the COSMOS archive
and we visually inspected them. Considering that the area covered by
the \textit{HST}/ACS follow-up is smaller than the whole COSMOS field
and excluding strongly contaminated sources, we worked with $649$
objects in total. We show a collection of the latter in Appendix
\ref{appendixB}. Given their clumpy morphology, 
we recentered the cutouts on the barycenter of the light found by
\textsc{SExtractor} \citep{bertin_1996}, allowing for a small
fragmentation and smoothing over large scales. The final results do
not change if we center the images on the peak of the light distribution.
We, then, stacked the cutouts computing their median to minimize the impact of asymmetries
and irregularities. We finally measure the effective radius
with a curve-of-growth, obtaining $R_{\rm e} = (0.48\pm0.01)$~arcsec ($\sim4$~kpc
at $z=1.55$, Figure
\ref{fig:stack}). The uncertainty is obtained bootstrapping $1,000$
times the stacking procedure and extracting the curve of growth.
To confirm this estimate, we used \textsc{GALFIT}
\citep{peng_2010} to model the 2D light distribution with a Sersi\'{c}
profile, leaving all the parameters free to vary. To
extract a meaningful size directly comparable with the
previous estimate, we measured the effective (half-light) radius
of the PSF-deconvolved profile, obtaining $R_{\rm GALFIT} =
0.46"$. 
The $R_{\rm e}$ value is comparable with the effective radius of star-forming
galaxies on the average mass-size relations in literature
(i.e., median circularized $R_{\mathrm{e, circ}}=3.4-3.0$ kpc,
semi-major axis $R_{\mathrm{semi-major}}\sim4.7-4.1$ kpc for late-type galaxies with
$\rm{log}(M_\star/M_\odot) = 10.75$ at $z=1.25-1.75$,
\citealt{vanderwel_2014}). 

\section{Number counts of line emitters}
\label{sec:hacounts}
We compute the projected cumulative number counts of line emitters at $z\sim1.5$
starting from the photometric samples in COSMOS and GOODS-S. We base
the counts on the \textit{predicted} \ha, \oiii, and \oii\ fluxes as detailed above.
Then, we model the evolution of the number
counts of \ha\ emitters with cosmic time, a crucial
step in preparation of forthcoming large spectroscopic surveys with
Euclid \citep{laureijs_2009} and WFIRST
\citep{green_2012,spergel_2015}. Our method has the advantage of fully
exploiting the large number statistics of current photometric surveys and it
complements the classical approach based on a spectroscopic dataset and the
modeling of the evolution with redshift of the \ha\ luminosity functions
\citep{geach_2010,pozzetti_2016}. A detailed analysis of the \ha\ LF
for the FMOS-COSMOS survey is deferred to future work (Le F\`{e}vre et al. in prep.)

\subsection{\ha\ emitters: the FMOS-COSMOS redshift range}
\label{sec:fmosnumbercounts}
\begin{figure*}
  \includegraphics[width=0.49\textwidth]{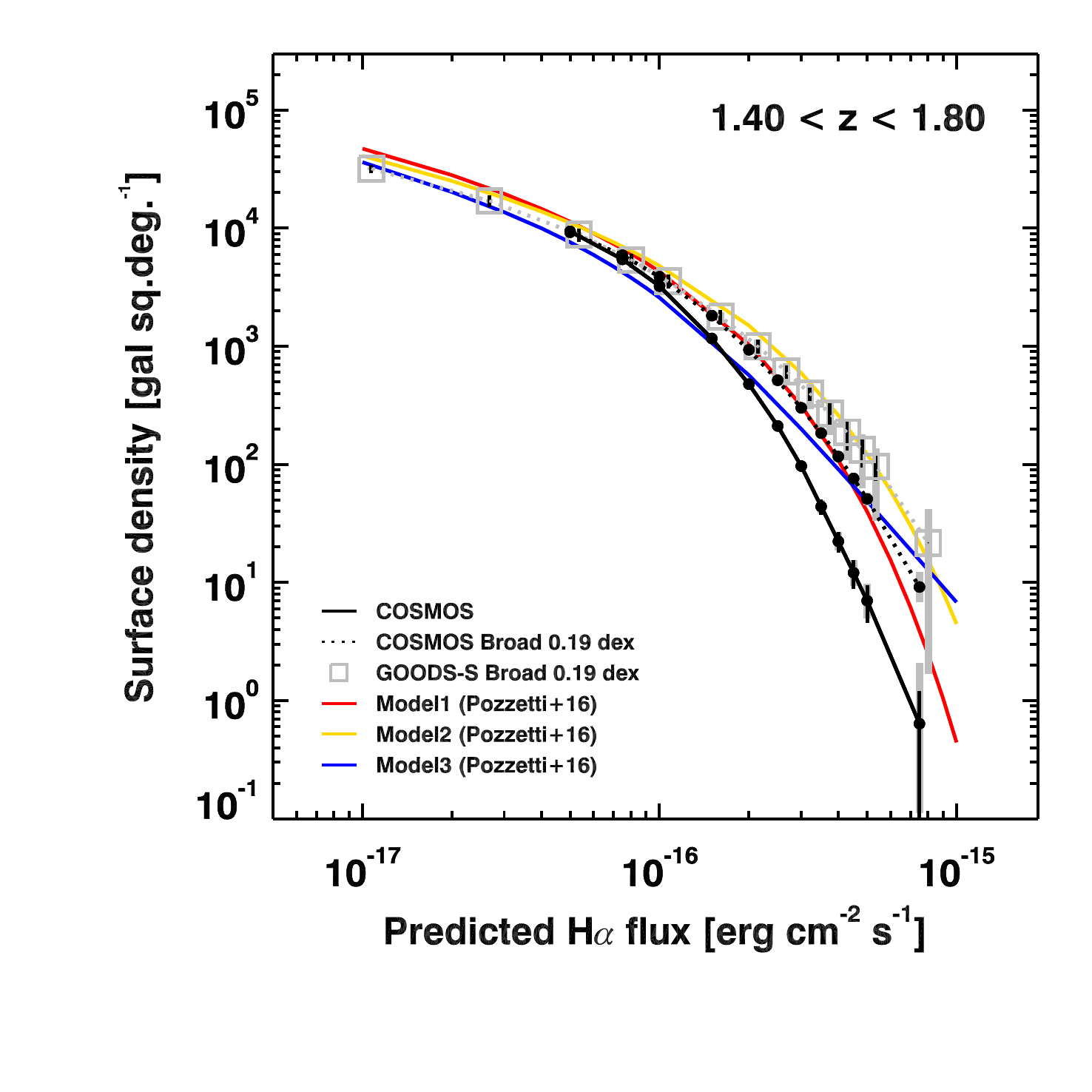}
  \includegraphics[width=0.49\textwidth]{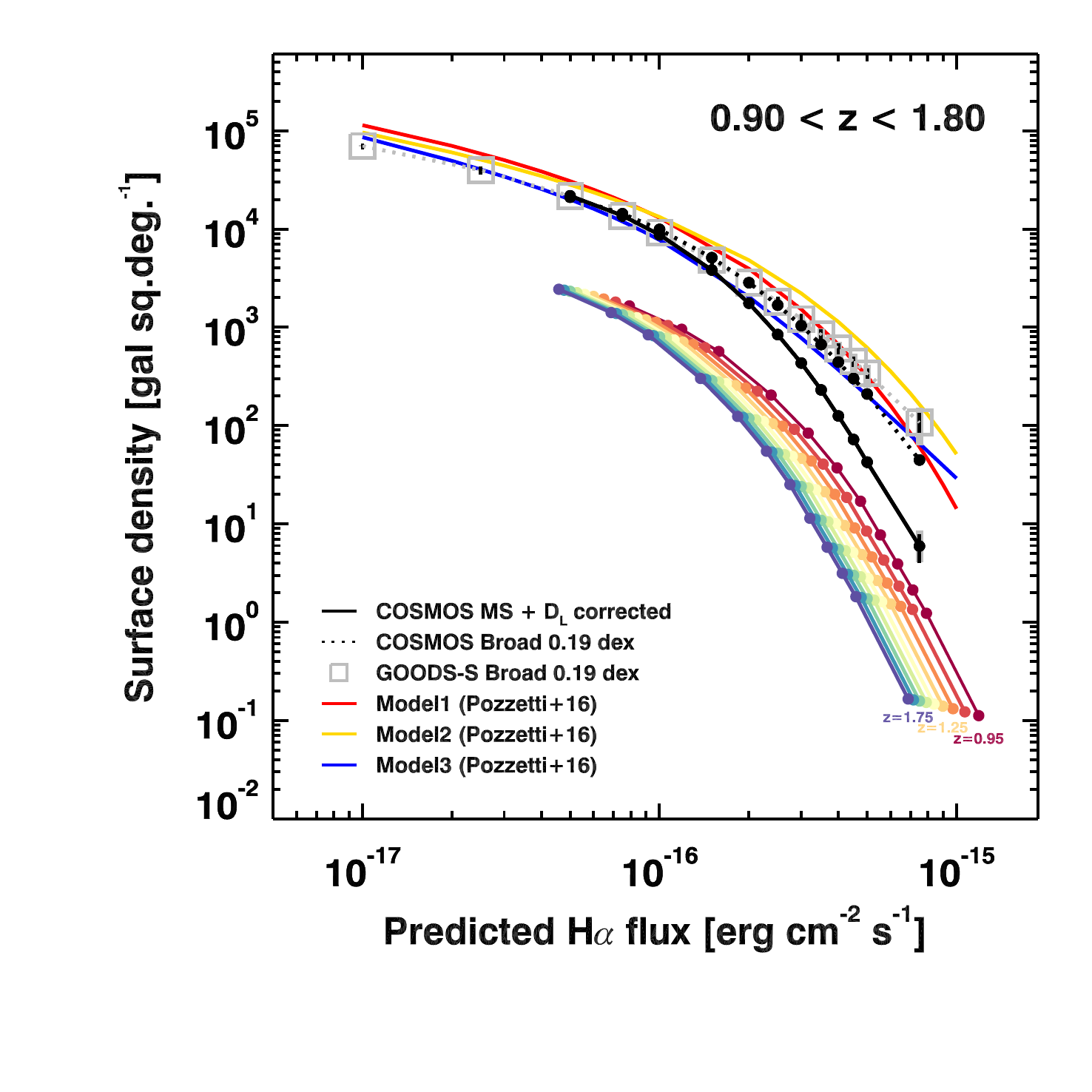}
  \caption{\textbf{Cumulative \ha\ counts.} \textit{Left:} Cumulative
    number counts in the redshift range $1.4<z<1.8$. The \ha\ fluxes
    are predicted from the photometry. The solid black line
  marks the cumulative counts integrated over the full redshift range.
  Grey bars indicate the Poissonian $68$\% confidence
  interval. Black bars show the $1\sigma$ uncertainty on cumulative counts
  from bootstrap and Monte Carlo simulations. The dotted
  black line marks the upper limit on cumulative counts, including the
  uncertainty on the predicted fluxes, causing a broadening
  of the original values. Grey squares and error bars show the
  upper limit on the cumulative counts for the GOODS-S photometric sample and
  their $1\sigma$ uncertainties. The red, golden, and blue solid
  lines mark Model 1, 2, and 3 by \citet{pozzetti_2016}. \textit{Right}:
  Cumulative number counts in the redshift range $0.9<z<1.8$ covered by the
   forthcoming Euclid mission. Colored lines mark the
  cumulative counts in $dz=0.1$ redshift slices (Section
  \ref{sec:redshiftevolution}). Other lines and symbols follow the same scheme of the
  left panel. \textit{Note: The lower
        and upper (convolved, ``broad'') counts can be obtained subtracting and adding the
        absolute error $\sigma_{\rm conv}$ to the ``average'' counts in
        Table \ref{tab:counts}.}}
\label{fig:counts}
\end{figure*}

 First, we computed the cumulative number counts for the redshift
range $1.4<z<1.8$ covered by the FMOS-COSMOS survey, starting from the
COSMOS and GOODS-S photometric samples spread over an area of
$1.57$~deg$^2$ and $0.054$~deg$^2$, respectively. The cumulative
number counts are reported in Table \ref{tab:counts} 
and shown in Figure \ref{fig:counts}.
We computed
the uncertainties on the cumulative counts both as Poissonian $68$\%
confidence intervals and from simulations. In order to capture
  the sample variance,
we bootstrapped 1,000 mock samples of the same size of the observed
one, randomly extracting objects from the photometric samples, allowing for any number of duplicates. We, then, recomputed the number
counts for each mock sample and estimated the uncertainties as the
standard deviation of their distribution for each flux. We further simulated the 
impact of the cosmic variance on small angular scales counting galaxies in areas of
0.26~deg$^2$ ($1/6$ of the total surface covered by the COSMOS
photometric sample) and $0.054$~deg$^2$, taken randomly
in the COSMOS field. We, then, added these contributions in
quadrature.\\

Furthermore, we included the effect of the uncertainties on the
predicted \ha\ fluxes on the final estimate of the number counts, as
necessary to fairly represent their scatter. These uncertainties
naturally spread out the counts in a flux bin to
the adjacent ones. In presence of an asymmetric distribution of
galaxies in the flux bins, this causes a net diffusion of objects in a
specific direction: in this case, from low towards high fluxes. This
happens because of the negative, steep slope reached in the brightest
flux bins, simply meaning that there are many more emitters at low
fluxes than at the high ones. Neglecting the uncertainties on the
predicted fluxes would, thus, result in an underestimate of the number
counts at high fluxes, since the low-flux population dominates over
the bright tail.  Note that this is relevant in our calculations,
given the relatively large uncertainty also in the brightest flux
tail, while this is generally not an issue for well determined total
fluxes (i.e., with narrow-band imaging or, in principle, grism
spectroscopy, but see Section \ref{sec:optimization}). The typical flux error is
$\sigma_{\rm pred}=0.1$~dex, obtained subtracting in quadrature the
error associated with the total observed
\ha\ flux from FMOS-COSMOS ($\sigma_{\rm obs}=0.17$~dex, dominated by
aperture corrections) from the dispersion of the distribution of $\rm{H\alpha_{obs}/H\alpha_{pred}}$
flux ratios ($\sigma=0.19$~dex, Figure \ref{fig:hapredictions}). Uncertainties related to SED
modeling and intrinsic scatter both contribute to this dispersion
(Section \ref{sec:hapredictions}). To simulate the diffusion of galaxies
from low to high fluxes, we convolved the counts per flux bin with a 
Gaussian curve of fixed width $\sigma_{\rm  broad}$ in the
logarithmic space, renormalizing for the initial counts per flux bin.
Finally, we recomputed the cumulative counts, now broadened by the errors
on predicted fluxes. Adopting the most conservative approach, we
set $\sigma_{\rm broad} = 0.19$~dex, as if all the dispersion of the
distribution of $\rm{H\alpha_{obs}/H\alpha_{pred}}$ were due to the
uncertainty on $\rm{H\alpha_{pred}}$. 
This procedures returns a strong upper limit on the cumulative number
count estimate, increasing the original values for the COSMOS
photometric sample by a factor of $\sim3$ at
\ha\ fluxes of $3\times10^{-16}$~\esc, as shown in Figure
\ref{fig:counts}. In the same figure we show the results of an identical analysis
applied to the GOODS-S photometric sample, along with the modeling
of the recent compilation of spectroscopic and narrow-band data and
LFs by \cite{pozzetti_2016}. All the curves refer to the same
redshift range $1.4<z<1.8$. The counts for the COSMOS and GOODS-S
samples are fully consistent within the uncertainties down to the
COSMOS completeness flux limit of $5\times10^{-17}$~\esc. The deeper
coverage of the rest-frame UV range available for GOODS-S allows us to
extend the number counts to \ha\ fluxes of
$1\times10^{-17}$~\esc. Below these limits, the convolved number
counts in the two fields are lower than the initial ones due to the
incompleteness. The cumulative counts are broadly consistent with the
empirical models by \cite{pozzetti_2016}, collecting several datasets
present in the literature. The agreement is fully reached when considering the effect the
uncertainties on the flux predictions. In
particular, our results best agree with Models 2 and 3,
the latter being derived from high-redshift data only, revising the number counts
towards lower values than previously estimated \citep{geach_2010}. 
Note that our selection includes only color-selected normal SFGs. 
Other potentially bright \ha\ emitters, such as low-mass starbursting galaxies and AGN, might
further enhance the final number counts (Section \ref{sec:discussion}). 
\begin{figure*}
  \includegraphics[width=\textwidth]{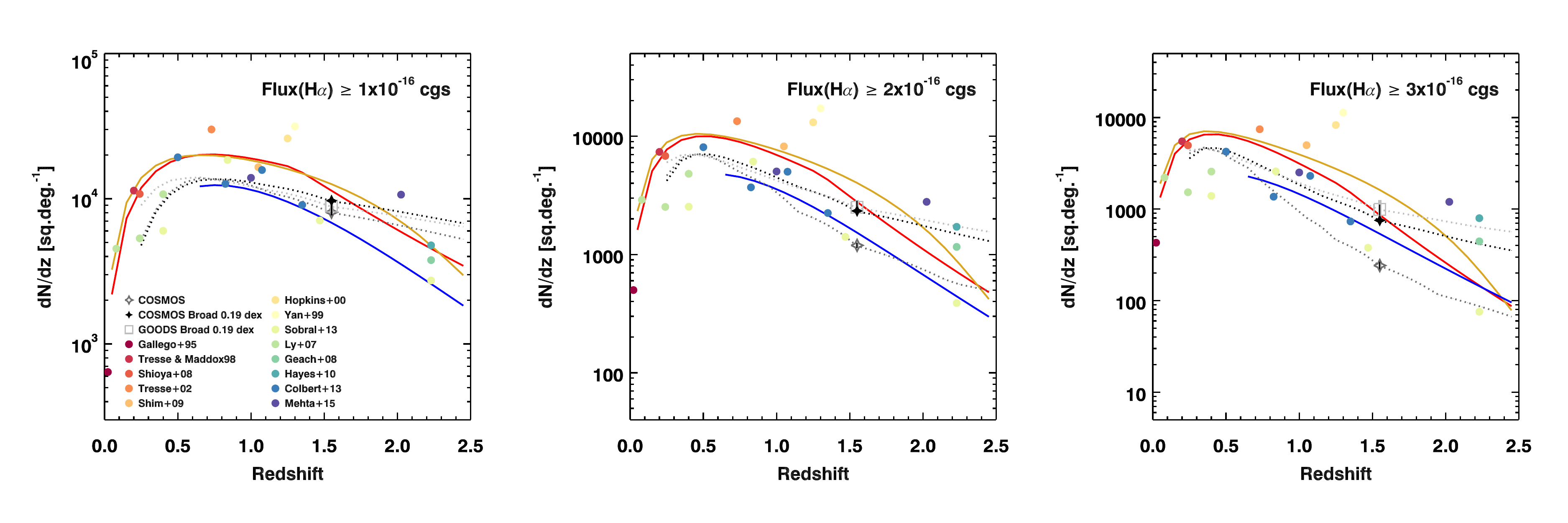}
  \caption{\textbf{Differential \ha\ counts.} The dark grey empty star 
    mark the differential counts $dN/dz$ from the photometric sample 
    in COSMOS. The black filled star corresponds to the upper limit on
    the counts, including the uncertainty on the predicted fluxes.
    The light grey empty square indicates the upper limit on
    the counts for the GOODS-S sample. Grey vertical bars indicate the Poissonian $68$\% confidence
    interval. Black bars show the $1\sigma$ uncertainty on cumulative counts
    from bootstrap and Monte Carlo simulations. The dark grey dotted
    line represents the $dN/dz$ counts for the COSMOS sample.
    The black and light grey dotted lines show the upper limits on the
    $dN/dz$ counts for the COSMOS and GOODS-S samples, respectively.
    The red, golden, and blue solid
    lines mark Model 1, 2, and 3 by \citet{pozzetti_2016}. Individual points from the literature
    in the compilation by \citet{pozzetti_2016} are displayed as
    filled circles. Differential counts for predicted \ha\ fluxes $\geq1\times10^{-16}$, 
    $\geq2\times10^{-16}$, and $\geq3\times10^{-16}$~\esc\ are shown in the left, central, and 
    right panels, respectively.}
\label{fig:dNdz}
\end{figure*}

\subsection{\ha\ emitters: redshift evolution}
\label{sec:redshiftevolution}
In order to compare our results with similar existing and forthcoming
surveys covering different redshift ranges, we modeled the time evolution
of expected \ha\ fluxes and counts. Our parametrization includes two
main effects regulating the \ha\ flux emerging from star
formation in galaxies:
\begin{itemize}
\item the increasing normalization of the Main Sequence with redshift as $(1+z)^{2.8}$
\citep{sargent_2014}: high-redshift sources are intrinsically
brighter in \ha\ due to higher SFRs at fixed stellar mass
\item fluxes decrease as the luminosity distance $D_\mathrm{L}^2(z)$
\end{itemize}
The mass-metallicity relation also evolves with redshift,
but its effects on the dust content of galaxies are compensated
by the increase of the gas fraction, so that the mass-extinction
relation mildly depends on redshift \citep{pannella_2015}. Moreover, the stellar
mass function of SFGs is roughly constant from $z\sim2$
\citep[i.e.,][]{ypeng_2010, ilbert_2013}. Therefore, these contributions and other secondary effects (i.e., a
redshift-dependent initial mass functions) are not included in the
calculation.\\

For reference, we computed the cumulative number counts integrated on
the redshift range $0.9<z<1.8$ that will be probed by the Euclid
mission. First, we assigned
the cumulative \ha\ counts from the COSMOS photometric 
sample to the redshift slice $1.5<z<1.6$, enclosing the average
redshift probed by the survey $\langle z \rangle = 1.55$, and we rescaled
them for the volume difference. Then, we split the calculation in redshift
steps of $dz = 0.1$, rescaling the \ha\ fluxes for each
redshift slice by $(1+z)^{2.8}/D_\mathrm{L}^2(z)$ and for the
volume enclosed. Note that 
rescaling the \ha\ fluxes effectively corresponds to a shift
on the horizontal axis of Figure \ref{fig:counts}, while the volume
term acts as a vertical shift. To compute the counts over the full
redshift range, we interpolated the values in the $dz=0.1$ slices on a
common flux grid and added them. We notice that modeling the evolution of the total
\ha\ fluxes with redshift increases by a factor of $\sim1.5$ the
cumulative counts for fluxes above $\geq2\times10^{-16}$~\esc\ obtained simply
rescaling for the volume difference the results for the COSMOS photometric sample to the
redshift range $0.9<z<1.8$. However, this increase might be partially
  balanced by an increasing fraction of massive galaxies becoming
  quiescent. Finally, we convolved the integrated counts
with a $0.19$~dex wide Gaussian to account for the uncertainty
  on the predicted \ha\ fluxes (assumed to be comparable with the one
  derived at $1.4<z<1.8$), obtaining an upper limit of the number counts. We calculated uncertainties as
Poissonian $68$\% confidence intervals and with bootstrap and Monte Carlo techniques as detailed in Section
\ref{sec:fmosnumbercounts}. 
We show the results of our modeling in Figure \ref{fig:counts}, along
with the empirical curves by \cite{pozzetti_2016} and the number
counts for the GOODS-S photometric sample, obtained applying the same
redshift rescaling as in COSMOS. When accounting for the uncertainties
on \ha\ fluxes, calculations for both COSMOS and
GOODS-S photometric samples are in agreement with the models by
\cite{pozzetti_2016} predicting the lowest counts over the $0.9<z<1.8$ redshift range. 
In this interval, we expect $\sim2,300$ galaxies~deg$^{-2}$ for \ha\ fluxes $\geq2\times10^{-16}$~\esc,
the nominal limit for the Euclid wide survey, and $8,500-9,300$ galaxies~deg$^{-2}$ from the GOODS-S
and COSMOS field, respectively, at a limit of $\geq1\times10^{-16}$~\esc, the
baseline depth for WFIRST. Integrating over $1.1<z<1.9$, similar to the formal limits of the WFIRST 
\ha\ survey, we expect $\sim6,200-6,800$ galaxies~deg$^{-2}$ above $\geq1\times10^{-16}$~\esc\ 
for the GOODS-S and COSMOS fields, respectively, in agreement with previous
estimates \citep{spergel_2015} within the uncertainties.\\

The consistency with empirical models and datasets in literature and the importance
of including the uncertainties of the predicted \ha\ fluxes are further 
confirmed by computing the differential counts $dN/dz$, shown in
Figure \ref{fig:dNdz}. These estimates are relevant for the forthcoming redshift surveys
and complement the cumulative counts shown in Figure \ref{fig:counts} and reported 
in Table \ref{tab:counts}. The three panels show the broad agreement
between the evolution of number counts we predict based on the simple
modeling of the MS and the public data at different \ha\ fluxes. For
these plots, we extended our calculations to the redshift interval
$0.2<z<2.5$. At lower redshift a large number
of the most massive and brighest \ha\ emitters are likely to quench with
time, causing an overestimate of counts. On the
other hand, the uncertainties on the evolution of the $f$ factor with
time and the increasing contribution of dust obscured SFGs to the
overall formation of new stars at $z>2.5$ limit the analysis above
this threshold. However, the evolution of the normalization of
the MS is enough to reproduce the growth and drop of the expected \ha\
counts over several Gyrs of cosmic time. Notice that we calculated the upper limits in each redshift slice
convolving with a Gaussian curves of fixed width of $0.19$~dex as
detailed in the previous section.
\begin{figure*}
  \includegraphics[width=\textwidth]{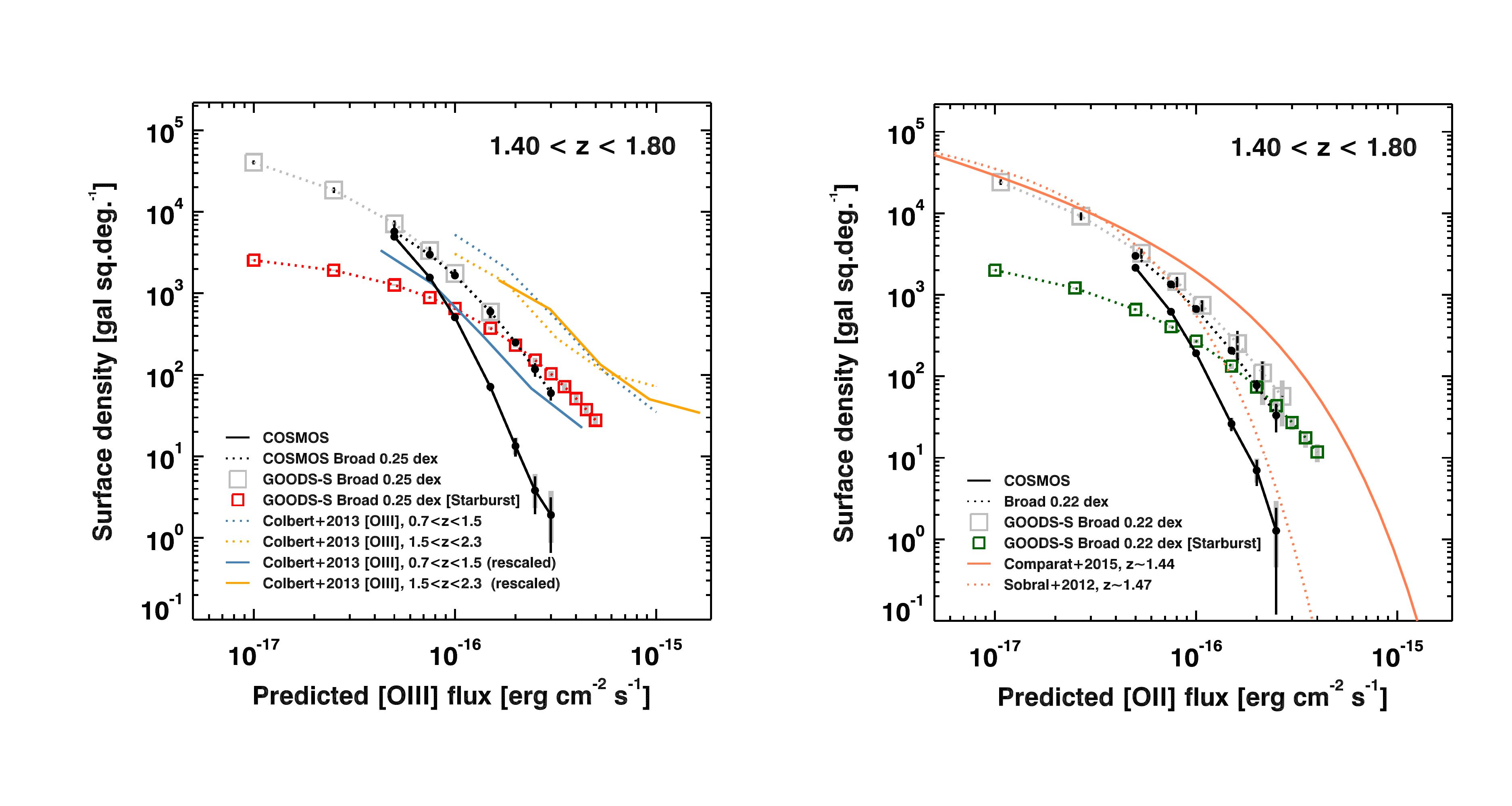}
  \caption{\textbf{Oxygen line emitters number counts.} \textit{Left}: Cumulative number counts of \oiii\ emitters in the redshift range
    $1.4<z<1.8$. The solid and dotted black lines
    mark the COSMOS cumulative counts and the upper limits keeping 
    into account the uncertainties on predicted fluxes. Grey squares
    indicate the upper limit on counts in GOODS-S. Red squares
    represent the upper limit on counts of simulated 
    starbursting galaxies in GOODS-S. Grey bars indicate the Poissonian $68$\% confidence
    interval. Black bars show the $1\sigma$ uncertainty on cumulative counts
    from bootstrap and Monte Carlo simulations. Yellow and
      blue dotted lines show the \oiii\ counts from the
    WISP survey by \citealt{colbert_2013}. Yellow and blue solid tracks
    show the same counts, but properly rescaled to match the cosmic
    volume within $1.4<z<1.8$ and the luminosity distance at $z\sim1.55$. 
    \textit{Right}: Cumulative number counts of \oii\ emitters in the redshift range
    $1.4<z<1.8$. The solid and dotted black lines
    mark the COSMOS cumulative counts and the upper limits keeping 
    into account the uncertainties on predicted fluxes. Grey squares
    indicate the upper limits counts in GOODS-S. Green squares
    represent the upper limit on counts of simulated 
    starbursting galaxies in GOODS-S. Error bars are
    coded as in the left panel. The orange solid and dotted lines indicate the estimate 
    derived integrating the luminosity functions in \citealt{comparat_2015} and \citealt{sobral_2012}
    at $z\sim1.45$ and assuming their validity over the redshift range
    $1.4<z<1.8$. \textit{Note: The lower
        and upper (convolved, ``broad'') counts can be obtained subtracting and adding the
        absolute error $\sigma_{\rm conv}$ to the ``average'' counts in
        Table 2.}}
\label{fig:oxygencounts}
\end{figure*}

\subsection{\textsc{[OII]} and \textsc{[OIII]} number counts at $1.4<z<1.8$}
\label{sec:oxygencounts}
We computed the number counts of oxygen line emitters based on the
\oii\ and \oiii\ flux predictions in the redshift range
$1.4<z<1.8$. We applied the same method described in Section
\ref{sec:fmosnumbercounts}, keeping into
account the uncertainties on the predicted fluxes convolving
  the number counts with Gaussian curves of fixed width. Results are shown in
Figure \ref{fig:oxygencounts} and reported in Table
2. The \oiii\ number counts are roughly
consistent with the results from the WISP survey presented in
\cite{colbert_2013}, once (i) rescaling for the volume and the
  luminosity distance is properly
taken into account, and (ii) low mass galaxies are included in the
calculation. Our estimates fall between the WISP counts in the
  $0.7<z<1.5$ and $1.5<z<2.3$ intervals. Given how we predict \oiii\ fluxes (Section
  \ref{sec:oiii}), the increase of the average \oiii/\hb\
ratios and of the MS normalization with redshift can explain
  the offset between our estimates and Colbert's et
  al. (2013). Moreover, low mass galaxies play a critical role, since they have
intrinsically higher \oiii/\hb\ ratios. In fact, bright \oiii\ emitters in the WISP survey are generally low mass ($M_\star \sim 10^{8.5}-10^{9.5}$~\msun, \citealt{atek_2011,
  henry_2013}). The low mass regime is also sensitive to the
presence of high sSFR, unobscured, starbursting galaxies, thus we
expect them to be relevant for the
\oiii\ number counts. We simulated their impact on the counts from the
GOODS-S sample as detailed in Section
\ref{sec:discussion:sb}, and we found a substantial extension of
counts above $1.5\times10^{-16}$~\esc, 
the limit we reach when counting normal Main-Sequence galaxies (Figure
\ref{fig:oxygencounts}). Starbursting galaxies are expected to
  reach \oiii\ fluxes of $3\times10^{-16}$~\esc. 
In the interval $1.4<z<1.8$, we expect $\sim1,100$ and $\sim150$ galaxies~deg$^{-2}$
above $\geq1\times10^{-16}$ and $\geq2\times10^{-16}$~\esc, averaging the results for the
COSMOS and GOODS-S fields. Including the effect of low-mass starburst, we expect 
$\sim1,700$ galaxies~deg$^{-2}$ for \oiii\ fluxes above $\geq1\times10^{-16}$.\\ 

For what concerns the number counts of \oii\
emitters, the contribution of low mass galaxies and the different
mass completeness limits explain the difference between the COSMOS and GOODS-S
samples. The number counts we derived fall in the range of recent estimates
at $z\sim1.45$ by \citet{sobral_2012} and \citet{comparat_2015}. We derived
these counts integrating their LFs assuming their validity over the redshift range 
$1.4<z<1.8$ and for fluxes up to $3\times10^{-16}$~\esc, the limit of our estimates. 
We divided the counts by \citet{comparat_2015} by ln$(10)$ to account for the different 
normalizations of the two LFs.
Our calculations are in agreement with the estimates by \citet{sobral_2012} up 
to $\sim1\times10^{-16}$~\esc, while we find higher counts above this
threshold (a factor $2-3.5\times$ at $\sim2\times10^{-16}$~\esc\ considering our
``average'' estimate reported in Table 2 for COSMOS and GOODS-S, respectively).
On the other hand, we systematically find less counts than in \citet{comparat_2015},
a factor $4.5-4\times$ ($3-2.5\times$) at
  $\sim1\times10^{-16}$~\esc\ and $11-6.5\times$ ($6-4.5\times$) at $\sim2\times10^{-16}$~\esc\
considering the ``average'' estimates (the broadened counts) for COSMOS and GOODS-S, respectively.
We note that the LF by \citet{comparat_2015} probes only the tail of the brightest emitters, finding a larger 
number of them than what extrapolated by a fit at lower fluxes by \citet{sobral_2012} (see Figure 13
in \citealt{comparat_2015}). Part of the discrepancy we find is due to the correction for the extinction
of the Galaxy that \citet{comparat_2015} applied, while we report purely observed
and dust reddened fluxes. Moreover, the different sample sizes
of \cite{sobral_2012},
\citet{comparat_2015}, and our work might affect the results in the poorly populated 
tail of bright emitters. Over the redshift range $1.4<z<1.8$, we
expect $2,600$ ($2,700$) and $\sim400$ ($\sim500$) 
galaxies~deg$^{-2}$ based on the COSMOS (GOODS-S) field ``average''
estimate for \oii\ fluxes of $\geq5\times10^{-17}$
and $\geq1\times10^{-16}$~\esc\ (Table 2). These fluxes correspond to $\sim8\sigma$ and $\sim15\sigma$ 
detection thresholds expected for the Prime Focus Spectrograph survey in the same redshift
range \citep{takada_2014}. When including the effect of
  low-mass, starbursting galaxies (Section \ref{sec:discussion:sb}),
  we, thus, expect $\sim3400$ and $\sim700$ galaxies~deg$^{-2}$ at fluxes of $\geq5\times10^{-17}$
and $\geq1\times10^{-16}$~\esc, as derived from the average counts in
GOODS-S in the range $1.4<z<1.8$.

\section{Discussion}
\label{sec:discussion}
In the previous sections we showed how it is possible to estimate
number counts of line emitters using solely the photometric information and a
calibration sample of spectroscopically confirmed objects, reaching a
precision at least comparable with the
one achieved with standard approaches, generally based on small spectroscopic
samples and extrapolations of the LFs. We
computed the number counts for the redshift slice $1.4<z<1.8$ covered
by our calibration sample from the FMOS-COSMOS survey and we
extended our calculation for the \ha\ emitters to the $0.9<z<1.8$ interval probed by the
Euclid mission, as a reference. We now envisage possible
 caveats and developments of this work. 

\subsection{The effect of [N II] lines on low resolution spectroscopy}
\label{sec:discussion:nii}
\begin{figure*}
  \includegraphics[width=0.49\textwidth]{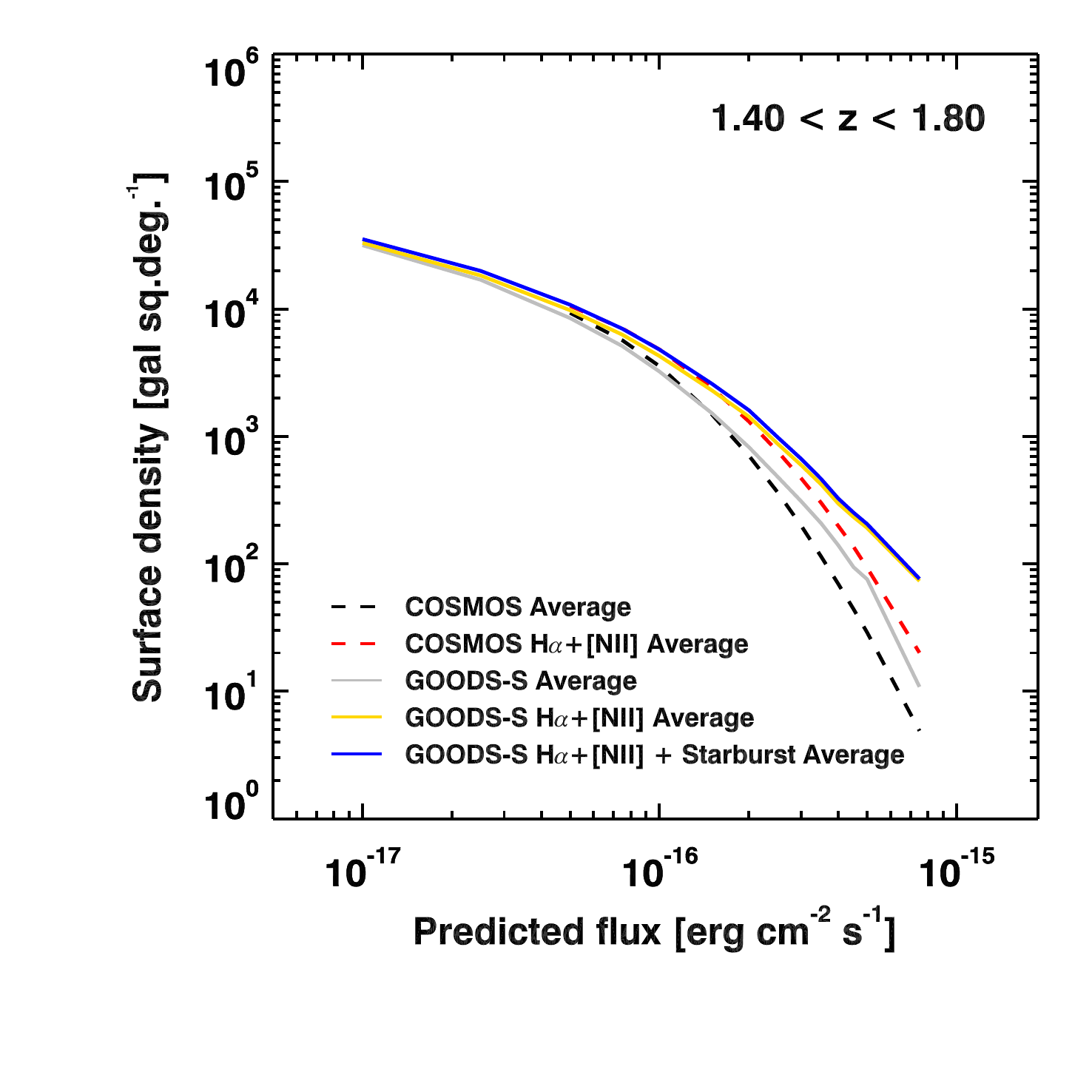}
  \includegraphics[width=0.49\textwidth]{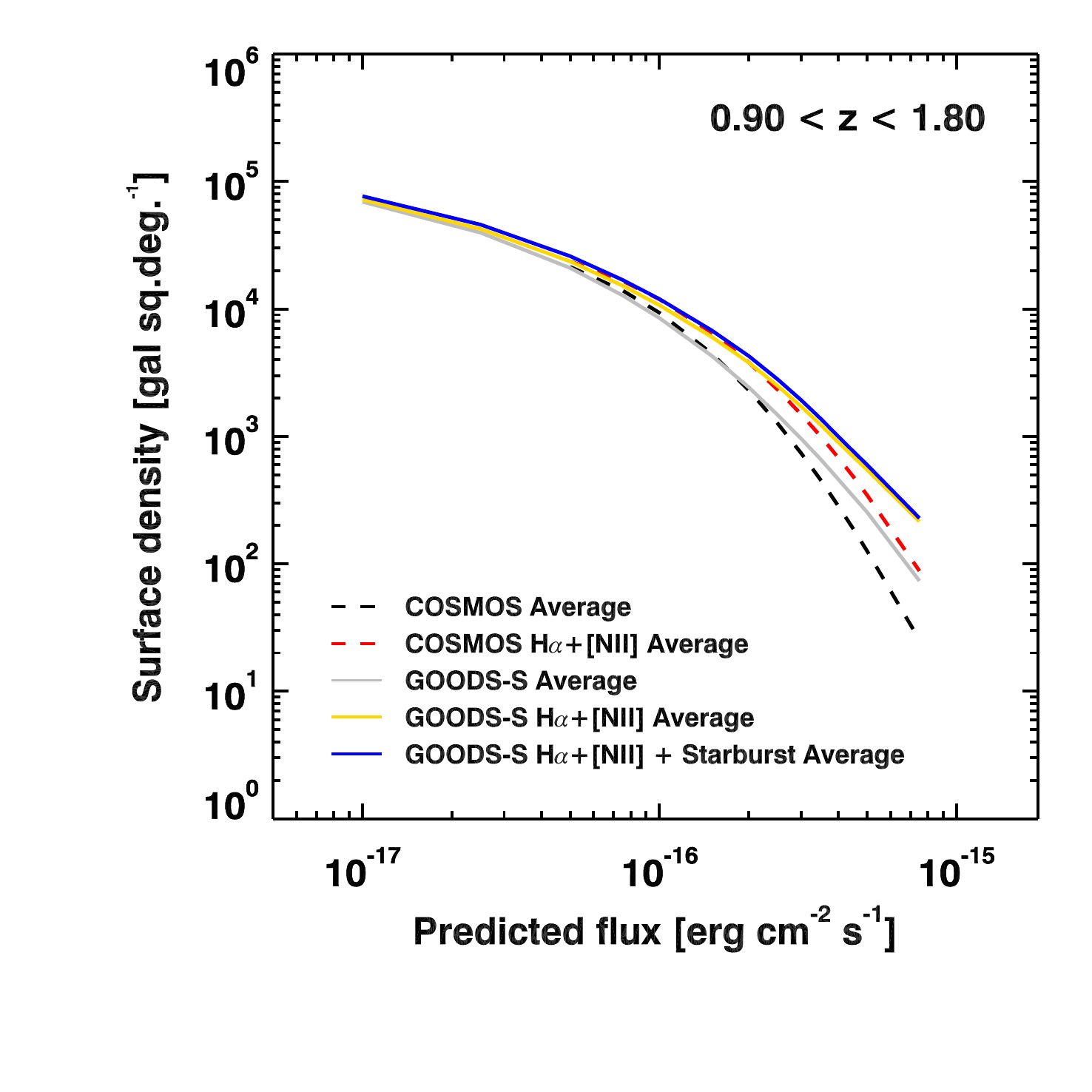}
  \caption{\textbf{Effect of unresolved} \nii\ emission and
      starbursting galaxies on \ha\ counts. The dashed black and
      red lines mark the \ha\ and \ha+\nii\ counts average estimate in
      the COSMOS field, respectively (Tables \ref{tab:counts} and \ref{tab:hanii}). The solid grey and
      yellow lines mark the \ha\ and \ha+\nii\ counts in
      the GOODS-S field.  The solid blue line
      indicates the \ha+\nii\ counts in the GOODS-S field,
      including the effect of starbursting galaxies (Table \ref{tab:sb}). \textit{Left:}
      FMOS-COSMOS redshift range of $1.4<z<1.8$. \textit{Right}:
      Full redshift range of $0.9<z<1.8$ covered by the
      forthcoming Euclid mission.}
\label{fig:niicorr}
\end{figure*}

In Section \ref{sec:hacounts}, we computed the galaxy number counts based on the
aperture-corrected \ha\ fluxes only. However, future slitless spectroscopy will not be able to
resolve the \nii-\ha\ complex, resulting in a boost of galaxy number
counts when the \nii\ flux is high. In Section \ref{sec:properties:spectroscopy} we
found an average line ratio of $\rm{log}($\nii/H$\alpha) \sim -0.5 $
for the bright emitters observable by Euclid, and we provided a simple
parametrization of the relation between log(\nii/\ha) and the total observed
\ha\ fluxes (Figure \ref{fig:specprop}). This relation can be 
extended at higher redshift, but it must be taken
  with caution, being naturally affected by observational biases
  \citep{kashino_2017}. 
We, thus, model the effect of the \nii\ flux boost fitting a first-order polynomial relation to the FMOS-COSMOS observed
  $\mathrm{log}(M_\star/M_\odot)$ - $\mathrm{log}($\nii$/$\ha$)$
  relation \citep[Sample-1, Table 2, Figure 14 in][]{kashino_2017} and
  applying a mass-dependent correction to each source. We show the
  results on the number counts in Figure \ref{fig:niicorr}. We extended the
  number counts to the $0.9<z<1.8$ interval assuming the same
  correction. Note that the redshift evolution of the mass-metallicity
  relation \citep[i.e.,][]{steidel_2014,sanders_2015} might impact
  this correction.\\
  
 We report in Table \ref{tab:hanii} the counts for
 \ha$+$\nii\ emitters. The flux boost due to unresolved \nii\
   emission increases by a factor of $\sim1.8\times$ ($\sim1.6\times$)
     the \ha\ number
     counts above $2\times10^{-16}$~\esc\ in the range $1.4<z<1.8$
     ($0.9<z<1.8$),  as derived from the average counts both in the
     COSMOS and GOODS-S fields. 

\subsection{The AGN contribution} 
\label{sec:discussion:agn}
Strong line emitters such as AGN or starbursting galaxies might
increase the number counts as well. We flagged and excluded from our
COSMOS sample known \textit{Chandra} detected sources in
the catalog by \cite{civano_2016}, since we could not reliably predict 
\ha\ fluxes based on their photometry. However, considering only the
\textit{Chandra} sources with an estimate of the photometric
redshift by Salvato et al. (in prep.), $\sim17$\% of the X-ray
detected sample by \cite{civano_2016} (671/4016 galaxies) lie at $1.4<z<1.8$, corresponding to
471 objects per deg$^{2}$ in this redshift range. This represents a 
minimal fraction of the overall population of SFGs composing our COSMOS
photometric sample ($31,193$ objects in total). On the other hand,
the color-selection we adopted does not prevent low luminosity or
obscured AGN to be included in the final sample. Moreover, the
FMOS-COSMOS selection function did include some X-ray detected AGN
\citep{silverman_2015}. However, only $11$ galaxies in the
\textit{Chandra} catalog by \cite{civano_2016} are detected as \ha\
emitters with fluxes $\geq2\times10^{-16}$~\esc, representing a
fraction of $8$\% of the overall bright FMOS-COSMOS sample. Therefore,
X-ray AGN should not provide a significant contribution to the \ha\
number counts at high fluxes in the redshift range $1.4<z<1.8$. 
\begin{figure}
  \includegraphics[width=0.5\textwidth]{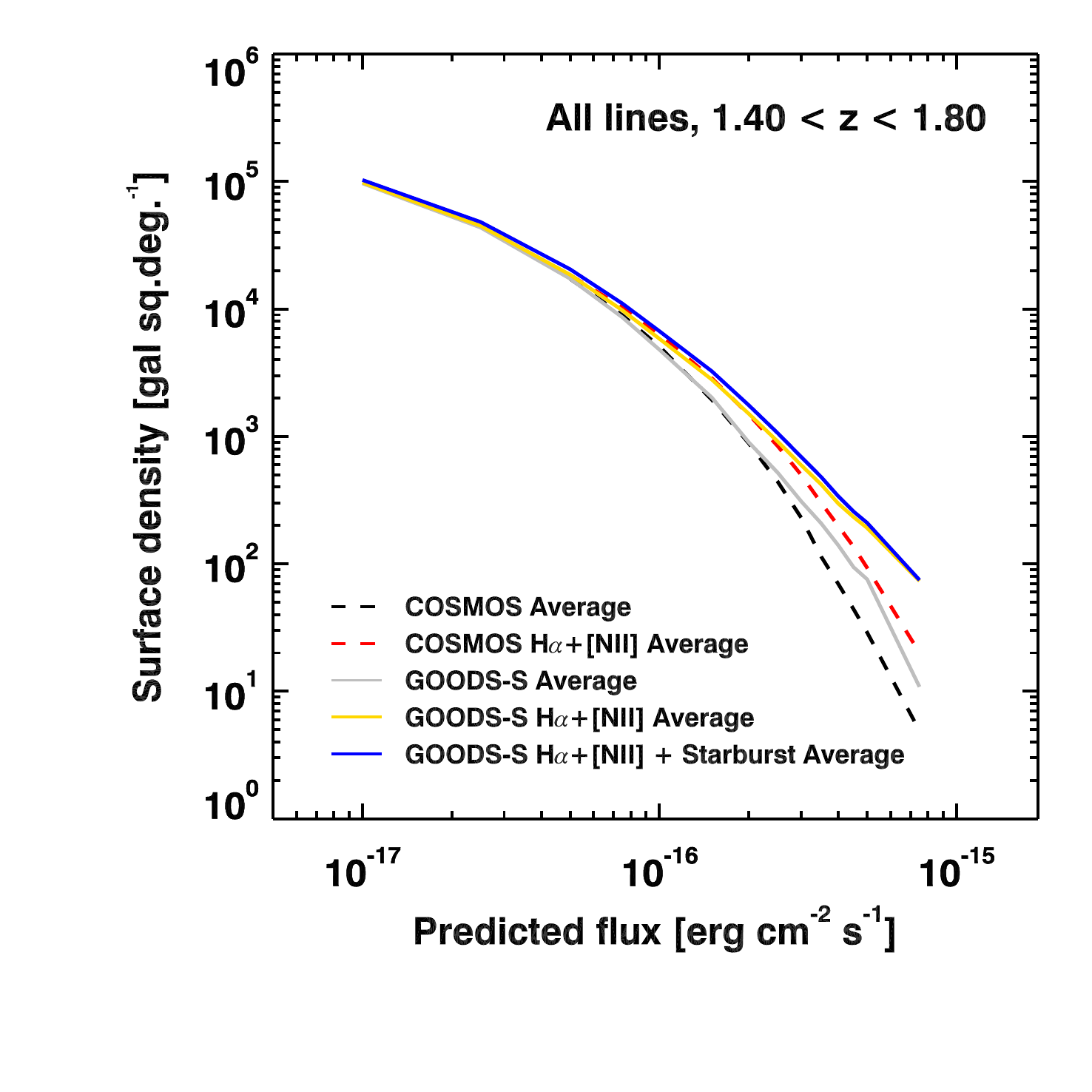}
  \caption{\textbf{Total cumulative number counts of line emitters at $1.4<z<1.8$.}
    The black dashed line indicates the cumulative number
      counts obtained adding the average estimates of the \ha,
    \oii, and \oiii\ emitter counts in the COSMOS field (Table
      \ref{tab:counts} and 2). The red dashed line
    shows the counts in COSMOS when taking into
    account the \nii\ unresolved emission (``average'' estimates
    in Table \ref{tab:hanii}).
  The grey, gold, and blue solid lines mark the
    cumulative counts for emitters in GOODS-S
    considering (i) \ha, \oii, and \oiii\ emitters;
    (ii) including the effect of \nii\ unresolved emission as for
    the COSMOS field; (iii) finally adding
    the population of low-mass starbursting galaxies (Table \ref{tab:sb}).}
\label{fig:allemitters}
\end{figure}

\subsection{Starbursting galaxies}
\label{sec:discussion:sb}
Given the large dust attenuation, only few \ha\ photons are expected
 to escape from massive starbursting galaxies (i.e, lying several times above
  the main sequence at fixed redshift). However, at moderate stellar masses ($M_\star
\lesssim 10^{9}-10^{10}$~\msun) galaxies showing high specific SFR
(sSFR) and extreme line EWs might contribute to the number counts
\citep{atek_2011}. To assess this effect on the cumulative counts of
\ha\ emitters, we simulated a population of starbursting galaxies at 
$M_\star < 10^{10}$~\msun\ artificially increasing their SFRs by a
factor of $\times4$ and considering a volume number density equal to
$4$\% of the one of main sequence SFGs \citep{rodighiero_2011}. 
Note
that the choice of a mass limit of $10^{10}$~\msun\ to simulate
starburst is conservative, as extreme sSFR and EW in existing slitless
spectroscopic surveys occur at $M_\star \sim 10^{8.5}-10^{9.5}$~\msun
\citep{atek_2011}. Since
more reliable SFRs are available at low stellar masses in GOODS-S
than in COSMOS, we used the GOODS-S for the experiment. We, then,
recalculated the \ha\ fluxes and the
number densities for the starburst population as in
Sections \ref{sec:hapredictions} and \ref{sec:hacounts}. We
  show the results in Figure \ref{fig:niicorr} and report the counts
  for stabursting galaxies in Table \ref{tab:sb}. The increase of the
\ha\ cumulative number counts due to the low mass,
starbursting population is of $\sim15$\% and $20$\% at
$1\times10^{-16}$~\esc\ and $2\times10^{-16}$~\esc, respectively,
at both $1.4<z<1.8$ and $0.9<z<1.8$. Therefore, our best
  estimates for \ha\ number counts including the starbursting
  population are $\sim3,800$ and $\sim1,000$ ($\sim9,700$ and $\sim2,900$)
  galaxies~deg$^{-2}$ in the redshift interval $1.4<z<1.8$
  ($0.9<z<1.8$) for \ha\ fluxes $\geq1\times10^{-16}$~\esc\ and
  $\geq2\times10^{-16}$~\esc, respectively, as evaluated from the
  average counts in GOODS-S (Tables \ref{tab:counts} and \ref{tab:sb}).\\

The impact of low-mass
  starburst on the number counts of \oii\ and \oiii\ emitters is 
  relevant (Figure \ref{fig:oxygencounts}, Table
  \ref{tab:sb}, and Section \ref{sec:oxygencounts}). In the redshift
range $1.4<z<1.8$, these galaxies increase by $\sim50$\% the number
counts derived from Main-Sequence objects at fluxes $\geq1\times10^{-16}$~\esc.\\

Finally, we underline that, in order to reach their main scientific
 goals in cosmology, future spectroscopic surveys need to map the
  highest possible number of spectroscopic redshifts, irrespectively
  of which lines are detected. We, thus, collected the cumulative
  number counts of \ha, \oii, and \oiii\ emitters in the redshift
  range $1.4<z<1.8$ at which we calibrated the predicted fluxes.
 The results are shown in Figure \ref{fig:allemitters}, where
   we also included the effect of a possible flux boost due to
   unresolved \nii\ emission and the impact of starbursting
   galaxies as detailed above. We did not
 attempt to extend these predictions to different redshift ranges given the
 uncertainty of the extrapolations of the recipes we adopted to
 estimate the oxygen emission lines.

\subsection{Estimating a survey effective depth and return}
\label{sec:optimization}
\begin{figure}
  \includegraphics[width=0.5\textwidth]{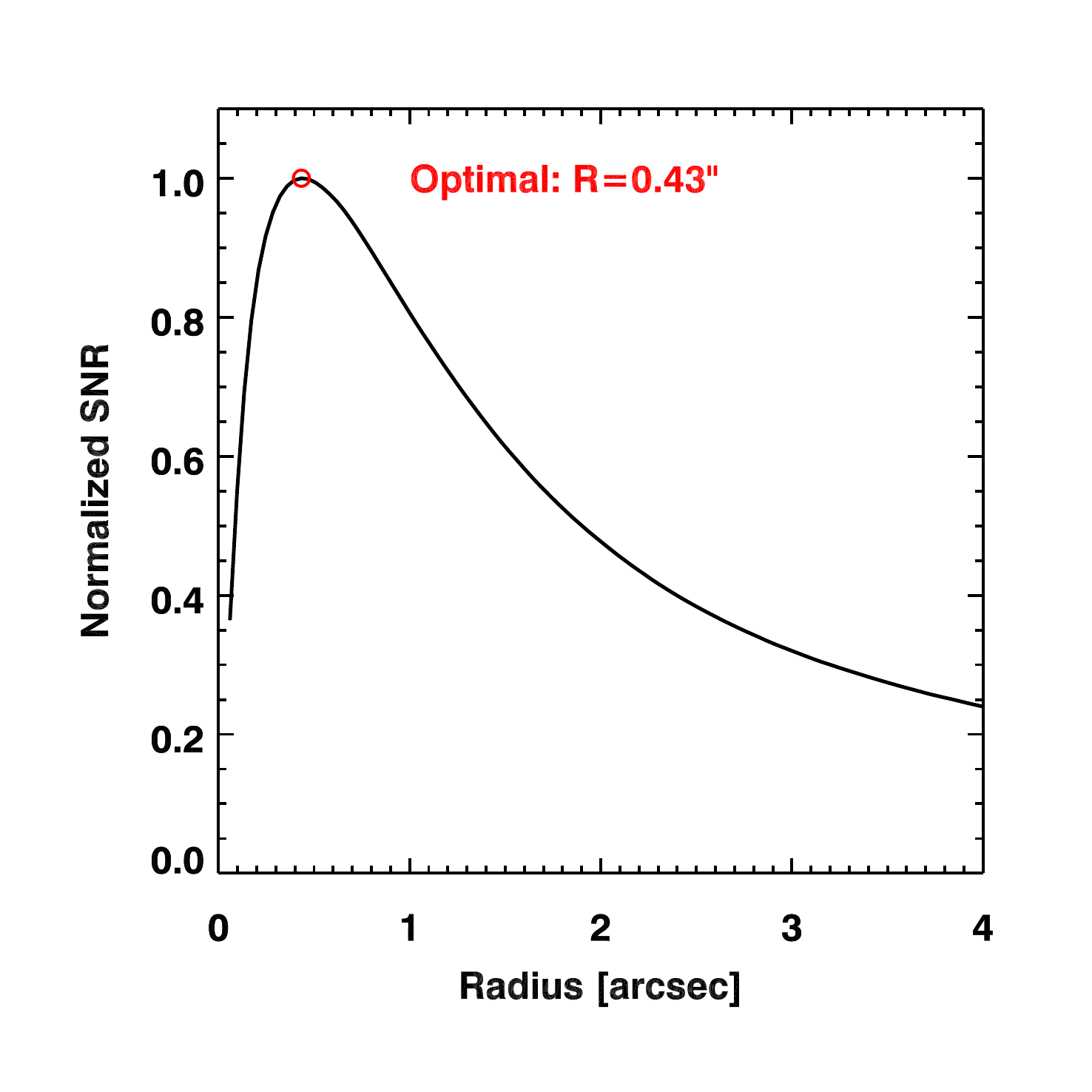}
  \caption{\textbf{Signal-to-noise ratio in circular aperture photometry of bright
      \ha\ emitters.} The black line represents the
    signal-to-noise ratio in circular apertures as a function of 
    their radius for the median \textit{HST}/$i_{814}$ image of
    bright \ha\ emitters in Figure \ref{fig:stack}. We normalized
    the curve to its peak. The red circle marks the radius maximizing
    the signal-to-noise ratio.} 
\label{fig:snr}
\end{figure}
In order to optimize the detectability and, thus, the number of detections for extended 
objects like galaxies, one has to reach a compromise
between (i) recovering as much as possible of
galaxies' flux, which requires large apertures; and (ii) limiting
the noise associated with the measurement, obtained minimizing the apertures.
This leads to a situation in which the \textit{optimal} aperture is driven by
the galaxy surface brightness profile, as discussed in the previous sections. 
Moreover, flux measurements are necessarily performed in
some apertures, and the ensuing flux losses must be taken into account
when analyzing the performances of a survey. For example,
spectroscopic surveys with multi-object
longslits or fibers with fixed diameters will be affected
by losses outside the physically pre-defined apertures. Aperture corrections
introduce further uncertainties on the total flux estimates,
thus the effective depth of a survey is shallower in terms
of total galaxy flux than what computed inside the aperture. A similar
effect also influences slitless spectroscopy: despite providing a
high-fidelity 2D map of each emission line in galaxies and allowing for recovering the
full flux under ideal circumstances, sources must be first robustly
identified before emission line fluxes can be
measured. The advantage of slitless spectroscopy is that the size and shape of
apertures might in principle be adjusted to the size of each object,
not being physically limited by a fiber or slit.  

Based on the stacked image of the \ha\ emitters with fluxes
$\geq2\times10^{-16}$~\esc\ shown in Figure
\ref{fig:stack}, we estimated the optimal radius for the circular
aperture that maximizes its signal-to-noise ratio (Figure \ref{fig:snr}).
This radius is $0.43$'' ($\sim0.9$~$R_{\rm e}$), 
causing an aperture loss of a factor of $\sim2.2\times$.
The flux losses ensuing any aperture measurement imply a higher
``effective'' flux limit of a survey -- defined as the 
minimum {\em total} emission line flux recoverable above a given
signal-to-noise detection threshold -- than the ``nominal'' limit
defined in a specific aperture. For example, observations
designed to provide secure detections down to a line flux $F_{\rm ap}$
within an aperture of radius $R=R_{\rm e}\sim0.5"$ 
(i.e., the ``nominal'' depth) would set an ``effective'' depth of
$F_{\rm eff}=2F_{\rm ap}$. This effective depth can be used to assess
the ``return'' of the survey, i.e., the number of recoverable
spectroscopic redshifts, by comparing with the cumulative number
counts of galaxies above $F_{\rm eff}$ as in Figures \ref{fig:counts} and
\ref{fig:oxygencounts}, and Tables
\ref{tab:counts} and 2. In fact, as common
practice, we derived the line fluxes in Section
\ref{sec:linepredictions} from \textit{integrated}, observed SED
properties, thus
not taking into account the size of the galaxies.
If neglected, aperture losses cause an increase of the
effective flux limit with respect to the nominal one and
a decrease of the return at any flux. However, given
the shape of the number counts, this effect is more
pronounced at high than at low fluxes. For reference, the total number
of detections for a nominal sensitivity $F_{\rm ap}
\geq2\times10^{-16}$~\esc\ inside a 0.5'' circular aperture would
correspond to a decrease by a factor of $\sim10$ of the return when
considering the effective depth $F_{\rm eff} = 2F_{\rm ap} \geq4\times10^{-16}$,
considering the case of \ha\ emitters in the COSMOS field (Table \ref{tab:counts}). On
the other hand, for $F_{\rm ap} \geq5\times10^{-17}$~\esc, the return
drops by a factor of $\sim3$ when estimating it at the corresponding
effective depth $F_{\rm eff}\geq1\times10^{-16}$~\esc. The smaller
factor at lower fluxes is due to the  flattening of the counts and it
could be overestimated, since such weaker emitters likely
have typical sizes smaller than we estimated in Section
\ref{sec:properties:sizes}, resulting in lower flux losses.
Note that, when computing counts within fixed apertures, we 
kept into account the evolution of the intrinsic sizes of SFGs
($R_{\rm e} \propto (1+z)^{-0.8}$, \citealt{vanderwel_2014,
  straatman_2015}) when assessing the effect for redshift intervals
larger than $1.4<z<1.8$. Moreover, the effect of the PSF of
\textit{HST}/ACS is negligible on the estimate of the optimal
aperture, while it may play a role for ground based and seeing-limited
observations. 

Adopting apertures larger than the optimal one, the flux losses and
the difference between nominal and effective depths are reduced. For
example, considering circular apertures of $2"$ diameter or,
equivalently, rectangular apertures of $1"\times3.4"$ ($\sim2R_{\rm
  e}\times7R_{\rm e}$) would reduce the aperture losses to only a
factor of $\sim1.2$, the pseudo-slit mimicking the longslit
spectroscopic case and a possible choice for the extraction of
slitless spectra. In this case, the effective depth would be only
$1.2\times$ shallower than the nominal depth, and the implied change
in return would also be fairly limited (a factor of $1.2 - 1.6$ at
$5\times10^{-17}$ and $2\times10^{-16}$~\esc,
respectively), if aperture losses are neglected. 
Note, however, that at fixed integration time,
using apertures of any shape, but larger - or smaller - than the optimal one
decreases the achievable nominal signal-to-noise ratio, further reducing
the return with respect to the optimal case presented above.
Doubling the aperture area does not come for free, as it requires a $4\times$ higher
integration time to reach the same flux limit with the same signal-to-noise ratio. 
Hence, adopting larger apertures for line detection to reduce aperture losses, without adjusting
accordingly the exposure time, is not a way to boost the return of a
survey, as it instead reduces the return with respect to the optimal case. 
Following the definitions of ``effective'' and ``nominal'' depths, any
possible combination of flux losses and corresponding survey returns
can be estimated using the profile given in Figure \ref{fig:stack} and the
cumulative number counts for total fluxes in Figure \ref{fig:counts},
\ref{fig:oxygencounts}, \ref{fig:niicorr}, and \ref{fig:allemitters} and Tables
\ref{tab:counts}-\ref{tab:sb}, according to the specific
apertures set in each survey. We emphasize that the optimal aperture suggested here
($R\sim0.5''$) is rather large by space standards, corresponding to
$\sim5\times$ the full width half maximum of \textit{HST}/ACS point
spread function.

We warn the reader that several other effects might reduce the possible impact
of these findings. First, our sizes are not directly measured on \ha\
emission line maps, but based on the UV rest-frame proxy, and it is
perhaps a surprising finding that aperture losses are so large even
with a $R\sim0.5"$ aperture on images with the typical \textit{HST} spatial
resolution. We cannot
rule out that individual bright emitters might be more compact than
the median we show in Figure  \ref{fig:stack}, although the
attenuation of UV continuum light is
expected to be fully comparable to that of \ha, and both are tracing
SFRs. Then, for low spectral resolution
observations, line blending (i.e., \nii+\ha) will boost the number
counts. On the other hand, resolving the emission lines, as it might
be expected for longslit or fiber spectroscopy from the ground, would
cause the opposite effect, reducing the signal-to-noise per resolution element.
Finally, AGN and starbursting galaxies can further increase the number counts in the
brightest tail, considering their expected compact emission and high
EW. We caution the reader that this is a simple experiment based on a specific class of bright \ha\ emitters,
with an average radially symmetric shape, a disk-like light
profile, and a typical \textit{HST}/ACS point spread function. The effect of seeing and the exact PSF shape of each set of observations
can be modeled convolving the profile in Figure \ref{fig:stack}, assessing its effect on the optimal aperture.
Future simulations might address several open issues
with detailed descriptions of the specific characteristics of each
survey, which is beyond the scope of this work.

\section{Conclusions}
\label{sec:conclusions}
We have shown that fluxes of rest-frame optical emission lines can be reliably
estimated for thousands of galaxies on the basis of good quality
multicolor photometry. We have further explored one of the
possible applications of having this information for large samples of
galaxies, namely to establish number counts and to investigate the
observable and physical properties of line emitters that will
be observed by cosmological surveys. In particular:
\begin{itemize}
\item We accurately predicted \ha\ fluxes for a sample of
  color-selected SFGs in COSMOS and GOODS-S at redshift $1.4<z<1.8$
  based on their SFRs and dust attenuation estimates from SED
  modeling. These galaxies fairly represent the normal main sequence
  population at this redshift. We calibrated the predicted fluxes against
  spectroscopic observations from the FMOS-COSMOS survey. The
  statistical uncertainty on the final
  predicted fluxes is $\sigma_{\rm Pred} \sim 0.1-0.2$~dex (Figure \ref{fig:hapredictions}).
\item We predicted the fluxes of the \hb, \oii, and \oiii\ lines
  applying simple empirical recipes and calibrating with
  spectroscopically confirmed galaxies from the FMOS-COSMOS survey and
  data publicly available.
\item We computed the cumulative number counts of \ha\ emitters in the
redshift range $1.4<z<1.8$, finding a broad agreement with existing data in literature
and the empirical curves by
\cite{pozzetti_2016} modeling the evolution of the \ha\ luminosity
function with redshift (Figure \ref{fig:counts}). We obtain fully consistent results when we
properly take into account the uncertainty on the predicted \ha\
fluxes, effectively enhancing the number counts at large fluxes.
\item We extended the \ha\ number counts to the redshift range $0.9<z<1.8$
  covered by future surveys such as Euclid and WFIRST. We adopted a
  physically motivated approach, modeling the evolution of the main
  sequence of galaxies with redshift and including the effect of the
  luminosity distance on the observed fluxes. This method provides
  results consistent with models and datasets in literature, while
  returning $\sim1.5\times$ higher counts for fluxes up to
  $\geq2\times10^{-16}$~\esc\ than a simple volume rescaling.
\item We argue that the evolution of the MS of galaxies is enough to reproduce the
  time evolution of the differential number counts $dN/dz$ in the range $0.2<z<2.5$, 
  in good agreement with the current data (Figure \ref{fig:dNdz}).  
\item We computed the number counts for \oii\ and \oiii\ emitters in
  the redshift range $1.4<z<1.8$, extending the predictions to lower
  fluxes (Figure \ref{fig:oxygencounts}). Our estimates of \oiii\ counts are in agreement with previous works once the
  effect of low-mass galaxies is taken into account. On the other hand, we revise towards 
  lower values the tail of the brightest \oii\ emitters at high redshift.
\item We investigated the properties of the typical \ha\ emitters
  visibile in future wide spectroscopic surveys with observed \ha\ fluxes
  $\geq2\times10^{-16}$~\esc. We find them massive
  ($\rm{log}(M_\star/M_\odot) \rangle = 10.7\pm 0.4$), luminous in
  observed optical and near-IR bands, and with extended UV sizes ($R_{\rm
    e}\sim0.48"=4.4$~kpc at $z\sim1.5$). We estimate average \nii/\ha\
 ratio and rest-frame
  EW(\ha) of log(\nii/\ha)$=-0.52\pm0.01$ and
  log[EW(\ha)]$=2.05\pm0.01$, respectively.
\item We examine caveats and possible extensions of this work,
  including potential counts boosting or decrease by several
  factors. Failing at resolving the \nii\ emission or the inclusion of AGN and
  low mass, unobscured, starbursting galaxies with large sSFR and EW
  might enhance the counts of bright emitters. The impact of low-mass,
  high-sSFR galaxies is particularly strong on the number counts of
  oxygen emitters ($\sim50$\% increase for fluxes
    $\geq1\times10^{-16}$~\esc). 
\item We further discuss the possible optimization of
  sources detection and explore the relation between the ``nominal''
  and ``effective'' depths of a set of observations. We show how the
  latter is relevant to estimate the ``return'' of a survey in terms
  of recoverable spectroscopic redshifts. We find that an ``optimal''
  circular aperture of $R\sim0.5$'' maximizes the signal-to-noise, causing 
  a factor of $\sim2\times$ flux losses that can correspond to a drop of the 
  return, if neglected. 
\item We release a catalog containing all the relevant photometric
  properties and the line fluxes used in this work. 
\end{itemize}

\section*{Acknowledgements}
We acknowledge the constructive comments from the anonymous referee,
which significantly improved the content and presentation of the
results. We thank Georgios Magdis for useful discussions throughout the elaboration of this work.
We also thank Melanie Kaasinen and Lisa Kewley for providing the total
\oii\ fluxes from their observing program. This work is based on data
collected at Subaru Telescope, which is operated by the National
Astronomical Observatory of Japan. The authors wish to recognize and
acknowledge the very significant cultural role and reverence that the
summit of Mauna Kea has always had within the indigenous Hawaiian
community.  We are most fortunate to have the opportunity to conduct
observations from this mountain. FV acknowledges the Villum
Fonden research grant 13160 ``Gas to stars, stars to
dust: tracing star formation across cosmic time''.
AC and LP acknowledge the grants ASI n.I/023/12/0 ``Attivit\`{a} relative alla fase B2/C
per la missione Euclid'' and MIUR PRIN 2015 ``Cosmology and Fundamental 
Physics: illuminating the Dark Universe with Euclid''.
%%%%%%%%%%%%%%%%%%%%%%%%%%%%%%%%%%%%%%%%%%%%%%%%%%



\bibliographystyle{mnras}
\bibliography{mnras_ha_counts_valentino_vfinal} % if your bibtex file is called example.bib

\begin{landscape}
  \centering
  \begin{table}
    \caption{Cumulative number counts of \ha\ emitters from the COSMOS and GOODS-S
      photometric samples.}
    \label{tab:counts}
    \begin{threeparttable}
      \begin{tabular}{ccccccccccccccccc}
        \hline
        Flux limit & \multicolumn{8}{c}{COSMOS}& \multicolumn{8}{c}{GOODS-S}\\
        \scriptsize[$10^{-16}$~\esc] & &\\
                   &\multicolumn{4}{c}{$1.4<z<1.8$}  & \multicolumn{4}{c}{$0.9<z<1.8$} &\multicolumn{4}{c}{$1.4<z<1.8$}  & \multicolumn{4}{c}{$0.9<z<1.8$}\\
        \cmidrule(lr){2-5}\cmidrule(lr){6-9}\cmidrule(lr){10-13}\cmidrule(lr){14-17} 
                   & Average\tnote{a} & $\sigma_{\rm conv}$\tnote{b}& $\sigma_{\rm P,68}$\tnote{c}& $\sigma_{\rm MC}$\tnote{d}& Average & $\sigma_{\rm conv}$ &$\sigma_{\rm P,68}$& $\sigma_{\rm MC}$& Average & $\sigma_{\rm conv}$ &$\sigma_{\rm P,68}$& $\sigma_{\rm MC}$& Average & $\sigma_{\rm conv}$ &$\sigma_{\rm P,68}$& $\sigma_{\rm MC}$\\
        \tiny & \scriptsize [deg$^{-2}$] & \scriptsize [deg$^{-2}$] & \scriptsize [deg$^{-2}$]   &  \scriptsize [deg$^{-2}$] & \scriptsize [deg$^{-2}$] & \scriptsize [deg$^{-2}$]  & \scriptsize [deg$^{-2}$]  & \scriptsize [deg$^{-2}$] &\scriptsize [deg$^{-2}$] & \scriptsize [deg$^{-2}$] & \scriptsize [deg$^{-2}$] &\scriptsize [deg$^{-2}$] & \scriptsize [deg$^{-2}$] & \scriptsize [deg$^{-2}$] & \scriptsize [deg$^{-2}$]& \scriptsize [deg$^{-2}$]\\
        \hline
        $\geq0.10$   &     --           &      --         &     --            &   --             &     --            &   --               &    --               &    --               &  $31635$ &   $\pm195$ &  $ \pm 776 $ &      $ \pm  2485 $ &  $69400$ &   $\pm449$ &    $ \pm 1140 $ &    $ \pm 5053 $ \\                                                    
        $\geq0.25$   &     --           &       --        &     --            &   --             &     --            &   --               &    --               &    --               &  $16936$ &   $\pm99$   &  $ \pm 573 $ &      $ \pm  2024 $ &  $39661$ &   $\pm18$   &   $ \pm 869 $   &    $ \pm 4239 $ \\                                                     
        $\geq0.50$   & $9318$       & $\pm78$   &     $\pm78$ &   $\pm250$ &   $21422$    &   $\pm357$  &   $ \pm118 $ &   $ \pm576 $ &   $8467$   &   $\pm344$ &  $ \pm 404 $ &      $ \pm  1207 $ &  $20999$ &   $\pm582$ &    $ \pm 629 $   &    $ \pm 2776 $ \\                                                       
        $\geq0.75$   & $5671$       & $\pm231$ &     $\pm59$ &   $\pm132$ &   $13994$    &   $\pm310$  &   $ \pm93 $   &   $ \pm342 $ &   $5107$   &   $\pm270$ &  $ \pm 316 $ &      $ \pm  675 $   &  $12809$ &   $\pm745$ &    $ \pm 488 $   &    $ \pm 1709 $ \\                                                        
        $\geq1.0$     & $3555$       & $\pm324$ &     $\pm46$ &   $\pm86$   &   $9328$      &   $\pm594$  &   $ \pm75 $   &   $ \pm216 $ &   $3225$   &   $\pm326$ &  $ \pm 249 $ &      $ \pm  425 $   &  $8487$   &   $\pm702$ &    $ \pm 396 $   &    $ \pm 1094 $ \\                                                          
        $\geq1.5$     & $1490$       & $\pm327$ &     $\pm28$ &   $\pm50$   &   $4468$      &   $\pm643$  &   $ \pm50 $   &   $ \pm118 $ &   $1524$   &   $\pm250$ &  $ \pm 172 $ &      $ \pm  216 $   &  $4283$   &   $\pm552$ &    $ \pm 280 $   &    $ \pm 554 $ \\                                                           
        $\geq2.0$     & $706$         & $\pm228$ &     $\pm18$ &   $\pm26$   &   $2301$      &   $\pm545$  &   $ \pm34 $   &   $ \pm68 $   &   $823$     &   $\pm177$ &  $ \pm 128 $ &      $ \pm  104 $   &  $2421$   &   $\pm429$ &    $ \pm 210 $   &    $ \pm 296 $ \\                                                            
        $\geq2.5$     & $364$         & $\pm152$ &     $\pm12$ &   $\pm15$   &   $1262$      &   $\pm419$  &   $ \pm24 $   &   $ \pm40 $   &   $481$     &   $\pm131$ &  $ \pm 100 $ &      $ \pm  63 $     &  $1471$   &   $\pm337$ &    $ \pm 163 $   &    $ \pm 165 $ \\                                                              
        $\geq3.0$     & $199$         & $\pm102$ &     $\pm8$   &   $\pm10$   &   $737$        &   $\pm304$  &   $ \pm17 $   &   $ \pm24 $   &   $309$     &   $\pm88$   &  $ \pm 84 $   &      $ \pm  38 $     &  $958$     &   $\pm253$ &   $ \pm 133 $   &    $ \pm 100 $ \\                                                                
        $\geq3.5$     & $114$         & $\pm70$   &     $\pm6$   &   $\pm7$     &   $449$        &   $\pm219$  &   $ \pm13 $   &   $ \pm16 $   &   $208$     &   $\pm60$   &  $ \pm 72 $   &      $ \pm  25 $     &  $655$     &   $\pm190$ &    $ \pm 112 $   &    $ \pm 64 $ \\                                                                  
        $\geq4.0$     & $69$           & $\pm47$   &     $\pm4$   &   $\pm4$     &   $284$        &   $\pm159$  &   $ \pm10 $   &   $ \pm11 $   &   $140$     &   $\pm47$   &  $ \pm 62 $   &      $ \pm  18 $     &  $460$     &   $\pm148$ &    $ \pm 95 $     &    $ \pm 43 $ \\                                                                   
        $\geq4.5$     & $44$           & $\pm32$   &     $\pm3$   &   $\pm3$     &   $187$        &   $\pm114$  &   $ \pm7 $     &   $ \pm8 $     &   $94$       &   $\pm39$   &  $ \pm 54 $   &      $ \pm  13 $     &  $336$     &   $\pm114$ &    $ \pm 84 $     &    $ \pm 32 $ \\                                                                    
        $\geq5.0$     & $29$           & $\pm22$   &     $\pm3$   &   $\pm2$     &   $126$        &   $\pm84$    &   $ \pm6 $     &   $ \pm6$      &   $76$       &   $\pm20$   &  $ \pm 54 $   &      $ \pm  8 $       &  $254$     &   $\pm87$   &    $ \pm 76 $     &    $ \pm 25 $ \\                                                                      
        $\geq7.5$     & $5$             & $\pm4$     &     $\pm1$   &   $\pm1$     &   $25$          &   $\pm19$    &   $ \pm3 $     &   $ \pm2 $     &   $11$       &   $\pm11$   &  --                 &       --                  &  $73$       &   $\pm34$   &   $ \pm 27 $    &    $ \pm 5 $ \\                                                                             
        \hline
      \end{tabular}
      \begin{tablenotes}
      \item[a] Mean of the convolved and unconvolved number counts
        (Section \ref{sec:fmosnumbercounts}). \textit{Note: The lower
        and upper (convolved, ``broad'') counts shown in Figure
        \ref{fig:counts} can be obtained subtracting and adding the
        absolute error $\sigma_{\rm conv}$ to the counts reported in
        this column.}
      \item[b] Absolute error associated with the convolution of the
        lower counts with a Gaussian curve $0.19$~dex wide ($\rm{[convolved\,counts - unconvolved\,counts]/2}$, Section
        \ref{sec:fmosnumbercounts}).
      \item[c] Poissonian $68$\% confidence interval of the unconvolved counts. The naturally
        asymmetric Poissonian uncertainties have been round up to the
        highest value between the lower and upper limits. 
      \item[d] Monte Carlo bootstrap uncertainties on the unconvolved counts.
      \end{tablenotes}
    \end{threeparttable}
  \end{table}
\end{landscape}

\begin{landscape}
  \begin{table}
    \begin{center}
      \caption{Cumulative number counts of \oii\ and \oiii\ emitters from the COSMOS and GOODS-S
        photometric samples in the redshift range $1.4<z<1.8$.}
    \end{center}
    \label{tab:oxygencounts}
    \begin{threeparttable}
      \begin{tabular}{ccccccccccccccccc}
        \hline
        Flux limit & \multicolumn{8}{c}{COSMOS}& \multicolumn{8}{c}{GOODS-S}\\
        \scriptsize[$10^{-16}$~\esc] & &\\
                   &\multicolumn{4}{c}{\oii$\lambda\,3727$~\AA}  & \multicolumn{4}{c}{\oiii$\lambda\,5007$~\AA} &\multicolumn{4}{c}{\oii$\lambda\,3727$~\AA}  & \multicolumn{4}{c}{\oiii$\lambda\,5007$~\AA}\\        
        \cmidrule(lr){2-5}\cmidrule(lr){6-9}\cmidrule(lr){10-13}\cmidrule(lr){14-17} 
                   & Average\tnote{a} & $\sigma_{\rm conv}$\tnote{b}& $\sigma_{\rm P,68}$\tnote{c}& $\sigma_{\rm MC}$\tnote{d}& Average & $\sigma_{\rm conv}$ &$\sigma_{\rm P,68}$& $\sigma_{\rm MC}$& Average & $\sigma_{\rm conv}$ &$\sigma_{\rm P,68}$& $\sigma_{\rm MC}$& Average & $\sigma_{\rm conv}$ &$\sigma_{\rm P,68}$& $\sigma_{\rm MC}$\\
        \tiny & \scriptsize [deg$^{-2}$] & \scriptsize [deg$^{-2}$] & \scriptsize [deg$^{-2}$]   & \scriptsize [deg$^{-2}$]  & \scriptsize [deg$^{-2}$] & \scriptsize [deg$^{-2}$]  & \scriptsize [deg$^{-2}$]  &  \scriptsize [deg$^{-2}$] &\scriptsize [deg$^{-2}$] & \scriptsize [deg$^{-2}$] & \scriptsize [deg$^{-2}$] & \scriptsize [deg$^{-2}$] & \scriptsize [deg$^{-2}$] & \scriptsize [deg$^{-2}$] & \scriptsize [deg$^{-2}$] &  \scriptsize [deg$^{-2}$] \\
        \hline

      $\geq0.10$   &    --          &   --           &    --              &    --              &   --          &   --              &    --                &    --                 &          $23727$ &         $\pm317$ & $ \pm 672 $ &  $ \pm 2016 $ & $41401$ & $\pm968$   & $ \pm 898 $ &  $ \pm 2415 $\\
      $\geq0.25$   &    --          &   --           &    --              &    --              &   --          &   --              &    --                &    --                 &          $8705$   &         $\pm527$ & $ \pm 405 $ &  $ \pm 1071 $ & $17874$ & $\pm668$   & $ \pm 579 $ &  $ \pm 1653 $\\
      $\geq0.50$   &   $2573$ &   $\pm426$ & $ \pm 37 $  &   $ \pm 77 $ &  $5342$   &   $\pm388$ &   $ \pm 57 $  &    $ \pm 129 $ &             $2729$   &         $\pm495$ & $ \pm 221 $ &  $ \pm 386 $   & $6084$   & $\pm1062$ & $ \pm 322 $ &  $ \pm 676 $\\              
      $\geq0.75$   &   $981$   &   $\pm362$ & $ \pm 20 $  &   $ \pm 28 $ &  $2272$   &   $\pm711$ &   $ \pm 32 $  &    $ \pm 40 $   &             $1093$   &         $\pm354$ & $ \pm 135 $ &  $ \pm 137 $   & $2385$   & $\pm963$   & $ \pm 180 $ &  $ \pm 271 $\\    
      $\geq1.0$     &   $431$   &   $\pm239$ & $ \pm 12 $  &   $ \pm 16 $ &  $1083$   &   $\pm574$ &   $ \pm 19 $  &    $ \pm 25 $   &             $482$     &         $\pm261$ & $ \pm 84 $   &  $ \pm 74 $     & $1079$   & $\pm673$   & $ \pm 106 $ &  $ \pm 133 $\\      
      $\geq1.5$     &   $116$   &   $\pm90$   & $ \pm 5 $    &   $ \pm 5$    &  $334$     &   $\pm262$ &   $ \pm 7 $    &    $ \pm 9 $     &             $163$     &         $\pm90$   & $ \pm 58 $   &  $ \pm 22 $     & $332$     & $\pm259$   & $ \pm 58 $   &  $ \pm 41 $\\       
      $\geq2.0$     &   $42$     &   $\pm35$   & $ \pm 3 $    &   $ \pm 2 $   &  $132$     &   $\pm118$ &   $ \pm 4 $    &    $ \pm 3 $     &             $73$       &         $\pm37$   & $ \pm 48 $   &  $ \pm 9 $       &  --          &     --            &   --                &   --\\
      $\geq2.5$     &   $17$     &   $\pm16$   & $ \pm 2$     &   $ \pm 1 $   &  $60$       &   $\pm57$   &   $ \pm 2 $    &    $ \pm 2 $     &             $38$       &         $\pm19$   & $ \pm 42 $   &  $ \pm 1 $       &  --          &     --            &   --                &   --\\      
      $\geq3.0$     &    --          &   --           &    --             &   --              &  $31$       &   $\pm29$   &  $ \pm 2 $     &    $ \pm 1 $     &             --            &         --              &   --              &   --                  &  --          &     --            &   --                &   --\\  

      \hline
      \end{tabular}
      \begin{tablenotes}
      \item[a] Mean of the convolved and unconvolved number counts
        (Section \ref{sec:fmosnumbercounts}). \textit{Note: The lower
        and upper (convolved, ``broad'') counts shown in Figure
        \ref{fig:oxygencounts} can be obtained subtracting and adding the
        absolute error $\sigma_{\rm conv}$ to the counts reported in
        this column.}
      \item[b] Absolute error associated with the convolution of the lower counts with Gaussian curves $0.22$~dex and $0.25$~dex
        wide for \oii\ and \oiii, respectively ($\rm{[convolved\,counts - unconvolved\,counts]/2}$, Section \ref{sec:fmosnumbercounts}).
      \item[c] Poissonian $68$\% confidence interval of the lower counts. The naturally
        asymmetric Poissonian uncertainties have been round up to the
        highest value between the lower and upper limits. 
      \item[d] Monte Carlo bootstrap uncertainties on the lower counts.      
      \end{tablenotes}
    \end{threeparttable}
  \end{table}
\end{landscape}

\begin{landscape}
  \begin{table}
    \centering
    \caption{Cumulative number counts of \ha$+$\nii\ emitters from the COSMOS and GOODS-S
      photometric samples.}
    \label{tab:hanii}
    \begin{threeparttable}
      \begin{tabular}{ccccccccccccc}
        \hline
        Flux limit & \multicolumn{6}{c}{COSMOS}& \multicolumn{6}{c}{GOODS-S}\\
        \scriptsize[$10^{-16}$~\esc] & &\\
                   &\multicolumn{3}{c}{$1.4<z<1.8$}  & \multicolumn{3}{c}{$0.9<z<1.8$} &\multicolumn{3}{c}{$1.4<z<1.8$}  & \multicolumn{3}{c}{$0.9<z<1.8$}\\
        \cmidrule(lr){2-4}\cmidrule(lr){5-7}\cmidrule(lr){8-10}\cmidrule(lr){11-13} 
                   & \ha+\nii\tnote{a} & $\sigma_{\rm conv}$\tnote{b}& $\sigma_{\rm P,68}$\tnote{c}& \ha+\nii\ & $\sigma_{\rm conv}$& $\sigma_{\rm P,68}$ & \ha+\nii\ & $\sigma_{\rm conv}$ & $\sigma_{\rm P,68}$ & \ha+\nii\ & $\sigma_{\rm conv}$ & $\sigma_{\rm P,68}$ \\ 
        \tiny & \scriptsize [deg$^{-2}$] & \scriptsize [deg$^{-2}$] &  \scriptsize [deg$^{-2}$] & \scriptsize [deg$^{-2}$]  & \scriptsize [deg$^{-2}$] & \scriptsize [deg$^{-2}$] &  \scriptsize [deg$^{-2}$]&  \scriptsize [deg$^{-2}$] & \scriptsize [deg$^{-2}$] & \scriptsize [deg$^{-2}$] & \scriptsize [deg$^{-2}$] & \scriptsize [deg$^{-2}$]\\
        \hline
        $\geq0.10$&	-- &		-- &              -- &            -- &		-- &               -- &             $33024$ &  $\pm144$ & $\pm793$ &  $72042$ & $\pm397$  & $\pm1162$ \\    
        $\geq0.25$&	-- &		-- &              -- &            -- &		-- &               -- &             $18304$ &  $\pm120$ & $\pm595$ &  $42435$ & $\pm30$    & $\pm898$\\     
        $\geq0.50$&	$10568 $&	$\pm122$& $\pm83$ &  $23705 $  & $\pm408$ &  $\pm124$ &    $9767$ &    $\pm333$ & $\pm434$ &  $23649$ & $\pm571$  & $\pm668$\\    
        $\geq0.75$&	$7019 $  &	$\pm157$& $\pm66$ &  $16620 $  & $\pm171$ &  $\pm102$ &   $6276$ &    $\pm257$ &  $\pm350$ &  $15250$ & $\pm715$  & $\pm534$\\    
        $\geq1.0$&	$4813 $  &	$\pm250$& $\pm54$ &  $11904 $  & $\pm441$ &  $\pm86$ &     $4271$ &    $\pm301$ &  $\pm288$ &  $10700$ & $\pm660$  & $\pm447$\\    
        $\geq1.5$&	$2408 $  &	$\pm304$& $\pm37$ &  $6530  $   & $\pm563$ &  $\pm62$ &     $2300$ &    $\pm251$ &  $\pm212$ &  $6027$ &   $\pm530$  & $\pm336$\\   
        $\geq2.0$&	$1308 $  &	$\pm262$& $\pm26$ &  $3796  $   & $\pm559$ &  $\pm46$ &     $1428$ &    $\pm154$ &  $\pm172$ &  $3789$ &   $\pm406$  & $\pm267$\\   
        $\geq2.5$&	$762  $  &	$\pm204$& $\pm19$ &  $2320  $   & $\pm491$ &  $\pm35$ &     $875$ &      $\pm173$ &  $\pm133$ &  $2490$ &   $\pm366$  & $\pm216$\\   
        $\geq3.0$&	$468  $  &	$\pm154$& $\pm15$ &  $1488  $   & $\pm398$ &  $\pm27$ &     $595$ &      $\pm133$ &  $\pm111$ &  $1717$ &   $\pm313$  & $\pm179$\\   
        $\geq3.5$&	$301  $  &	$\pm114$& $\pm11$ &  $992   $    & $\pm315$ &  $\pm21$ &     $419$ &      $\pm106$ &  $\pm96$ &    $1224$ &   $\pm268$  & $\pm151$\\   
        $\geq4.0$&	$198  $  &	$\pm88$&  $\pm9$ &     $680   $    & $\pm249$ &  $\pm17$ &     $297$ &      $\pm94$ &    $\pm81$ &    $898$ &     $\pm230$  & $\pm130$\\  
        $\geq4.5$&	$136  $  &	$\pm65$&  $\pm7$ &     $480   $    & $\pm196$ &  $\pm14$ &     $233$ &      $\pm67$ &    $\pm75$ &    $690$ &     $\pm182$  & $\pm116$\\  
        $\geq5.0$&	$93   $   &	$\pm52$&  $\pm6$ &     $346   $    & $\pm155$ &  $\pm12$ &     $191$ &      $\pm44$ &    $\pm72$ &    $547$ &     $\pm142$  & $\pm106$\\  
        $\geq7.5$&	$20   $   &	$\pm15$&  $\pm3$ &     $88    $     & $\pm52$   &  $\pm5$ &       $73$ &        $\pm18$ &    $\pm32$ &    $215$ &     $\pm54$   & $\pm55$\\   		       
        \hline
      \end{tabular}
      \begin{tablenotes}
      \item[a] Mean of the convolved and unconvolved number counts of \ha+\nii\ emitters (Section \ref{sec:discussion:nii}).
      \item[b] Absolute error associated with the convolution of the
        lower counts with a Gaussian curve $0.19$~dex wide ($\rm{[convolved\,counts - unconvolved\,counts]/2}$, Section
        \ref{sec:fmosnumbercounts}).
      \item[c] Poissonian $68$\% confidence interval of the lower counts. The naturally
        asymmetric Poissonian uncertainties have been round up to the
        highest value between the lower and upper limits. 
      \end{tablenotes}
    \end{threeparttable}
  \end{table}
\end{landscape}
  
\begin{landscape}
  \begin{table}
    \centering
    \caption{Cumulative number counts of starbursting emitters from the GOODS-S photometric sample.}
    \label{tab:sb}
    \begin{threeparttable}
      \begin{tabular}{ccccccccccccc}
        \hline
        Flux limit & \multicolumn{12}{c}{GOODS-S Starburst}\\
        \scriptsize[$10^{-16}$~\esc] & \\
                   &\multicolumn{9}{c}{$1.4<z<1.8$}  & \multicolumn{3}{c}{$0.9<z<1.8$}\\
        \cmidrule(lr){2-10}\cmidrule(lr){11-13} 
                   & \ha\tnote{a} & $\sigma_{\rm conv}$\tnote{b}& $\sigma_{\rm P,68}$\tnote{c}& \oii\tnote{d} & $\sigma_{\rm conv}$\tnote{e}& $\sigma_{\rm P,68}$& \oiii\tnote{f} & $\sigma_{\rm conv}$\tnote{g} & $\sigma_{\rm P,68}$& \ha\  & $\sigma_{\rm conv}$ & $\sigma_{\rm P,68}$\\ 
        \tiny & \scriptsize [deg$^{-2}$] & \scriptsize [deg$^{-2}$] &  \scriptsize [deg$^{-2}$] & \scriptsize [deg$^{-2}$]  & \scriptsize [deg$^{-2}$] & \scriptsize [deg$^{-2}$] &  \scriptsize [deg$^{-2}$]&  \scriptsize [deg$^{-2}$] & \scriptsize [deg$^{-2}$] & \scriptsize [deg$^{-2}$] & \scriptsize [deg$^{-2}$] & \scriptsize [deg$^{-2}$] \\
        \hline
        $\geq0.10$&	 $2273$ &   $\pm18$ &     $\pm42$ &   $2040$ &        $\pm30$ &        $\pm40$ &    $2606$ &  $\pm59$ &       $\pm45$ &  $4758$ &    $\pm40$ &   $\pm60$\\                               
        $\geq0.25$&	 $1592$ &   $\pm14$ &     $\pm35$ &   $1212$ &        $\pm3$ &          $\pm31$ &    $1968$ &  $\pm48$ &       $\pm39$ &  $3469$ &    $\pm35$ &   $\pm51$\\  
        $\geq0.50$&	 $1025$ &   $\pm5$ &       $\pm28$ &   $636$ &          $\pm23$ &        $\pm22$ &    $1271$ &  $\pm2$ &         $\pm31$ &  $2324$ &    $\pm1$ &     $\pm42$\\ 
        $\geq0.75$&	 $718$ &     $\pm13$ &     $\pm23$ &   $379$ &          $\pm28$ &        $\pm17$ &    $861$ &    $\pm29$ &       $\pm25$ &  $1667$ &    $\pm28$ &   $\pm35$\\  
        $\geq1.0$&	 $529$ &     $\pm12$ &     $\pm20$ &   $236$ &          $\pm34$ &        $\pm13$ &    $602$ &    $\pm47$ &       $\pm21$ &  $1256$ &    $\pm31$ &   $\pm31$\\  
        $\geq1.5$&	 $294$ &     $\pm24$ &     $\pm15$ &   $107$ &          $\pm26$ &        $\pm 8$ &     $327$ &    $\pm48$ &       $\pm15$ &  $756$ &      $\pm40$ &   $\pm24$\\ 
        $\geq2.0$&	 $177$ &     $\pm22$ &     $\pm11$ &   $55$ &            $\pm19$ &        $\pm 6$ &     $184$ &    $\pm48$ &       $\pm11$ &  $476$ &      $\pm45$ &   $\pm18$\\ 
        $\geq2.5$&	 $109$ &     $\pm20$ &     $\pm9$ &     $30$ &            $\pm14$ &        $\pm4$ &      $110$ &    $\pm42$ &       $\pm8$ &    $311$ &      $\pm44$ &   $\pm15$\\ 
        $\geq3.0$&	 $71$ &       $\pm16$ &     $\pm7$ &     $18$ &            $\pm9$ &          $\pm3$ &      $70$ &      $\pm33$ &       $\pm6$ &    $210$ &      $\pm39$ &   $\pm12$\\ 
        $\geq3.5$&	 $44$ &       $\pm16$ &     $\pm5$ &     $11$ &            $\pm7$ &          $\pm2$ &      $44$ &      $\pm27$ &       $\pm4$ &    $142$ &      $\pm36$ &   $\pm10$\\ 
        $\geq4.0$&	 $28$ &       $\pm14$ &     $\pm4$ &     $6$ &              $\pm 6$ &         $\pm2$ &      $31$ &      $\pm20$ &       $\pm4$ &    $98$ &        $\pm32$ &   $\pm8$\\ 
        $\geq4.5$&	 $19$ &       $\pm11$ &     $\pm3$ &     -- &                -- &                   -- &               $21$ &      $\pm17$ &       $\pm2$ &    $70$ &        $\pm27$ &   $\pm6$\\            
        $\geq5.0$&	 $14$ &       $\pm8$ &       $\pm3$ &     -- &                -- &                   -- &               $15$ &      $\pm13$ &       $\pm2$ &    $51$ &        $\pm22$ &   $\pm5$\\              
        $\geq7.5$&	 $3$ &         $\pm2$ &       $\pm1$ &     -- &                -- &                   -- &               -- &          -- &                 -- &             $13$ &        $\pm8$ &     $\pm2$\\        
        \hline
      \end{tabular}
      \begin{tablenotes}
      \item[a] Mean of the convolved and unconvolved number counts of \ha\ starbursting emitters (Section \ref{sec:discussion:sb}).
      \item[b] Absolute error associated with the convolution of the
        \ha\ unconvolved counts with a Gaussian curve $0.19$~dex wide ($\rm{[convolved\,counts - unconvolved\,counts]/2}$, Section \ref{sec:fmosnumbercounts}).
      \item[c] Poissonian $68$\% confidence interval of the lower counts. The naturally
        asymmetric Poissonian uncertainties have been round up to the
        highest value between the lower and upper limits. 
      \item[d] Mean of the convolved and unconvolved number counts of \oii\ starbursting emitters.
      \item[e] Absolute error associated with the convolution of the
        \oii\ lower counts with a Gaussian curve $0.22$~dex wide.
      \item[f] Mean of the convolved and unconvolved number counts of \oiii\ starbursting emitters.
      \item[g] Absolute error associated with the convolution of the
        \oiii\ lower counts with a Gaussian curve $0.25$~dex wide.
      \end{tablenotes}
    \end{threeparttable}
  \end{table}
\end{landscape}

%%%%%%%%%%%%%%%%%%%%%%%%%%%%%%%%%%%%%%%%%%%%%%%%%%

%%%%%%%%%%%%%%%%% APPENDICES %%%%%%%%%%%%%%%%%%%%%

\appendix

\section{Individual bright \ha\ emitters}
\label{appendixB}
\begin{figure*}
  \includegraphics[width=0.95\textwidth]{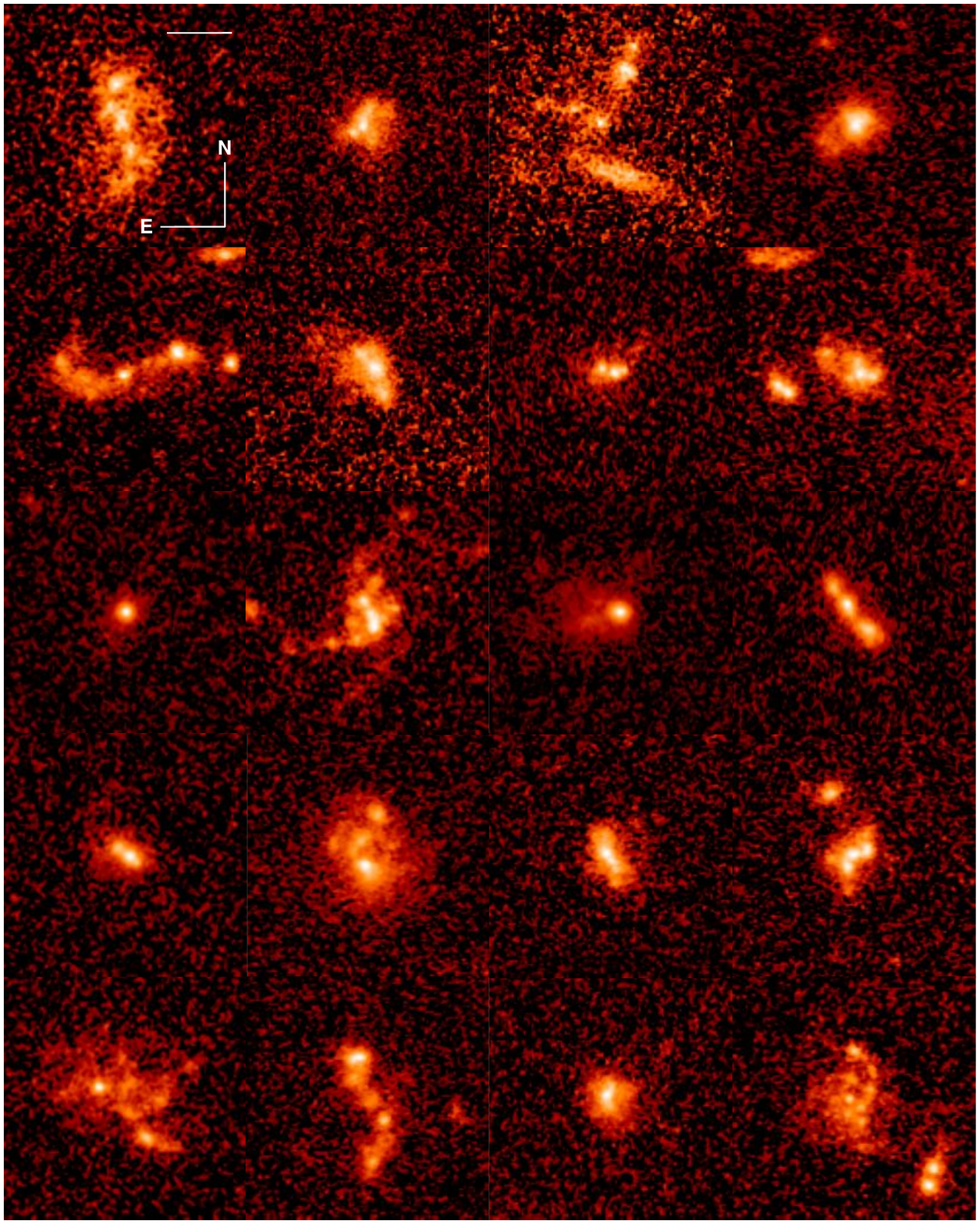}
  \caption{\textbf{\textit{HST}/$i_{814}$ cutouts of bright \ha\
      emitters in COSMOS.} The cutouts show a random sample of
    emitters with predicted \ha\ fluxes $\geq2\times10^{-16}$~\esc\ at
    $1.4<z<1.8$. The size is $3.75"\times3.75"$. The images are
    aligned North-East and they are scaled to the same background
    level. The white bar shown in the top-left panel is 1'' long.} 
\label{fig:emitters}
\end{figure*}

We show in Figure \ref{fig:emitters} a random selection of \ha\
emitters with predicted \ha\ fluxes $\geq2\times10^{-16}$~\esc. The
images are in the \textit{HST}/$i_{814}$ band.

\section{Catalog of line fluxes predictions}
\begin{landscape}
  \begin{table}
    \begin{center}
      \caption{Catalog of relevant SED-derived quantities and emission
      line flux predictions for the COSMOS sample (a machine-readable version is available online).}
    \end{center}
    \label{tab:predictioncosmos}
    \begin{threeparttable}
      \begin{tabular}{cccccccccccc}
        \hline
        ID\tnote{a} & RA & DEC & log$_{10}(M_\star)$& log$_{10}(\rm{SFR})$& $z_{\rm  phot}$& $A_{\rm V}$& \ha\ & \oii\ &  \oiii& \hb\ & $f_{\rm H\beta}$\tnote{b}\\  
        & [deg]& [deg]& [\msun] & [\msun~yr$^{-1}$]& & [mag]& \scriptsize [$10^{-16}$~\esc] & \scriptsize [$10^{-16}$~\esc]& \scriptsize [$10^{-16}$~\esc]& \scriptsize [$10^{-16}$~\esc]& \\
        \hline
         219860&   150.322350&     1.61483180&           9.69&       0.92&               1.64&           0.70&           0.29&           0.16&           0.27&           0.08&           1.00\\
         219985&   149.984920&     1.61497840&           9.46&       0.58&               1.62&           0.20&           0.23&           0.19&           0.31&           0.07&           1.00\\
         220037&   149.890400&     1.61494710&           9.63&       0.94&               1.63&           0.70&           0.31&           0.17&           0.30&           0.08&           1.00\\
         220136&   150.353750&     1.61500880&          10.45&       1.88&              1.60 &          2.00 &          0.76 &          0.13 &          0.19 &          0.11 &          1.15\\
         220152&   149.759090&     1.61519360&           9.67&       0.78&              1.51&           0.80&           0.24&           0.12&           0.22&           0.06&           1.00\\
          \dots&            \dots&             \dots&               \dots&     \dots&             \dots&         \dots &         \dots &        \dots &      \dots&           \dots&         \dots\\
      \hline
      \end{tabular}
      \begin{tablenotes}
      \item[a] Galaxy ID from \citep{laigle_2016}.
      \item[b] Stellar absorption correction factor (Section \ref{sec:hb}).     
      \end{tablenotes}
    \end{threeparttable}
  \end{table}
\end{landscape}
\begin{landscape}
  \begin{table}
    \begin{center}
      \caption{Catalog of relevant SED-derived quantities and emission
      line flux predictions for the GOODS-S sample (a machine-readable version is available online).}
    \end{center}
    \label{tab:predictiongoods}
    \begin{threeparttable}
      \begin{tabular}{cccccccccccc}
        \hline
        ID & RA & DEC & log$_{10}(M_\star)$& log$_{10}(\rm{SFR})$& $z_{\rm  phot}$& $A_{\rm V}$& \ha\ & \oii\ &  \oiii& \hb\ & $f_{\rm H\beta}$\tnote{a}\\  
        & [deg]& [deg]& [\msun] & [\msun~yr$^{-1}$]& & [mag]& \scriptsize [$10^{-16}$~\esc] & \scriptsize [$10^{-16}$~\esc]& \scriptsize [$10^{-16}$~\esc]& \scriptsize [$10^{-16}$~\esc]& \\
        \hline
        1& 53.08488800&  -27.95581300&      9.51&      1.40&      1.59&      1.00&      0.71&      0.29&      0.66&      0.16&      1.00\\
        2& 53.09927000&  -27.95315400&      9.51&      0.55&      1.70&      0.40&      0.15&      0.11&      0.19&      0.05&      1.00\\
        3& 53.07999000&  -27.95205100&     10.13&      1.02&      1.47&      1.60&      0.20&      0.05&      0.08&      0.03&      1.06\\
        4& 53.10614800&  -27.95160700&      9.34&      0.96&      1.66&      0.20&      0.52&      0.43&      0.76&      0.17&      1.00\\
        5& 53.10593400&  -27.95165800&      9.54&      1.14&      1.73&      0.40&      0.58&      0.40&      0.69&      0.17&      1.00\\
        6& 53.09929700&  -27.94932000&      8.74&      0.17&      1.55&      0.60&      0.07&      0.04&      0.11&      0.02&      1.00\\
       \dots&    \dots&             \dots&   \dots&     \dots& \dots&  \dots&  \dots&  \dots&  \dots&  \dots&    \dots\\
      \hline
      \end{tabular}
      \begin{tablenotes}
      \item[a] Stellar absorption correction factor (Section \ref{sec:hb}).     
      \end{tablenotes}
    \end{threeparttable}
  \end{table}
\end{landscape}


%%%%%%%%%%%%%%%%%%%%%%%%%%%%%%%%%%%%%%%%%%%%%%%%%%

% Don't change these lines
\bsp	% typesetting comment
\label{lastpage}
\end{document}